\documentclass[useAMS,usenatbib,a4paper]{mn2e}
\usepackage{graphicx,natbib}
\pdfoutput=1
\usepackage{rotating}
\usepackage{subfig}
\usepackage{lscape}
\usepackage{amssymb}
\usepackage{longtable}
\usepackage[pass]{geometry}

\title[Eruptive variable protostars from VVV]{A population of eruptive variable protostars in VVV}
\author[C. Contreras Pe\~{n}a et al.: Extreme Infrared Variables from VVV.]{C. Contreras Pe\~{n}a$^{1,4,2}$,\thanks{E-mail:cecontrep@gmail.com (CCP)} P. W. Lucas$^{2}$, D. Minniti$^{1,7}$, R. Kurtev$^{3,4}$, W. Stimson$^{2}$, \newauthor 
C. Navarro Molina$^{4,3}$, J. Borissova$^{3,4}$, M. S. N. Kumar$^{2}$,  M.A. Thompson$^{2}$, T. Gledhill$^{2}$, \newauthor
R. Terzi$^{2}$, D. Froebrich$^{5}$, and A. Caratti o Garatti$^{6}$\\
$^{1}$Departamento de Ciencias Fisicas, Universidad Andres Bello, Republica 220, Santiago, Chile\\
$^{2}$Centre for Astrophysics Research, University of Hertfordshire, Hatfield, AL10 9AB, UK\\
$^{3}$Instituto de F\'{i}sica y Astronom\'{i}a, Universidad de Valpara\'{i}so, ave. Gran Breta\~{n}a, 1111, Casilla 5030, Valpara\'{i}so, Chile\\
$^{4}$Millennium Institute of Astrophysics, Av. Vicuna Mackenna 4860, 782-0436, Macul, Santiago, Chile\\
$^{5}$Centre for Astrophysics and Planetary Science, University of Kent, Canterbury CT2 7NH, UK\\
$^{6}$Dublin Institute for Advanced Studies, School of Cosmic Physics, Astronomy \& Astrophysics Section,
31 Fitzwilliam Place, Dublin 2, Ireland \\
$^{7}$Vatican Observatory, V00120 Vatican City State, Italy}
\begin{document}

\date{\today}

\pagerange{\pageref{firstpage}--\pageref{lastpage}} \pubyear{2002}

\maketitle

\label{firstpage}

\begin{abstract}
We present the discovery of 816 high amplitude infrared variable stars ($\Delta K_{\rm s} >$ 1 mag) in 119 deg$^2$ of the Galactic midplane covered by the Vista Variables in the Via Lactea (VVV) survey. Almost all are new discoveries and about 50$\%$ are YSOs. This provides further evidence that YSOs are the commonest high amplitude infrared variable stars in the Galactic plane. In the 2010-2014 time series of likely YSOs we find that the amplitude of variability increases towards younger evolutionary classes (class I and flat-spectrum sources) except on short timescales ($<$25 days) where this trend is reversed. Dividing the likely YSOs by light curve morphology, we find 106 with eruptive light curves, 45 dippers, 39 faders, 24 eclipsing binaries, 65 long-term periodic variables (P$>$100 days) and 162 short-term variables. Eruptive YSOs and faders tend to have the highest amplitudes and eruptive systems have the reddest SEDs. Follow up spectroscopy in a companion paper verifies high accretion rates in the eruptive systems. Variable extinction is disfavoured by the 2 epochs of colour data. These discoveries increase the number of eruptive variable YSOs by a factor of at least 5, most being at earlier stages of evolution than the known FUor and EXor types. We find that eruptive variability is at least an order of magnitude more common in class I YSOs than class II YSOs. Typical outburst durations are 1 to 4 years, between those of EXors and FUors. They occur in 3 to 6\% of class I YSOs over a 4 year time span.
\end{abstract}


\section{Introduction}\label{vvv:intro}
   
          The VISTA Variables in the Via Lactea
\citep[VVV,][]{2010Minniti} survey has mapped a 560 deg$^{2}$ area
containing $\sim 3 \times 10^{8}$ point sources with multi-epoch
near-infrared photometry. The surveyed area includes the Milky Way
bulge and an adjacent section of the mid plane. The survey has already
produced a deep near-infrared Atlas in 5 bandpasses ($Z$, $Y$, $J$, $H$, $K_{\rm s}$), and the final
product will include a 2nd epoch of the multi-filter data and a catalogue of
more than $10^{6}$ variable sources.
           
           One of the main scientific goals expected to arise from the
final product of VVV is the finding of rare variable sources such as
Cataclysmic Variables and RS CVn stars, among others \citep[see][for a
discussion on classes of near-infrared variable stars that are being
studied with VVV]{2013Catelan}. One of the most important outcomes
is the possibility of finding eruptive variable Young Stellar Objects
(YSOs) undergoing unstable accretion. Such objects are usually
assigned to one of  two sub-classes: FUors, named after FU Orionis,
have long duration outbursts (from tens to hundreds of years); EXors, named for EX Lupi, have outbursts
of much shorter duration (from few weeks to several months). Both categories were  optically defined in
the first instance and fewer than 20 are known in total \citep[see e.g.,][]{2010Reipurth,2013Scholz,2014Audard}, very likely
because YSOs with high accretion rates tend to suffer high optical
extinction by circumstellar matter. For thorough reviews of the theory
and observations in this subject see \citet{1996Hartmann,2010Reipurth,2014Audard}. Given that VVV is the first near
infrared time domain survey of a large portion of the Galaxy, it is
reasonable to hope for a  substantial yield of new eruptive variable
YSOs in the dataset. In particular, we would expect the survey to
probe for high amplitude variability that occurs on  typical
timescales of up to a few years, which corresponds more to EXors (or
their younger, more obscured counterparts) than to FUors. Eruptive
variable YSOs are  important because it is thought that highly
variable accretion may be common amongst protostars, though rarely
observed owing to a duty cycle consisting of long  periods of slow
accretion and much shorter periods of unstable accretion at a much
higher rate. If this is true, it might explain both the observed
under-luminosity of low-mass, class I YSOs (the ``Luminosity problem''
\citep[see e.g][]{1990Kenyon,2009Evans,2012Caratti}) and the wide
scatter seen in the Hertzsprung-Russell (HR) diagrams of pre-main-sequence (PMS) clusters
\citep*{2009Baraffe,2012Baraffe}.

         In the search for this rare class of eruptive variable stars, \citet{2014Contreras} studied near-infrared high-amplitude variability in the Galactic plane using the two epochs of UKIDSS Galactic Plane Survey (UGPS) K band data \citep{2007Lawrence,2008Lucas}. \citeauthor{2014Contreras} found that $\sim 66\%$ of high-amplitude variable stars selected from UGPS data releases DR5 and DR7 are located in star forming regions (SFRs) and have characteristics of YSOs. They concluded that YSOs likely dominate the Galactic disc population of high-amplitude variable stars in the near-infrared. Spectroscopic follow-up confirmed four new additions to the eruptive variable class. These objects showed a mixture of the characteristics of the optically-defined EXor and FUor subclasses. Two of them were deeply embedded sources with very steep 1 to 5~$\mu$m spectral energy distributions (SEDs), though showing 
``flat spectrum'' SEDs at longer wavelengths. Such deeply embedded eruptive variables are regarded as a potentially distinct additional sub-class, though only a few had been detected previously: OO Ser, V2775 Ori, HOPS 383 and GM Cha \citep[see][]{1996Hodapp, 2007Kospal, 2007Persi, 2011Caratti, 2015Safron}.
         With the aims of determining the incidence of eruptive variability among YSOs and characterising the phenomenon, we have undertaken a search of the multi-epoch VVV dataset. In contrast to UGPS, the ongoing VVV survey offers several dozen epochs of K$_{\rm s}$ data over a time baseline of a few years. We expect that the VVV survey will also be used to identify YSOs by the common low
amplitude variability seen in nearly all such objects \citep[e.g.][]{2012Rice}. This
will complement studies in nearby star formation regions and in external galaxies, such as
the {\it Spitzer} YSOVAR programme \citep[e.g. ][]{2015Wolk} and a 2 epoch study of the LMC with
{\it Spitzer} SAGE survey data \citep{2009Vijh}.

         We have divided the results of this work in two publications. In this first study we present the method of the search and a general discussion on the photometric characteristics of the whole sample of high amplitude variables in the near-infrared. We present the follow-up and spectroscopic characteristics of a large sub-sample of candidate eruptive variable stars in a companion publication (hereinafter referred to as paper II ).
         
         In Sect. 2 of this work we describe the VVV survey, the data and the method used to select high amplitude infrared variables. Section 3 describes the make up and general properties of the sample, the evidence for clustering and the apparent association with SFRs. In this section we also classify the light curves of variables found outside SFRs and use this
information to estimate the contamination of our high amplitude YSO sample by other types of variable star.
We then estimate the high amplitude YSO source density from our sample and compare the average space density with those of other high amplitude infrared variables. In Sect. 4 we discuss the physical mechanisms that drive variability in YSOs and classify our YSOs via light curve morphology. This yields some ideas concerning which of the known mechanisms might be responsible for the observed variability. We test these mechanisms using two epoch $JHK_{\rm s}$ data.  Then we discuss the trends in the likely YSOs as a function of evolutionary status based on their spectral energy distribution. Finally we discuss the large sample of likely eruptive variables. Section 6 presents a summary of our results. 

\section{VVV}
          
           The regions covered by the VVV survey comprise the Bulge region within $-10^{\circ} < l < +10^{\circ} $ and $-10^{\circ} < b < +5^{\circ}$ and the disc region in $295^{\circ} < l < 350^{\circ} $ and $-2^{\circ} < b < +2^{\circ}$ \citep[see e.g.,][]{2010Minniti}.


           The data is collected by the Visible and Infrared Survey Telescope for Astronomy (VISTA). The 4m telescope is located at Cerro Paranal Observatory in Chile and is equipped with a near-infrared camera (VIRCAM) consisting of an array of sixteen 2048$\times$2048 pix detectors, with a typical pixel scale of 0.\arcsec 339, with each detector covering 694$\times$694 arcsec$^{2}$. The detectors are set in a 4$\times$4 array and have large spacing along the X and Y axis. Therefore a single pointing, called a ``pawprint'', covers 0.59$^{\circ}$ giving partial coverage of a particular field of view. A continuous coverage of a particular field is achieved by combining six single pointing with appropriate offsets. This combined image is called a tile. The VVV survey uses the five broad-band filters available in VIRCAM, $Z(\lambda_{eff}=0.87\mu$m), $Y(\lambda_{eff}=1.02\mu$m), $J(\lambda_{eff}=1.25\mu$m), $H(\lambda_{eff}=1.64\mu$m) and $K_{\rm s}(\lambda_{eff}=2.14\mu$m).
           
           The VVV survey area is comprised of 348 tiles, 196 in the bulge and 152 in the disc area. Each tile was observed in a single near-contemporaneous multi-filter (ZYJHK$_{\rm s}$) epoch at the beginning of the campaign, with an exposure time of 80 s per filter. A second epoch of contemporaneous JHK$_{\rm s}$ was observed in 2015. The variability monitoring was performed only in $K_{\rm s}$ with an exposure time of 16 s.  
                 
           The images are combined and processed at the Cambridge Astronomical Survey Unit (CASU). The tile catalogues are produced from the image resulting from combining six pawprints. The catalogues provide parameters such as positions and fluxes from different aperture sizes. A flag indicating the most probable morphological classification is also  provided, with ``-1'' indicating stellar sources, ``-2'' borderline stellar, ``1'' non-stellar , ``0'' noise, ``-7'' indicating sources containing bad pixels and finally class=-9 related to saturation \citep[for more details on all of the above, see][]{2012Saito}.  

	Quality control (QC) grades are also given by the European Southern Observatory (ESO) according to requirements provided by the observer. The constraints for VVV $K_{\rm s}$ variability data are: seeing $<$2 arcsec, sky transparency defined as ``thin cirrus'' or better. The ``master epoch'' of multi-filter data taken for each tile in a contemporaneous $JHK_{\rm s}$ observing block and a separate $ZY$ observing block have more stringent constraints: seeing $<$1.0, 1.0, 0.9, 0.9, 0.8 in $Z$, $Y$, $J$, $H$, $K_{\rm s}$ respectively and sky transparency of "clear" or better. According to whether observations fulfil the constraints established by the observer, these are classified as fully satisfied (QC A), almost satisfied, where for example some of the parameters are outside the specified constraints by $<10\%$ (QC B) and finally not satisfied (QC C).

\subsection{Selection method}\label{sec:vvvselec}

In order to search for variable stars we used the multi-epoch database of VVV comprising the observations of disc tiles with $|b| \leq 1^{\circ}$ taken between 2010 and 2012. We added the 2013, 2014 and 2015\footnote{We included a single $K_{\rm s}$ datapoint from 2015 observations, corresponding to the epoch with contemporaneous JHK$_{\rm s}$ photometry. Note that our analysis of  the light curve morphologies and periods is based on the 2010-2014 data only (Sect. \ref{vvv:sec_lcmorp}). The 2015 data became available only after that was complete but they were used in the colour variability analysis  (see Sect. \ref{vvv:sec_nirchange}).} data later to assist our analysis but they were not used in the selection. The catalogues were requested and downloaded from the CASU. We used catalogues of observations with QC grades A, B or C. Catalogues with QC grades C are still considered in order to increase the number of epochs. Some of them were still useful for our purposes. However, a small number of catalogues still presented some issues (e.g. zero point errors, bad seeing) making them useless, and as such were eliminated from the analysis. The number of catalogues in each tile varied from 14 to 23 epochs, with a median of 17 epochs per tile. When the 2013-15 data were 
added, the number of epochs available for the light curves rose to between 44 and 59. $K_{\rm s}$ photometry is derived from {\it apermag3} aperture fluxes (2\arcsec\ diameter aperture).

For each tile, the individual catalogues are merged into a single master catalogue. The first catalogue to be used as a reference was selected as the catalogue with the highest number of sources on it. In every case this corresponded to the catalogue from the deep $K_{\rm s}$ observation (80~s on source), which was taken contemporaneously with the $J$ and $H$ band data (in 2010). For all other epochs the time on source was 16~s. 

\begin{figure}
\centering
\resizebox{\columnwidth}{!}{\includegraphics{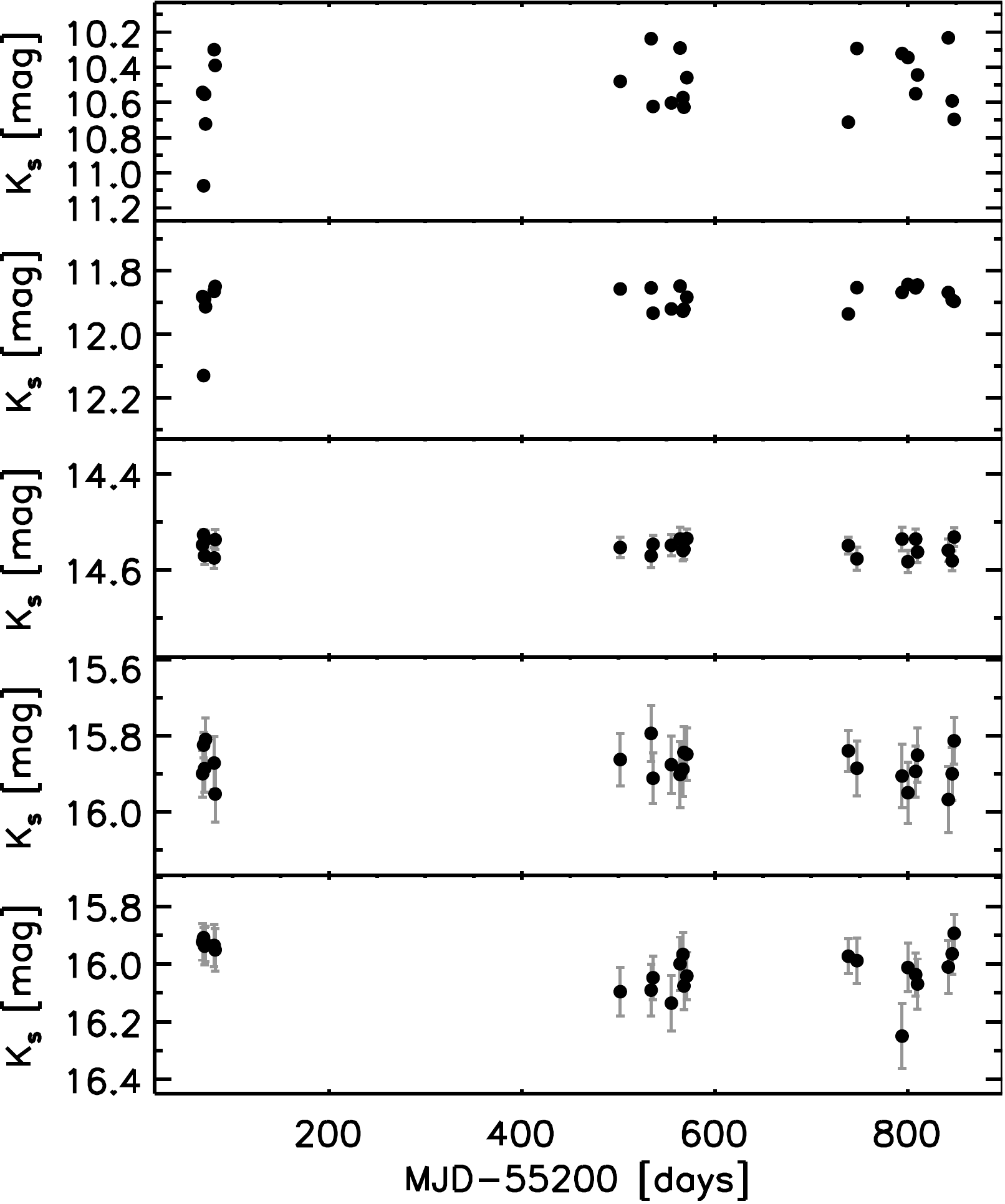}}
\caption{2010-2012 light curves of ``non-variable'' VVV objects (i.e. not classified as high-amplitude variables in our analysis). These are presented to show the typical scatter in magnitude across the analysed magnitude range.  We note that photometry for the brightest star is the standard CASU pipeline photometry.}
\label{selec:vvv3}
\end{figure}

Figure \ref{selec:vvv3} shows the typical scatter shown by stars at different magnitudes across the analysed range. In here we can see considerable scatter at bright magnitudes, due to effects of saturation, and the faint end of the distribution, which is dominated by photon noise. High amplitude variable star candidates are selected from the master catalogue from stars which fulfilled the following criteria in the 2010-12 data:

\begin{enumerate}
\item Detection with a stellar morphological classification (class$=-1$) in every available epoch.
\item Ellipticity with ell$<0.3$ in every epoch.
\item The absolute difference ($\Delta K_{\rm s}$) between the brightest ($K_{s,max}$) and faintest point ($K_{s,min}$) in the light curve of the source to be larger than 1 magnitude \citep[similar to the analysis in][]{2014Contreras}.

\begin{figure*}
\centering
\resizebox{0.6\textwidth}{!}{\includegraphics{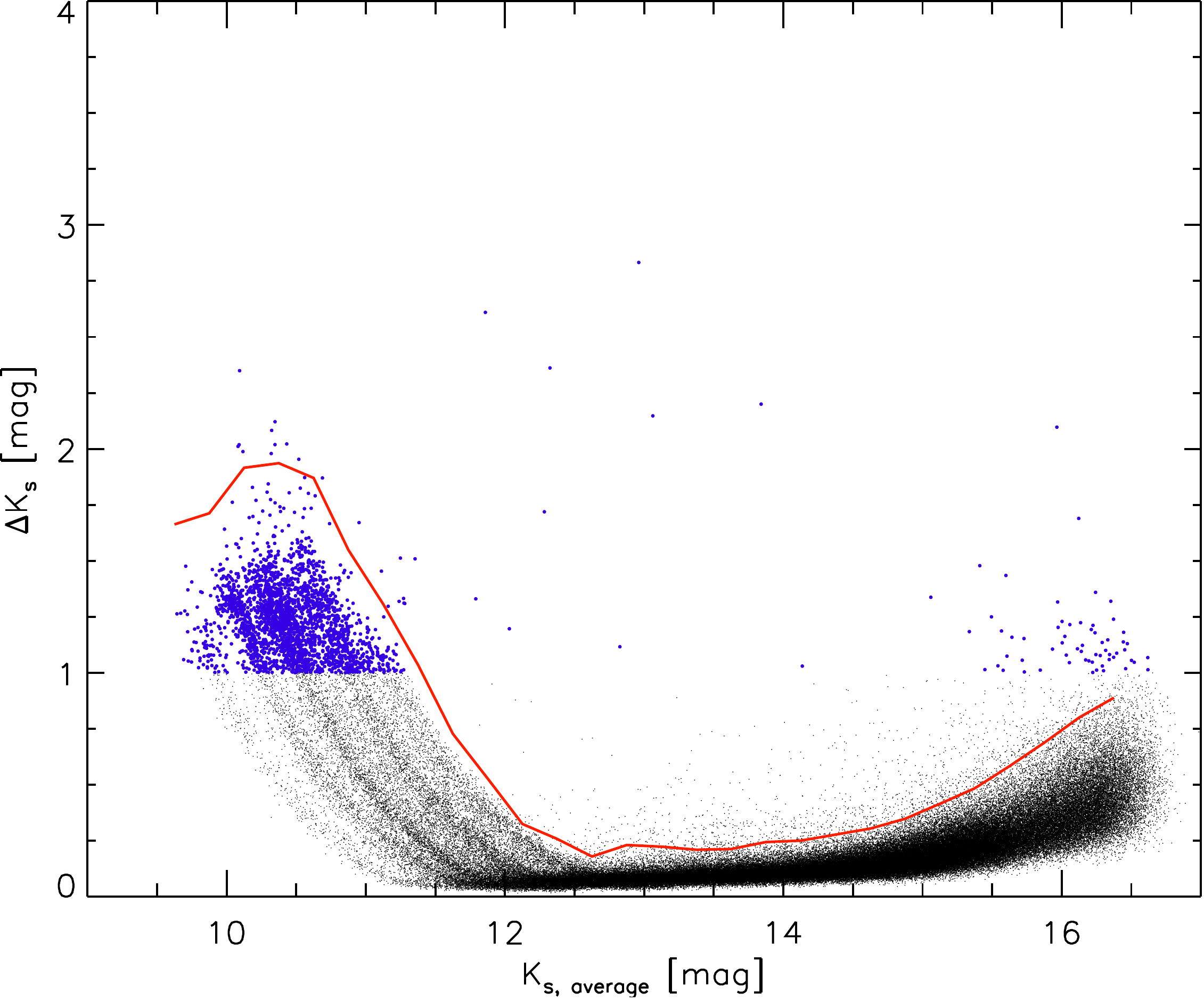}}
\caption{$K_{\rm s}$ vs $\Delta K_{\rm s}$ for one of the VVV tiles studied in this work, showing stars with class$=-1$ and ellipticity $<0.3$ in every epoch ( black circles). Variable star candidates which fulfil the condition $\Delta K_{\rm s} > 1$~magnitude are shown as blue circles. The red solid line marks the additional 3$\sigma$ cut applied to the objects as explained in the text. Stars above this line are selected for subsequent visual inspection.}
\label{selec:vvv1}
\end{figure*}

      The requirement for a detection at every available epoch in the 2010-12 interval was designed to exclude most transient objects such as novae, as well as 
reducing the number of false positives. This was the initial classification scheme. However, we observed that for each tile we were selecting a large number of sources as variable star candidates. Figure \ref{selec:vvv1} shows the average $K_{\rm s}$ magnitude vs $\Delta K_{\rm s}$ for variable stars in one of the VVV tiles. The figure shows that the majority of stars selected in the original classification scheme are located at the bright and faint ends of the distribution. The latter arise due to unreliable photometry at this faintest part. The VISTA detectors also become increasingly non-linear when reaching the saturation level. This non-linearity is corrected for in the creation of the catalogues, but differences between the magnitudes of the same object can still be observed, even for objects classified as stellar sources \citep[][]{2012Saito}. Figure 10 in \citet{2012Saito} shows that when comparing the $K_{\rm s}$ magnitudes of stellar sources found in overlapping regions of adjacent disc tiles, stars found at the brighter end show an increasing difference in magnitude \citep[an effect also observed in][]{2011Cioni, 2011Gonzalez}. This effect would explain the large differences observed at the brighter end of Fig.  \ref{selec:vvv1}. This part of the distribution also shows marked ``finger-like'' sequences. Each of the sequences can be explained by the fact that the VISTA detectors have different saturation levels. In order to minimize these effects we applied an additional cut. 

\item We separated the average $K_{\rm s}$ distribution of Fig. \ref{selec:vvv1} into bins of 0.5 magnitudes and derived the mean and standard deviation, $\sigma$, on $\Delta K_{\rm s}$ for each bin. In order to select an object as a candidate variable star we required its $\Delta K_{\rm s}$ to be 3$\sigma$ above the mean $\Delta K_{\rm s}$ at the corresponding magnitude level. This 3$\sigma$ line is shown in red in Fig. \ref{selec:vvv1} where we can see that it is able to account for the non-linearity effects at the bright end of the distribution.

\begin{figure*}
\centering
\resizebox{0.7\textwidth}{!}{\includegraphics{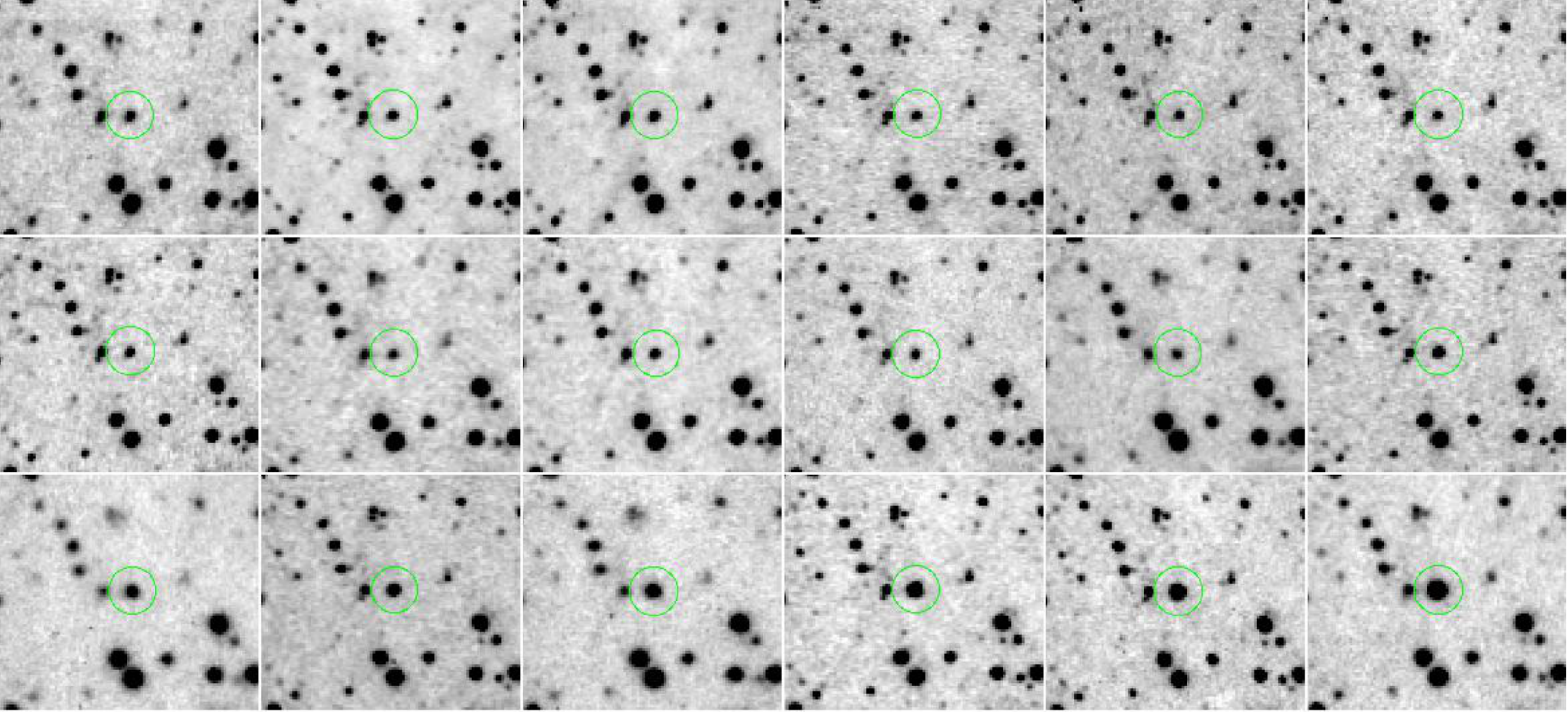}}
\caption{Example of the images used to visually inspect variable star candidates. In this case we show the images, taken between 2010 and 2012, of variable star VVVv322. Each image has a size of 1\arcmin$\times$1\arcmin. The star gets brighter towards the end of the sequence.}
\label{selec:vvv2}
\end{figure*}


      This additional constraint reduced the number of variable star candidates by a large factor. The initial requirements yield 158478 stars; the additional cut reduced this to 5085 stars. After the catalogue-based selection, we constructed 1\arcmin$\times$1\arcmin\ cut-out images around each candidate for every available epoch. Variable stars were confirmed as real through visual inspection of the individual images (an example is shown in Fig. \ref{selec:vvv2}). In some cases we performed manual photometry with IRAF in order to confirm the variability of the star. The most common causes for the appearance of false positives were, bad pixels in the images, saturation of bright sources, diffraction spikes and stars that were found on the edge of tiles. In the case of saturation, if this effect was present it was quite evident in individual images. In most cases saturation was observed in every single epoch, thus the variability observed in the light curve plots was not real, and the source was marked as such.

This selection method yielded a total of 840 real variable stars. However, 25 of them are found twice as they are covered by adjacent tiles in VVV.
The final list of VVV high-amplitude infrared variables consists of 816 stars. This includes one variable star, VVVv815, that showed large variability in 2010 but
did not meet all the selection criteria (see below).
The average magnitude for objects in the selected sample was found in the range $10.3<K_{\rm s}<16.9$~mag.
\end{enumerate}

\begin{figure}
\centering
\resizebox{\columnwidth}{!}{\includegraphics{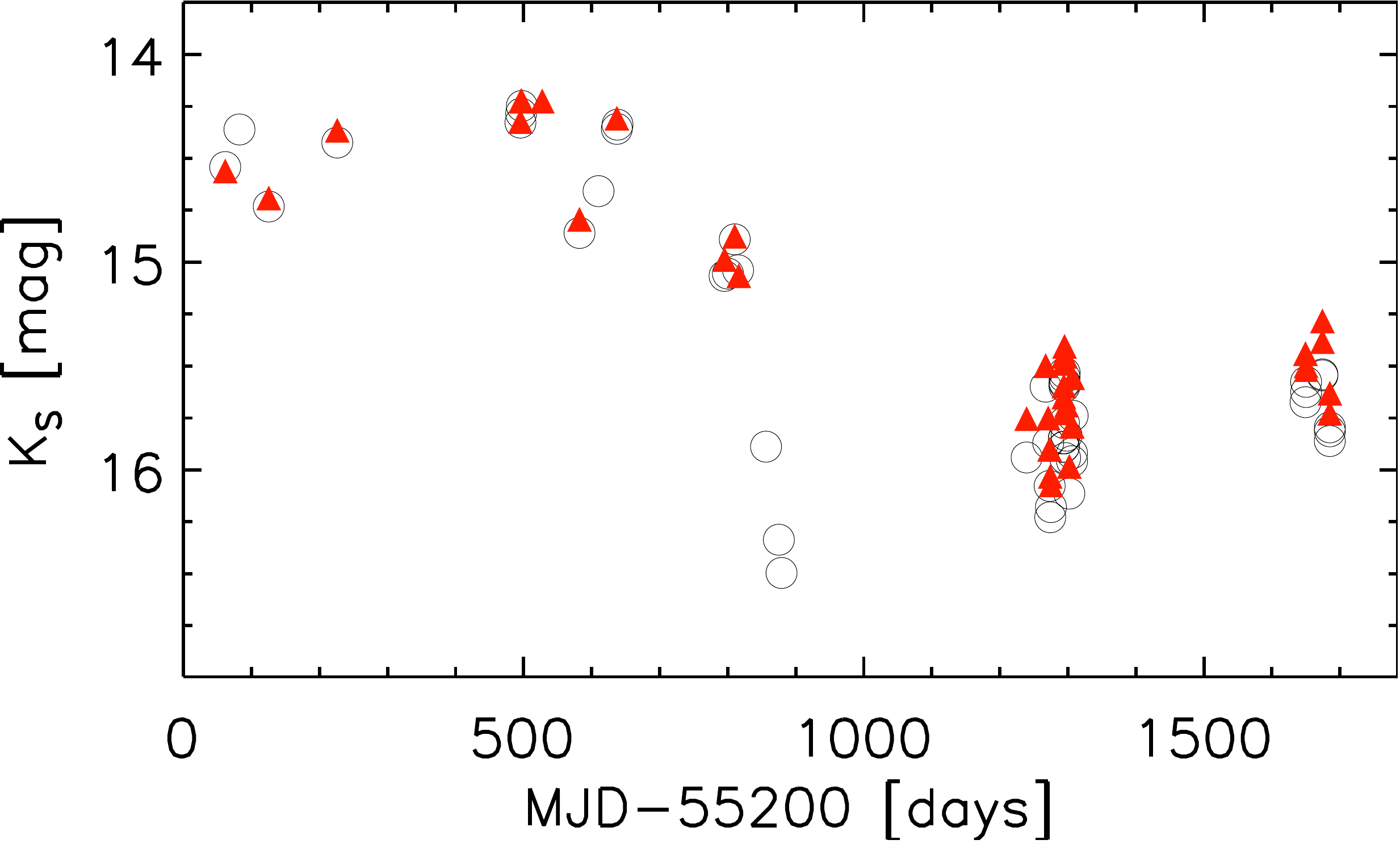}}
\caption{PSF (red triangles) vs aperture photometry (open circles) of star VVVv316. PSF photometry is shown only for data points classified as ``good'' or ``good but faint'' by {\sc DoPHOT}.}
\label{selec:dophotcom}
\end{figure}

Our requirement for a high quality detection at every epoch between 2010 and 2012 (see items (i) and (ii)) is bound to cause us to miss some real variables, very likely including some of the faintest or highest amplitude variables if they dropped 
below $K_{\rm s} \sim 16$ during that time, or if they became saturated and were therefore no longer classified as point sources. A significant fraction of all VVV sources are blended with adjacent stars and they can fail to pass our cuts on the morphological 
class and ellipticity at one or more epochs in consequence. The same can be true for YSOs with extended reflection nebulae or strong H$_{2}$ jets, as they might have slightly extended morphologies and fail to be classified as point sources \citep[see e.g.][]{2009Chen,2016Ward}. However, sources that pass these quality cuts are likely to be unblended and therefore to have reliable
photometry \citep[photometry from the VISTA pipeline is not always reliable for faint stars in crowded fields, see e.g.][]{2012Saito}. In order to check the 
reliability of the pipeline photometry, especially for faint sources with $K_{\rm s}=15-16$ magnitudes, we obtained point spread function (PSF) fitting photometry of all stars in tile d069 with 
{\sc DoPHOT} \citep{1993Schechter}. The results confirmed that the variables found by our selection have reliable pipeline photometry. This is illustrated in 
Fig. \ref{selec:dophotcom} for variable star VVVv316, where the comparison of {\sc DoPHOT} and VISTA pipeline photometry shows close agreement.

We investigated the incompleteness of our selection by examining two widely separated VVV disc tiles (d064 and d083), in which we removed our 
class and ellipticity cuts and required a minimum of only one detection in each year from 2010 to 2012 (with a stellar profile classification). This continues to 
select against transients and perhaps the most extreme variable YSOs but it allows us to assess incompleteness due to blending,  which can cause sources to be absent 
or to have different profile classifications at different epochs. We found that this more relaxed selection added over 400 additional candidates in the two tiles 
down to (mean) $K_{\rm s}$=15.5, an increase of more than a factor of 10. Following visual inspection (see below), we found that the number of real high 
amplitude variables was increased by a factor of $\sim$2, up to a limit of $K_{\rm s}$= 15.5. At fainter mean magnitudes the completeness of our selection with criteria
(i) to (iv) falls more steeply because most high amplitude variables will not satisfy the quality cuts at every epoch as the sensitivity limit is approached.


A case of this selection effect is found in a variable star VVVv815 mentioned above. It showed a large variation ($\Delta K_{\rm s} >$ 1 magnitude) in the analysis of 
an early release of 2010 data. However, the star does not show up as a variable star candidate in the analysis described above. Inspection of the master catalogue 
for the respective tile shows that the star has a classification different from stellar in 3 out of 18 epochs available for tile d090 in the 2010-2012 period. This 
star is included in our final list of VVV high-amplitude variables because it is also part of the sample that has follow up spectroscopic observations.     

The number of stars in the analysed VVV area that fulfil criteria (i) and (ii) above is 12 789 000 stars. Considering the number of real variable stars we see that high-amplitude infrared variability was observed in approximately 1 out of 15000 stars in the Galactic mid-plane at $295 < l < 350^{\circ}$.

\subsection{Issues with saturation}

The aforementioned problems of saturation still affect a small number of stars in our sample. This effect can become important when individual epochs of stars in our sample are brighter than $K_{\rm s}=12$ magnitudes. Saturation will reduce the flux at the inner core of the star, thus the magnitude of the star derived by using smaller diameter aperture than the default 2\arcsec\ aperture, will be fainter than the magnitude obtained from the default aperture.

In order to check whether the star is saturated we first obtain the magnitudes from aperture photometry using the measure fluxed from five different diameters for the apertures. These diameters are 1\arcsec\ (Apermag1), 1.41\arcsec\ (Apermag2), 2\arcsec\ (Apermag3), 2.82\arcsec\ (Apermag4) and 4\arcsec\ (Apermag5). We find that saturated stars show relatively large differences between the magnitudes from the first three apertures, and we set a threshold for saturation as stars having both Apermag1-Apermag3$>$0.05 mag and Apermag2-Apermag3$>$0.02 mag. Thus any individual epoch of a star with Apermag3$<12$~mag (in the K$_{\rm s}$ passband) and having these differences is flagged as saturated.

In order to correct for saturation, we follow \citet{2009Irwin} and defined a ring outside the saturated core, to obtain a new flux estimate. We then determine an aperture correction for the ring from a set of bright, unsaturated stars found within 5\arcmin\ of our object of interest. In our analysis we derived new fluxes using the ring defined by Apermag2 and Apermag4.  Comparison with 2MASS $K_s$ photometry indicates that this choice of apertures extends the dynamic range by 2.5 magnitudes, relative to the pipeline photometry. We 
caution that comparison with 2MASS $K_s$ photometry indicates that while this approach provides correct magnitudes, the uncertainties are large, typically 0.2 magnitudes.

In Table \ref{table:vvvfullphot} we present the 2010-2015 photometry for the 816 high-amplitude variable stars from VVV.

\begin{table}
\begin{center}
\begin{tabular}{@{}l@{\hspace{0.3cm}}c@{\hspace{0.25cm}}c@{\hspace{0.2cm}}c@{\hspace{0.2cm}}}
\hline
ID & MJD-55200 & $K_{\rm s}$  & $K_{\rm s,err}$ \\
& (days) & (mag) & (mag) \\ 
\hline
VVVv1 &  69.1171875 & 14.400 & 0.016\\
VVVv1 &  70.1484375 & 14.436 & 0.010\\
VVVv1 &  71.1445312 & 14.381 & 0.016\\
VVVv1 &  72.1406250 & 14.225 & 0.014\\
VVVv1 &  81.0781250 & 14.606 & 0.022\\
VVVv1 &  82.1015625 & 14.680 & 0.024\\
VVVv1 & 501.9882812 & 14.297 & 0.017\\
VVVv1 & 533.9648438 & 15.271 & 0.044\\
VVVv1 & 535.9726562 & 15.133 & 0.033\\
VVVv1 & 554.9726562 & 14.853 & 0.029\\
VVVv1 & 564.0117188 & 14.223 & 0.019\\
\hline
\end{tabular}
\caption{$K_{\rm s}$ photometry of the 816 high-amplitude variable stars from VVV. The full version of the table is available online.}\label{table:vvvfullphot}
\end{center}
\end{table}

\section{High Amplitude Infrared Variables from VVV}

\subsection{General characteristics}\label{sec:vvvsearch}

\begin{landscape}
\clearpage
\pagestyle{empty}
\setlength\LTleft{0pt}
\setlength\LTright{0pt}
\setlength\topmargin{-30pt}
\setlength\textwidth{702pt}
\setlength\textheight{38pc}

\begin{table*}
\begin{flushleft}
\begin{tabular}{@{}l@{\hspace{0.15cm}}c@{\hspace{0.15cm}}c@{\hspace{0.15cm}}c@{\hspace{0.15cm}}c@{\hspace{0.15cm}}c@{\hspace{0.15cm}}c@{\hspace{0.15cm}}c@{\hspace{0.15cm}}c@{\hspace{0.15cm}}c@{\hspace{0.15cm}}c@{\hspace{0.15cm}}c@{\hspace{0.15cm}}c@{\hspace{0.15cm}}c@{\hspace{0.15cm}}c@{\hspace{0.15cm}}c@{\hspace{0.15cm}}c@{\hspace{0.15cm}}c@{\hspace{0.15cm}}c@{\hspace{0.15cm}}l@{\hspace{0.15cm}}c@{}}
\hline
Object ID & VVV Designation & $\alpha$  & $\delta$  & $l$ & $b$ & $Z$ & $Z_{err}$ & $Y$ & $Y_{err}$ &  $J$ & $J_{err}$ & $H$ & $H_{err}$ & $K_{\rm s}$ & $K_{\rm s,err}$ &  $\Delta K_{\rm s}$ & $\alpha_{class}$ & SFR & Class$^{\mathrm{a}}$ & Period $^{\mathrm{b}}$ \\
 &  & (J2000) & (J2000) & (degrees) & (degrees) & (mag) & (mag) & (mag) & (mag) & (mag) & (mag) & (mag) & (mag) & (mag) & (mag) & (mag) & & & & (days) \\
\hline
VVVv1 & VVV J114135.16-622055.51 & 11:41:35.16 & -62:20:55.51  &  294.92603 & -0.56770  &  -- & -- & -- & -- & 17.99 & 0.06 & 15.95 & 0.02 & 14.44 & 0.01 & 1.34 & -0.29 & y & STV &   72\\ 
VVVv2 & VVV J114412.94-623449.09 & 11:44:12.94 & -62:34:49.09  &  295.28005 & -0.71146  &  -- & -- & -- & -- & -- & -- & 18.78 & 0.23 & 15.71 & 0.03 & 2.52 &  1.22 & y & STV & --\\
VVVv3 & VVV J115113.03-623729.29 & 11:51:13.03 & -62:37:29.29  &  296.07199 & -0.55784  &  13.17 & 0.01 & 12.93 & 0.01 & 12.90 & 0.01 & 12.70 & 0.01 & 12.24 & 0.01 & 2.21 & -- & n & Known & --\\
VVVv4 & VVV J115808.69-630708.60 & 11:58:08.69 & -63:07:08.60  &  296.95057 & -0.86785  &  -- & -- & -- & -- & 18.23 & 0.08 & 16.62 & 0.04 & 15.32 & 0.02 & 1.10 & -0.24 & y & STV & --\\
VVVv5 & VVV J115959.68-622613.20 & 11:59:59.68 & -62:26:13.20  &  297.02026 & -0.15716  &  17.69 & 0.02 & 16.62 & 0.01 & 15.87 & 0.01 & 15.25 & 0.01 & 13.53 & 0.01 & 1.30 & -- & n & LPV & --\\
VVVv6 & VVV J115937.81-631109.77 & 11:59:37.81 & -63:11:09.77  &  297.12836 & -0.89953  &  19.02 & 0.05 & 18.08 & 0.04 & 16.80 & 0.02 & 15.95 & 0.02 & 15.50 & 0.02 & 1.02 & -- & n & EB &    1.64\\
VVVv7 & VVV J120202.67-623615.60 & 12:02:02.67 & -62:36:15.60  &  297.28472 & -0.27538  &  -- & -- & -- & -- & -- & -- & -- & -- & 17.22 & 0.12 & 1.60 &  2.39 & y & Eruptive & --\\
VVVv8 & VVV J120059.11-631636.18 & 12:00:59.11 & -63:16:36.18  &  297.29582 & -0.95838  &  -- & -- & -- & -- & -- & -- & -- & -- & 16.86 & 0.09 & 1.38 &  0.64 & y & Eruptive & --\\
VVVv9 & VVV J120217.23-623647.83 & 12:02:17.23 & -62:36:47.83  &  297.31381 & -0.27888  &  -- & -- & -- & -- & 18.29 & 0.08 & 16.33 & 0.03 & 14.64 & 0.01 & 2.78 & -0.39 & y & Dipper & --\\
VVVv10 & VVV J120250.85-622437.62 & 12:02:50.85 & -62:24:37.62  &  297.33912 & -0.06749  &  18.57 & 0.04 & 18.23 & 0.05 & 16.97 & 0.03 & 16.35 & 0.03 & 16.07 & 0.04 & 1.19 & -- & n & STV & --\\
VVVv11 & VVV J120436.62-625704.60 & 12:04:36.62 & -62:57:04.60  &  297.63741 & -0.56188  &  20.20 & 0.19 & 19.01 & 0.13 & 17.87 & 0.07 & 16.79 & 0.05 & 16.08 & 0.08 & 1.10 & -- & n & STV & --\\
VVVv12 & VVV J121033.19-630755.71 & 12:10:33.19 & -63:07:55.71  &  298.33185 & -0.62611  &  -- & -- & -- & -- & -- & -- & 16.30 & 0.03 & 15.04 & 0.03 & 1.74 &  0.28 & y & Fader & --\\
VVVv13 & VVV J121216.83-624838.32 & 12:12:16.83 & -62:48:38.32  &  298.47603 & -0.27814  &  -- & -- & -- & -- & -- & -- & -- & -- & 16.72 & 0.14 & 1.40 &  1.39 & y & STV & --\\
VVVv14 & VVV J121218.13-624904.48 & 12:12:18.13 & -62:49:04.48  &  298.47958 & -0.28495  &  19.48 & 0.10 & 18.88 & 0.12 & 17.84 & 0.06 & 16.74 & 0.05 & 15.56 & 0.05 & 1.29 &  0.88 & y & LPV-YSO &  124\\
VVVv15 & VVV J121226.09-624416.97 & 12:12:26.09 & -62:44:16.97  &  298.48252 & -0.20371  &  19.00 & 0.07 & 17.72 & 0.04 & 16.57 & 0.02 & 15.60 & 0.02 & 15.04 & 0.03 & 1.09 & -- & y & EB &    2.27\\
VVVv16 & VVV J121329.76-624107.74 & 12:13:29.76 & -62:41:07.74  &  298.59498 & -0.13364  &  18.01 & 0.03 & 17.13 & 0.02 & 16.06 & 0.01 & 14.72 & 0.01 & 13.66 & 0.01 & 1.29 &  0.92 & y & Eruptive & --\\
VVVv17 & VVV J121352.08-625549.90 & 12:13:52.08 & -62:55:49.90  &  298.67278 & -0.36986  &  -- & -- & -- & -- & -- & -- & 17.79 & 0.12 & 16.41 & 0.10 & 1.22 &  0.44 & y & STV & --\\
VVVv18 & VVV J121950.31-632142.24 & 12:19:50.31 & -63:21:42.24  &  299.39868 & -0.70695  &  17.82 & 0.02 & 17.29 & 0.02 & 16.20 & 0.01 & 15.52 & 0.01 & 15.32 & 0.02 & 1.04 & -- & n & STV & --\\
VVVv19 & VVV J122255.30-632352.56 & 12:22:55.30 & -63:23:52.56  &  299.74594 & -0.70270  &  19.56 & 0.08 & 18.67 & 0.06 & 17.55 & 0.04 & 16.40 & 0.03 & 15.61 & 0.03 & 1.82 & -- & n & EB &   16.88\\
VVVv20 & VVV J122827.97-625713.97 & 12:28:27.97 & -62:57:13.97  &  300.32402 & -0.19849  &  -- & -- & -- & -- & 17.38 & 0.04 & 14.10 & 0.01 & 11.70 & 0.01 & 1.71 &  0.60 & y & Eruptive & --\\
VVVv21 & VVV J122902.24-625234.10 & 12:29:02.24 & -62:52:34.10  &  300.38193 & -0.11533  &  -- & -- & -- & -- & -- & -- & 17.12 & 0.05 & 15.80 & 0.03 & 1.79 &  0.86 & y & LPV-YSO &  603\\ 
VVVv22 & VVV J123105.60-624457.34 & 12:31:05.60 & -62:44:57.34  &  300.60547 &  0.03057  &  -- & -- & -- & -- & 18.81 & 0.14 & 16.94 & 0.05 & 15.55 & 0.03 & 1.73 & -0.34 & y & STV & --\\
VVVv23 & VVV J123128.53-624433.10 & 12:31:28.53 & -62:44:33.10  &  300.64855 &  0.04070  &  19.44 & 0.07 & 18.37 & 0.05 & 17.14 & 0.03 & 15.57 & 0.02 & 14.40 & 0.01 & 1.51 & -0.20 & y & Fader & --\\
VVVv24 & VVV J123235.68-634319.61 & 12:32:35.68 & -63:43:19.61  &  300.84794 & -0.92662  &  17.17 & 0.01 & 16.13 & 0.01 & 14.05 & 0.01 & 12.95 & 0.01 & 12.23 & 0.01 & 1.20 & -- & n & LPV &  430\\
VVVv25 & VVV J123514.37-624715.63 & 12:35:14.37 & -62:47:15.63  &  301.08129 &  0.02587  &  -- & -- & -- & -- & -- & -- & 16.01 & 0.02 & 12.34 & 0.01 & 1.68 &  0.22 & y & Eruptive & --\\
VVVv26 & VVV J123845.66-631136.03 & 12:38:45.66 & -63:11:36.03  &  301.50320 & -0.35674  &  -- & -- & -- & -- & 19.67 & 0.29 & 16.70 & 0.04 & 14.71 & 0.01 & 2.45 &  1.07 & y & Eruptive & --\\
VVVv27 & VVV J123848.33-633939.15 & 12:38:48.33 & -63:39:39.15  &  301.53114 & -0.82347  &  -- & -- & -- & -- & -- & -- & -- & -- & 12.04 & 0.01 & 2.73 & -- & n & LPV & 1329\\
VVVv28 & VVV J123911.54-630524.76 & 12:39:11.54 & -63:05:24.76  &  301.54688 & -0.25138  &  -- & -- & -- & -- & -- & -- & 18.91 & 0.32 & 16.78 & 0.09 & 1.39 &  0.54 & y & STV &    3.68\\
VVVv29 & VVV J123931.48-630720.38 & 12:39:31.48 & -63:07:20.38  &  301.58593 & -0.28170  &  -- & -- & -- & -- & -- & -- & -- & -- & 16.94 & 0.10 & 2.13 &  1.32 & y & Eruptive & --\\
VVVv30 & VVV J124140.56-635033.57 & 12:41:40.56 & -63:50:33.57  &  301.85616 & -0.99128  &  17.92 & 0.02 & 17.22 & 0.02 & 16.40 & 0.02 & 15.66 & 0.02 & 15.23 & 0.02 & 1.22 & -- & n & EB &    1.91\\
VVVv31 & VVV J124140.15-635918.05 & 12:41:40.15 & -63:59:18.05  &  301.86093 & -1.13689  &  13.91 & 0.01 & 13.27 & 0.01 & 12.46 & 0.01 & 12.08 & 0.01 & 11.69 & 0.01 & 1.15 & -- & n & LPV &  760\\
VVVv32 & VVV J124357.15-625445.09 & 12:43:57.15 & -62:54:45.09  &  302.07991 & -0.05314  &  19.58 & 0.07 & 18.08 & 0.04 & 16.07 & 0.01 & 14.07 & 0.01 & 12.45 & 0.01 & 2.49 &  0.33 & y & LPV-YSO & --\\
VVVv33 & VVV J124425.05-631355.76 & 12:44:25.05 & -63:13:55.76  &  302.14153 & -0.37116  &  -- & -- & -- & -- & -- & -- & 18.10 & 0.16 & 16.43 & 0.07 & 1.37 & -- & n & STV & --\\
VVVv34 & VVV J125029.87-625124.93 & 12:50:29.87 & -62:51:24.93  &  302.82470 &  0.01465  &  19.22 & 0.05 & 18.32 & 0.05 & 17.86 & 0.06 & 16.70 & 0.04 & 15.71 & 0.03 & 1.30 & -0.34 & y & STV & --\\
VVVv35 & VVV J125206.52-635711.52 & 12:52:06.52 & -63:57:11.52  &  303.00557 & -1.08152  &  17.62 & 0.01 & 16.85 & 0.01 & 17.13 & 0.03 & 16.21 & 0.03 & 15.97 & 0.04 & 1.24 & -- & n & EB &    1.12\\
VVVv36 & VVV J125917.72-633008.44 & 12:59:17.72 & -63:30:08.44  &  303.80825 & -0.64394  &  14.38 & 0.01 & 13.84 & 0.01 & 13.27 & 0.01 & 12.56 & 0.01 & 11.81 & 0.01 & 1.03 & -0.11 & y & Eruptive & --\\
VVVv37 & VVV J130243.05-631130.00 & 13:02:43.05 & -63:11:30.00  &  304.20331 & -0.34774  &  18.80 & 0.03 & 17.75 & 0.03 & 16.74 & 0.02 & 15.89 & 0.02 & 15.38 & 0.03 & 1.34 & -- & n & STV & --\\
VVVv38 & VVV J130311.38-631439.09 & 13:03:11.38 & -63:14:39.09  &  304.25411 & -0.40259  &  -- & -- & -- & -- & -- & -- & 17.79 & 0.13 & 16.32 & 0.07 & 1.41 &  1.20 & y & STV &   48.96\\
VVVv39 & VVV J130440.98-635313.45 & 13:04:40.98 & -63:53:13.45  &  304.38893 & -1.05277  &  18.67 & 0.03 & 17.89 & 0.03 & 17.57 & 0.05 & 16.38 & 0.04 & 15.49 & 0.03 & 1.51 & -- & n & STV & --\\
VVVv40 & VVV J130600.43-630144.40 & 13:06:00.43 & -63:01:44.40  &  304.58298 & -0.20394  &  20.71 & 0.20 & 19.49 & 0.14 & 16.52 & 0.02 & 15.42 & 0.02 & 14.69 & 0.01 & 1.24 & -- & n & EB &   10.29\\
\hline
\multicolumn{21}{l}{a. LPV:long-period variable; LPV-Mira:long-period variable found in SFR but with Mira-like characteristics; LPV-YSO:long-period variable found in SFR with YSO-like characteristics.}\\
\multicolumn{21}{l}{  STV:short-term variable; EB:eclipsing binary}\\
\multicolumn{21}{l}{b. We note that in some objects classified as long-period variables (LPV, LPV-MIRA and LPV-YSO), the light curve shows a clear periodic behaviour over long timescales ($>100$~days).}\\
\multicolumn{21}{l}{   However, we are unable to measure an exact period. In those cases this column is left blank.}\\
\hline
\end{tabular}
\caption{Parameters of the high-amplitude variables from VVV. For the description of the columns see Sect. \ref{sec:vvvsearch}. Here we show the first 40 sources in the list. The complete list is available online.}\label{table:vvvpar}
\end{flushleft}
\end{table*}

\end{landscape}

The selection method of Sect.\ref{sec:vvvselec} yields 816 high amplitude ($\Delta K_{\rm s} > $1 mag) infrared variables. In order to study the properties of these stars, we searched for additional information in available public databases. This search can be summarized as follows:

\begin{itemize}

\item {\bf SIMBAD} We query the SIMBAD database \citep{2000Wenger} for astronomical objects within a radius of 5\arcmin\ centred on the VVV object. 

\item {\bf Vizier} Additional information was provided with the use of the Vizier database \citep{2000Ochsenbein}. In here we queried whether the VVV object was found  within 2\arcsec\ of objects found in astronomical catalogues that were not available in SIMBAD.

\item {\bf The NASA/IPAC Infrared Science Archive (IRSA)} In here we queried for additional information in several near- and mid-infrared public surveys, which include 2MASS \citep{2006Skrutskie}, DENIS \citep{1994Epchtein}, {\it Spitzer}/GLIMPSE surveys \citep[see e.g.,][]{2003Benjamin}, WISE \citep{2010Wright}, Akari \citep{2007Murakami} and MSX6C \citep{2001Price}. The search was done automatically using the IDL scripts provided at the IRSA website. The catalogues were queried for objects found within a 10\arcsec\ radius of the VVV object. In most cases, several objects are found at these distances, then we only selected the nearest object to our star. In order to confirm whether these detections correspond to our variable star, 1\arcmin$\times$1\arcmin\ VVV images around the star were visually inspected. 
 
In addition, we used the WISE image service within IRSA to inspect multi-colour images of the areas around our variable stars, in order to establish a possible association of the the VVV object with a SFR.

\end{itemize}

The general properties of the sample can be found in Table \ref{table:vvvpar}. Column $1$ presents the original designation given to the sources. Column $2$ corresponds to the full VVV designation for the source. Coordinates for the objects are given in Columns $3$ and $4$, with columns $5$ and $6$ presenting the Galactic coordinates of the sources. In columns $7$, $8$, $9$, $10$ and $11$ we present the nearly contemporaneous $Z$, $Y$, $J$, $H$, $K_{\rm s}$ photometry from VVV. Column $12$ gives $\Delta K_{\rm s}$, the  absolute value of the peak-to-trough difference from the 2010-2014 light curves from VVV. Column $13$ presents $\alpha_{class}$, the 2 to 23~$\mu$m spectral index parameter that relates to the evolutionary class of sources that appear to be associated with SFRs (the method and data used to estimate this parameter is explained in Sect. \ref{vvv:sec_alphaclass}). In column $14$ we present whether the object is likely associated with SFRs or not, whilst column 15 presents the classification of the object from its light curve. The latter is discussed throughout the text. Finally, in column 16 we present the approximate period for variable stars where we are able to measure this parameter.

Most of the variable stars are unknown from searches in SIMBAD and Vizier ($\sim$ 98 per cent). Among the known variables there are 2 novae, Nova Cen 2005 and Nova Cen 2008 \citep{2010Hughes,2013Saito}, 2 eclipsing binaries (EBs), EROS2-GSA J132758-631449 and PT Cen \citep{2002Derue, 2004Budding}, 1 high-mass X-ray binary, BR Cir \citep[see e.g.][]{2008Tudose} and 9 OH/IR stars. Among the objects not previously classified as variable stars, 159 are found in the Spitzer/GLIMPSE catalogue of intrinsically red sources from \citet{2008Robitaille}, with the majority being classified as likely YSOs from their mid-IR colours and brightness.
  
\subsection{YSO population}

At this point, the reader should note that most of the variables are listed as spatially associated with SFRs ($\sim$65\%, falling to $\sim$54\% after allowing for chance projection by non-YSOs, see Sect. \ref{vvv:sec_sfrass}, \ref{sec:cont} and \ref{vvv:sec_physmec}) and these stars have spectral indices that indicate a class I or flat spectrum evolutionary status, therefore they are likely in an early evolutionary stage. They are 
usually sufficiently red to be undetected ($i > 21$~mag) in sensitive panoramic optical surveys such as VPHAS$+$ \citep{2014Drew}
unlike most of the known FUor and EXors. The spectral indices of the YSOs are discussed later in Sect.\ref{vvv:sec_alphaclass} following
classification of the light curves and an attempt to decontaminate chance projections of other variables in SFRs.

\begin{figure*}
\centering
\resizebox{0.7\textwidth}{!}{\includegraphics{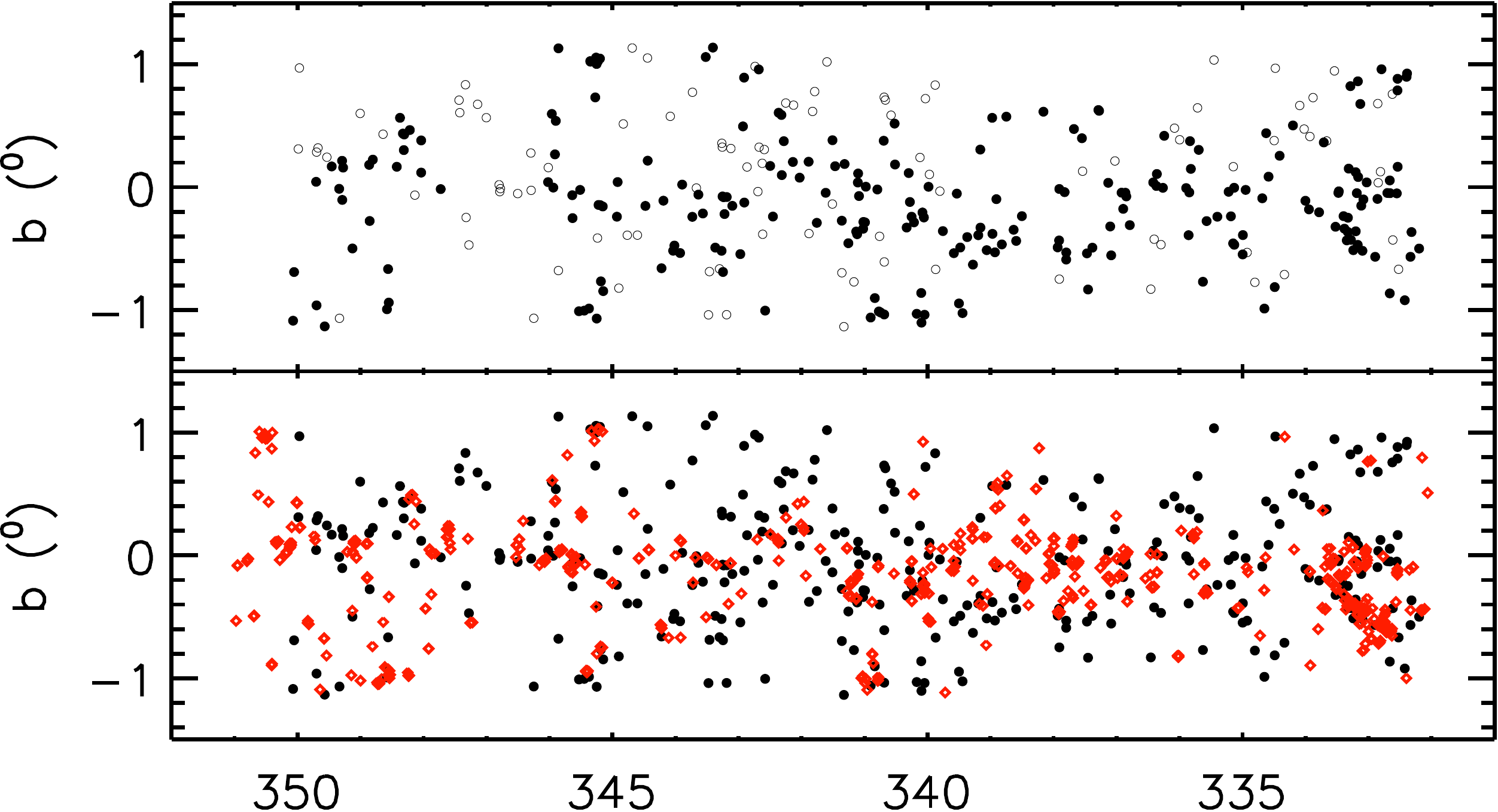}}
\resizebox{0.7\textwidth}{!}{\includegraphics{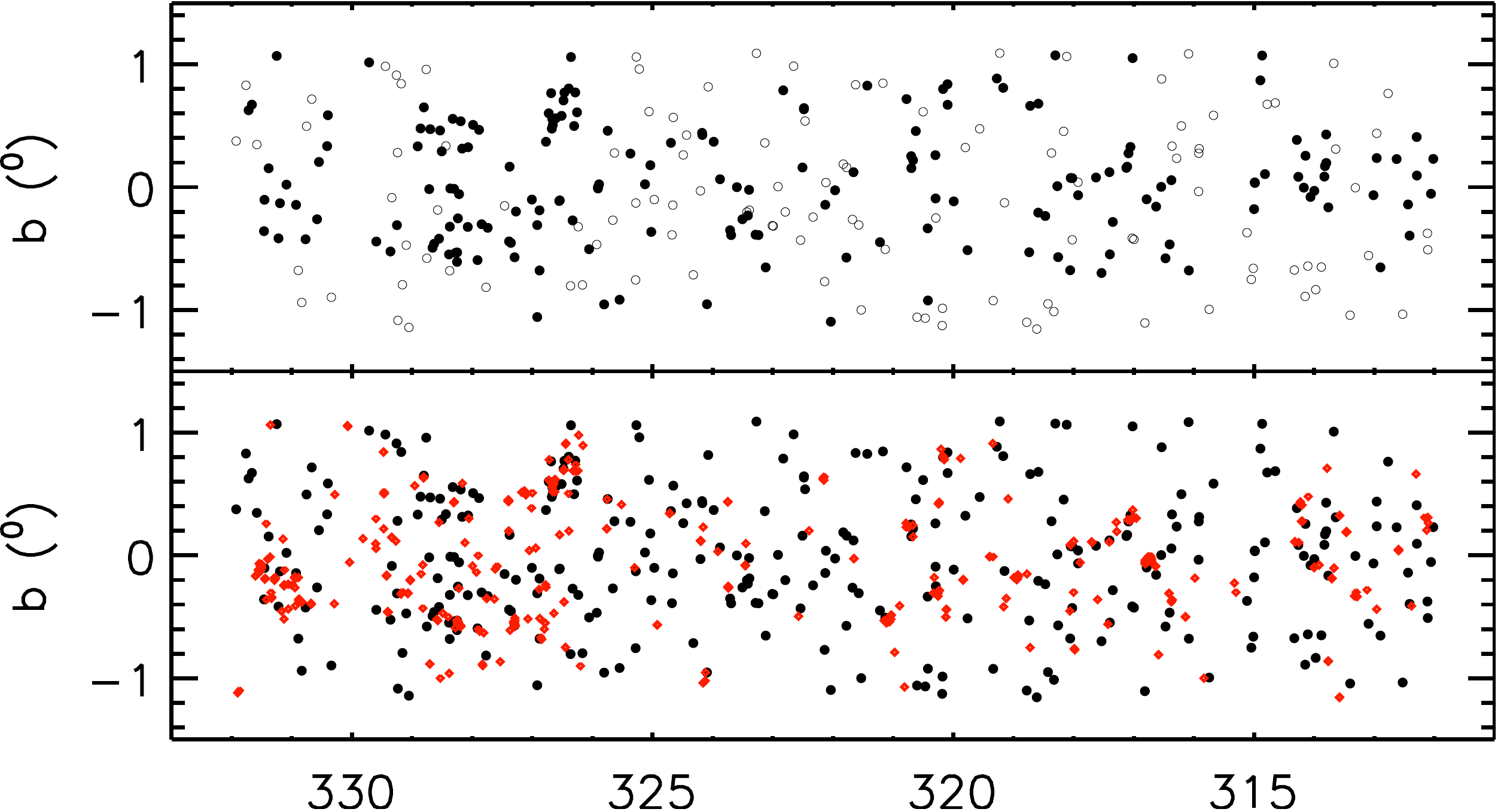}}
\resizebox{0.7\textwidth}{!}{\includegraphics{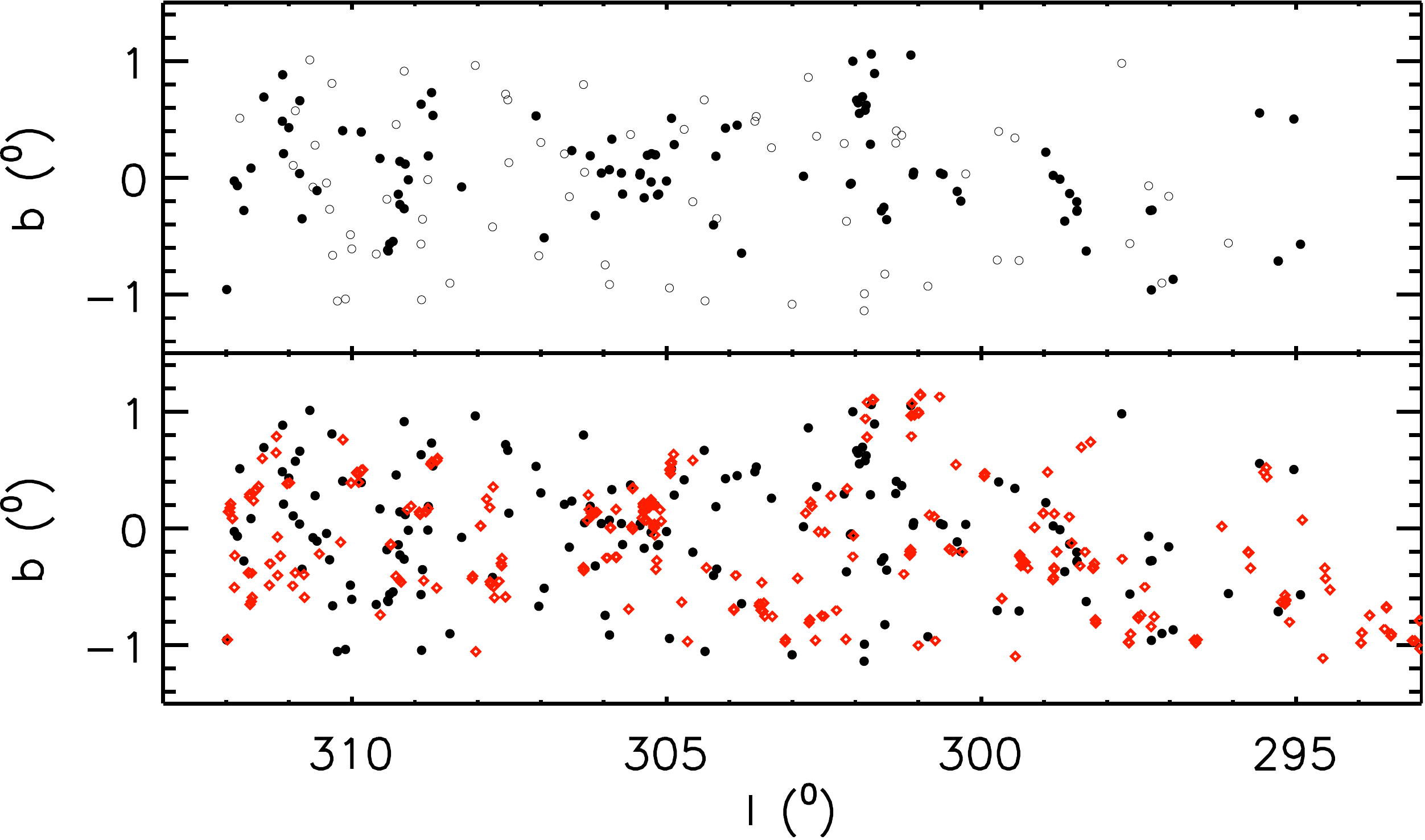}}
\caption{(top) Galactic distribution of high amplitude variable stars selected from VVV. These are divided into objects likely associated with SFRs (black filled circles) and those that are found outside these regions (black open circles) . The bottom graph shows the same distribution for the 816 high amplitude variables (black circles), but this time including the areas of star formation from the \citet{2002Avedisova} catalogue (red diamonds).}
\label{vvv:lb_dist}
\end{figure*}

\begin{figure*}
\centering
\resizebox{0.47\textwidth}{!}{\includegraphics{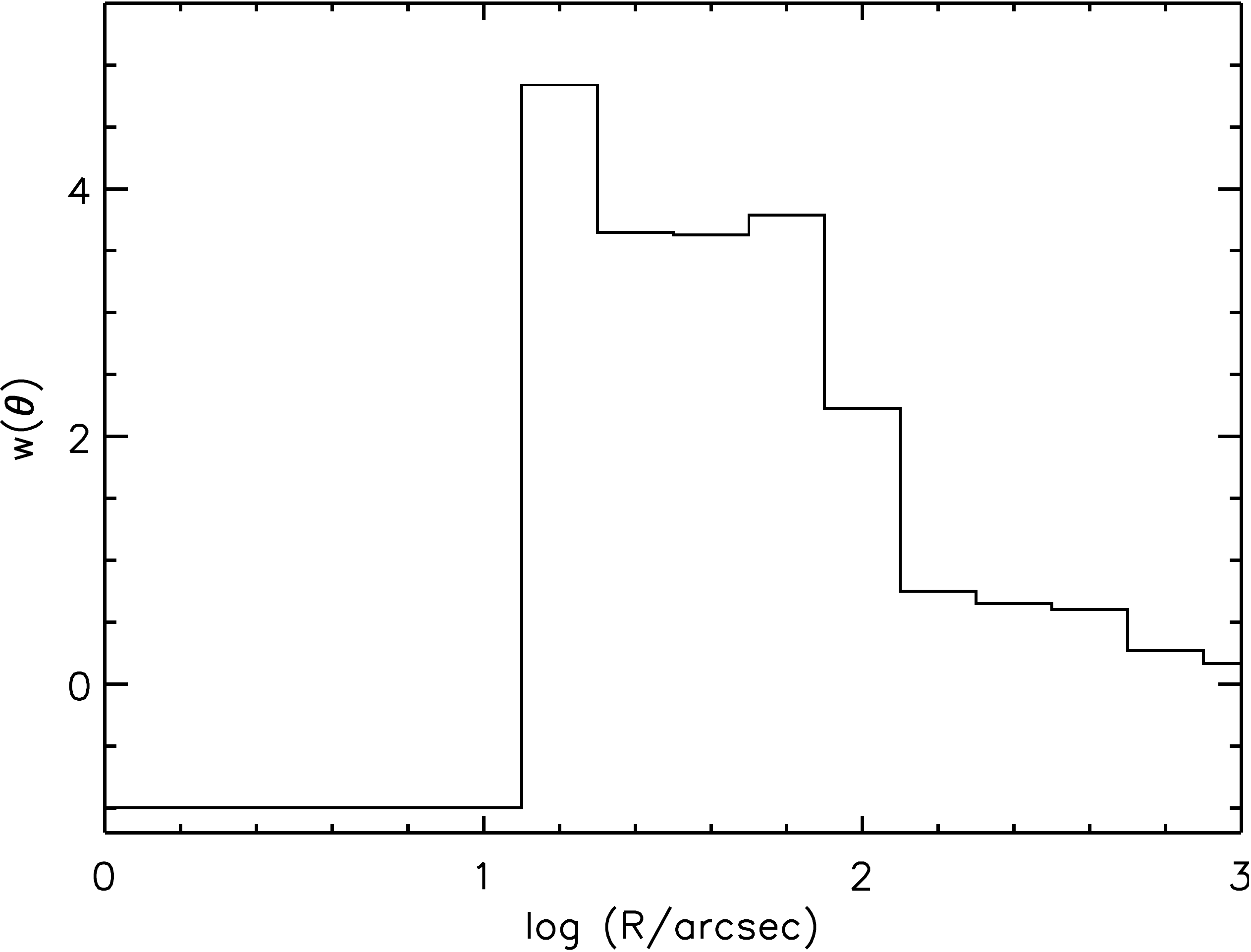}}
\resizebox{0.47\textwidth}{!}{\includegraphics{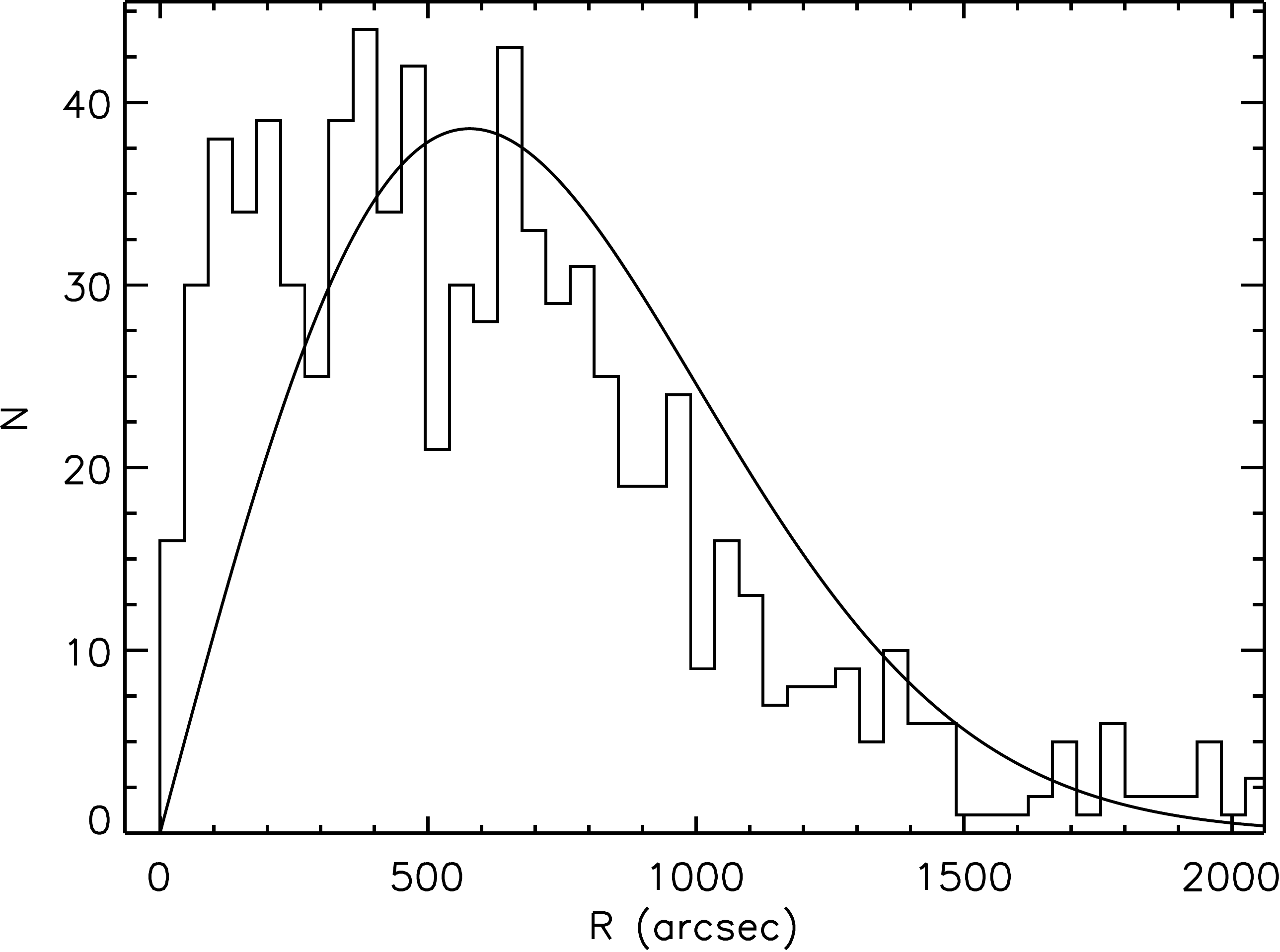}}
\caption{(left) Two-point angular correlation function of the sample of VVV high-amplitude variables. (right) Nearest neighbour distribution for the same sample. The smooth curve represents the expected distribution of a random (Poisson) distribution.}
\label{vvv:neighist}
\end{figure*}



\subsection{Association with SFRs}\label{vvv:sec_sfrass}

Figure \ref{vvv:lb_dist} shows the distribution for the 816 VVV variables across the Galactic midplane. It can be seen that our objects appear to be highly clustered, with their distribution following that of the SFRs from the Avedisova et al.(2002) catalogue (red diamonds in the bottom plot of Fig. \ref{vvv:lb_dist}). To study the apparent clustering, we derive the two-point angular correlation function, $w(\theta)$ and the nearest neighbour distribution of the sample of high amplitude variables. To derive $w(\theta)$, we follow \citet{1998Bate} and first estimate the mean surface density of companions (MSDC). For each star, we compute the angular separation, $\theta$, to all other stars, and bin the separation into annuli of separation $\theta$~to $\theta+\delta \theta$. The MSDC results from dividing the number of pairs found, $N_{p}$, at a given separation 
by the area of the annulus, and dividing by the total number of stars, $N_{\ast}$, or 

\begin{eqnarray*}
MSDC=\frac{N_{P}}{2\pi\theta\delta \theta N_{\ast}}
\end{eqnarray*}

The latter relates to $w(\theta)$ as

\begin{eqnarray*}
w(\theta) = MSDC \times \frac{A}{N_{\ast}} - 1
\end{eqnarray*}

\noindent where $A$ is the area covered by the survey \citep[][and references therein]{1998Bate}. This correlation function is valid as long as the separations $\theta$ are smaller than the angular length of the sample. We show the two-point correlation function in Fig. \ref{vvv:neighist}. We can see that we do not find any pairs for separations $\theta < 20\arcsec$, hence $w(\theta)=-1$. For separations between 20\arcsec\ and 100\arcsec\, $w(\theta)$ is larger than the values expected for random pairings ($w(\theta)=0$) and it remains somewhat above zero for separations up to a few hundred arcseconds. This nearest neighbour distribution of Fig. \ref{vvv:neighist}, also shows an excess of close neighbours at distances $R<200\arcsec$, compared to the expected number from a random (Poisson) distribution. Thus we are confident that we are tracing clustering in the VVV sample, on a spatial scale similar to that of distant Galactic clusters and star formation 
regions.

\begin{figure}
\centering
\resizebox{\columnwidth}{!}{\includegraphics{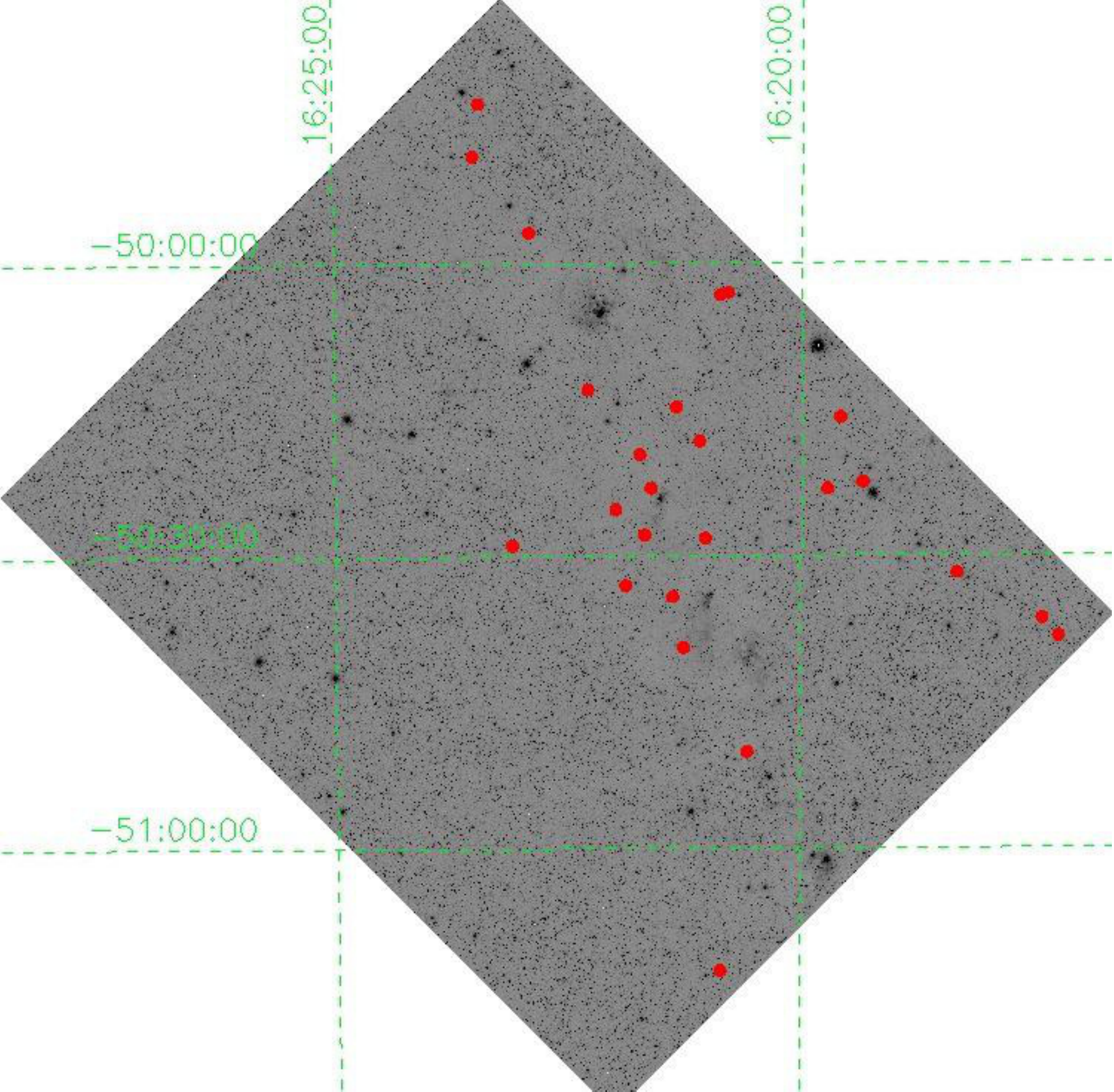}}\\
\resizebox{\columnwidth}{!}{\includegraphics{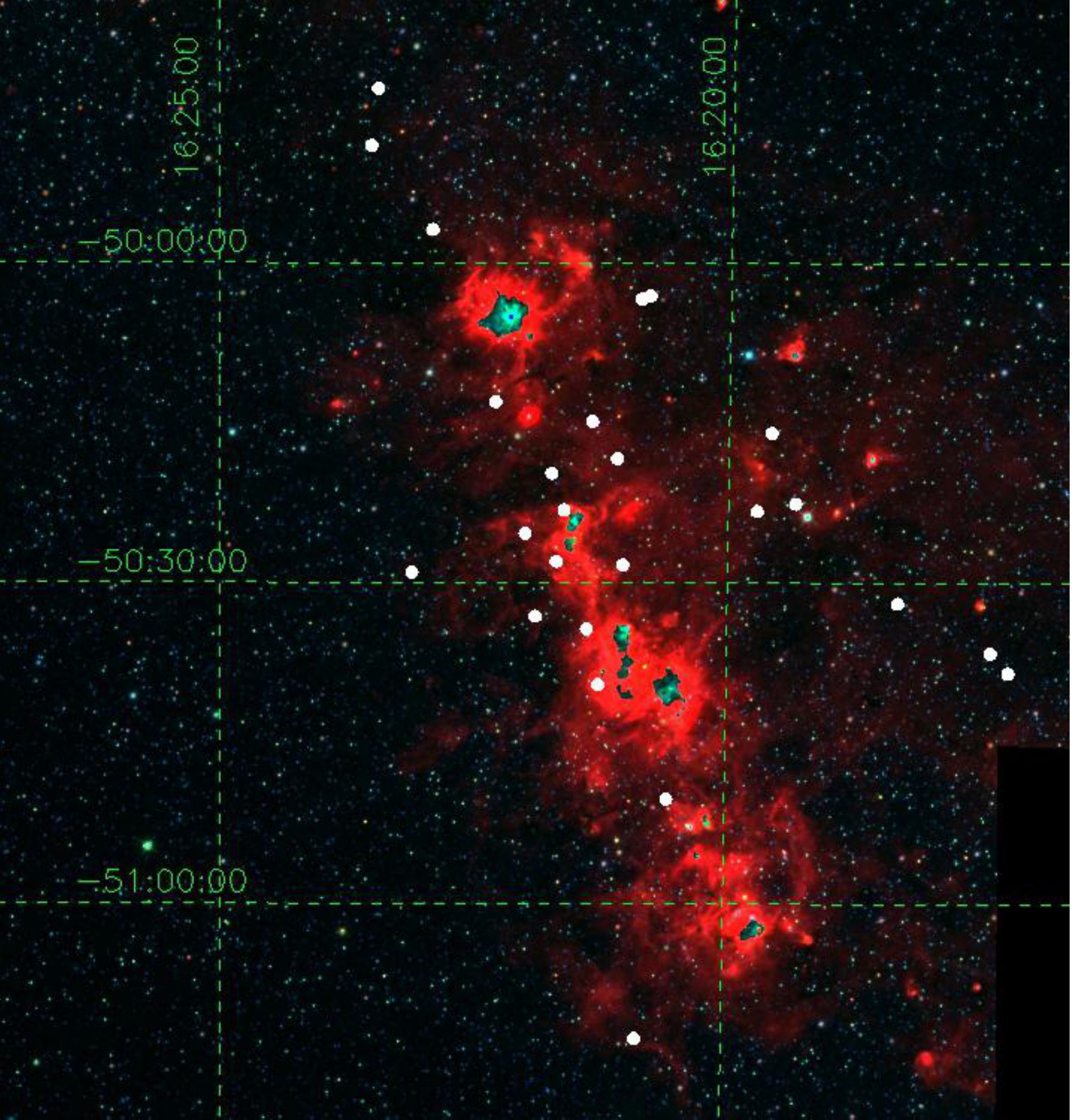}}
\caption{The top graph shows the $K_{\rm s}$ image of tile d065 along with the high amplitude variable stars found in this region. The clustering of the variable stars is already apparent in this image. The bottom graph shows the WISE false colour image of the same region (blue=3.5 $\mu$m, green=4.6 $\mu$m, red=12 $\mu$m). In here, the fact that variable stars preferentially locate around areas of star formation can be better appreciated.}
\label{vvv:d065}
\end{figure}

As an illustration of how variable stars in VVV are preferentially located in areas of star formation, Fig. \ref{vvv:d065} shows the $K_{\rm s}$ image of the area covered in tile d065. Twenty five highly variable stars are found in this tile, and it is clear that these are not evenly distributed along the area covered in d065, and instead are found clustered around an area of star formation, which is better appreciated in the cut-out image from WISE \citep{2010Wright}.

To establish a likely association with a SFR we used the criteria established in the UGPS search \citep[see][]{2014Contreras}, which were based on entries in the SIMBAD database and the Avedisova catalogue within a 5\arcmin\ 
radius of each high amplitude variable.
In addition we also check WISE images for evidence of star formation in the area of the object, e.g. evidence of bright extended 12 $\mu$m emission near the location of the objects or the finding of several stars with red W1-W2 colours (sources appearing green, yellow or red in 
WISE colour images) around the VVV object. We find that 
530 of our variable stars are spatially associated with SFRs, which represents 65$\%$ of the sample, remarkably 
similar to the observed association in UGPS objects \citep{2014Contreras}.

\subsection{Contamination by non-YSOs}\label{sec:cont}

In \citet{2014Contreras}, we estimated that about 10$\%$ of objects are probable chance associations with SFRs. This number is likely to be larger in our current analysis given that: (i) we are sampling mid-plane sightlines across the Galactic disc; (ii) the higher extinction in the Galactic midplane and the brighter saturation limit of VVV than UGPS, allows for a larger number of bright evolved variable stars to show up in our results. To determine the percentage of objects that might be catalogued as likely associated with SFRs by chance, we used the following method:

\begin{itemize}

\item[] a) Create a master catalogue of objects in the 76 tiles that were classified as stars in each of the epochs from the 2010-2012 analysis.
\item[] b) Select 816 stars randomly from this catalogue and query SIMBAD for objects found in a 5\arcmin\ radius.
\item[] c) Count the number of objects within this radius that were classified in any of the categories that could relate to star formation. This categories included T Tauri and Herbig Ae/Be stars, HII regions, Dark clouds, dense cores, mm and submm sources, FU Orionis stars, among others.
\item[] d) Repeat items b through c 40 times.

\end{itemize}

\begin{figure}
\centering
\resizebox{\columnwidth}{!}{\includegraphics{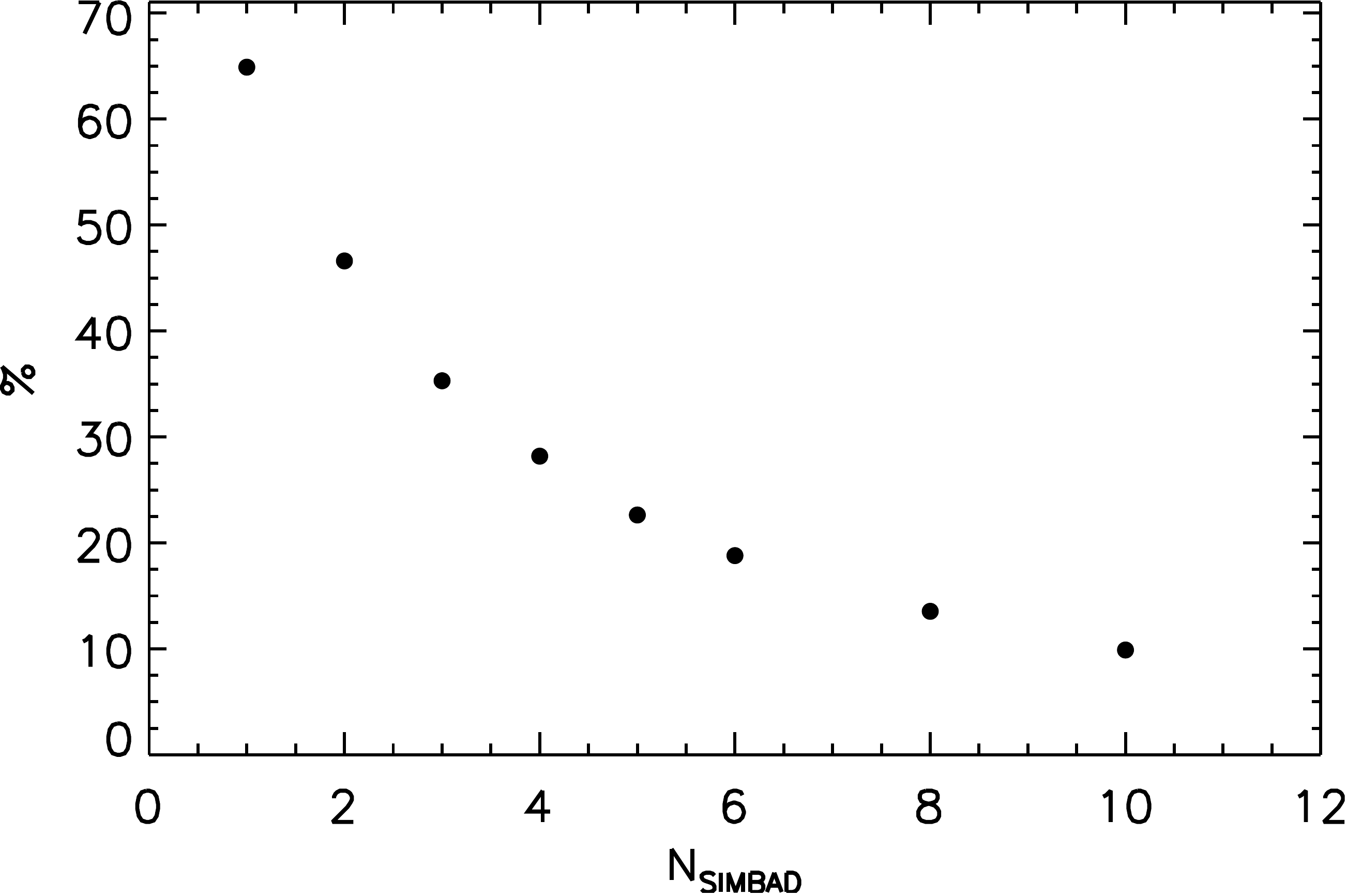}}
\caption{Percentage of objects flagged as likely associated with SFRs as a function of the number of objects classified in the categories that could relate to star formation found within a 5\arcmin\ radius query in SIMBAD.}
\label{vvv:simbcont}
\end{figure}

Figure \ref{vvv:simbcont} shows the number of stars found within 5\arcmin\ of the VVV object and that were classified in the categories shown above, N$_{simbad}$ vs the percentage of VVV objects with this number. It is already apparent that the percentage of chance selection will be higher than that estimated from GPS. However, we note that in order for an object to be flagged as associated with an SFR in Table \ref{table:vvvpar}, we needed at least 4 SIMBAD objects to appear within the 5\arcmin\ radius, thus giving us an estimate that $\sim$30$\%$ of the non-YSO population is spatially associated with an SFR by chance. Inspection of WISE images of 100 randomly selected sources yields a similar fraction of chance associations but most of these were also identified as SFR-associated from the SIMBAD results, so
the WISE data only slightly increased the chance association fraction. The Avedisova catalogue added
an even smaller fraction of chance associations not indicated by the SIMBAD and WISE data, so the final 
chance association probability for non-YSOs with star formation regions is 35\%. 


The number of non-YSOs in the SFR-associated sample is less than 35\% because non-YSOs do 
not dominate the full high amplitude sample but constitute about half of it. We found 286/816 (35\%) 
of variables outside SFRs, i.e. in 65\% of the area, suggesting that 54\% (35/0.65) of the sample is composed of 
objects other than YSOs but this neglects the fact that some YSOs will be members of SFRs that are not known in 
the literature nor visible in WISE images (see Sect. \ref{vvv:sec_varsnonsfrs}). Consequently, random addition of 35\% of half of the 
total sample of 816 variables to the SFR-associated sample would be expected to cause only 27\% contamination of 
the SFR-associated subsample by non-YSOs. This conclusion that the SFR-associated population of variables is 
dominated by bona-fide YSOs is supported by the two colour diagrams (figures \ref{vvv:nonsfrs_prop} and 
\ref{vvv:gc}) and light curves of the population (see Sect. \ref{vvv:sec_varsnonsfrs} and Sect. \ref{vvv:sec_lcmorp}), which differ from those outside SFRs.
We note that the results of spectroscopic follow-up of a subsample of VVV objects associated with SFRs (Paper II) show a figure of 25$\%$, a consistent figure despite some additional selection effects in that subsample.

\subsection{Properties of variables outside SFRs}\label{vvv:sec_varsnonsfrs}

To establish the nature of the objects that could be contaminating our sample of likely 
YSOs, and may also be interesting variable stars, we study the properties of objects 
found outside areas of star formation. Many of these are listed in SIMBAD as IR sources (from the IRAS and
MSX6C catalogues) and associated with OH masers, as well as being catalogued as evolved stars in previous surveys. Visual inspection of their light curves also shows that a large percentage of objects have periodic variability, with  $P >$ 100 days, whilst the remainder of the sample shows variability over shorter timescales in the order of $P<$ 20 days. We use the phase dispersion minimization \citep[{\sc pdm},][]{1978Stellingwerf} algorithm  found in the {\sc noao} package in {\sc iraf}  to search for a period in the light curve of these objects. This allows us to derive the periods or at least the approximate timescale of the variability of objects found outside areas of star formation. To provide a comparison with the {\sc pdm} results we also 
used the {\sc gatspy LombScargleFast} implementation of the Lomb Scargle algorithm, which benefits from automatic 
optimisation of the frequency grid so that significant periods are not missed. We found that {\sc pdm} was generally better
for the purpose of this initial investigation. The Lomb Scargle algorithm is designed to detect sinusoidal variations 
whereas {\sc pdm} makes no assumptions about the form of the light curve and is therefore much more sensitive to the periods 
of eclipsing binaries, for example. The Lomb Scargle method did help with the classification of a small number of long 
period variables (LPVs).

\begin{figure}
\centering
\resizebox{\columnwidth}{!}{\includegraphics{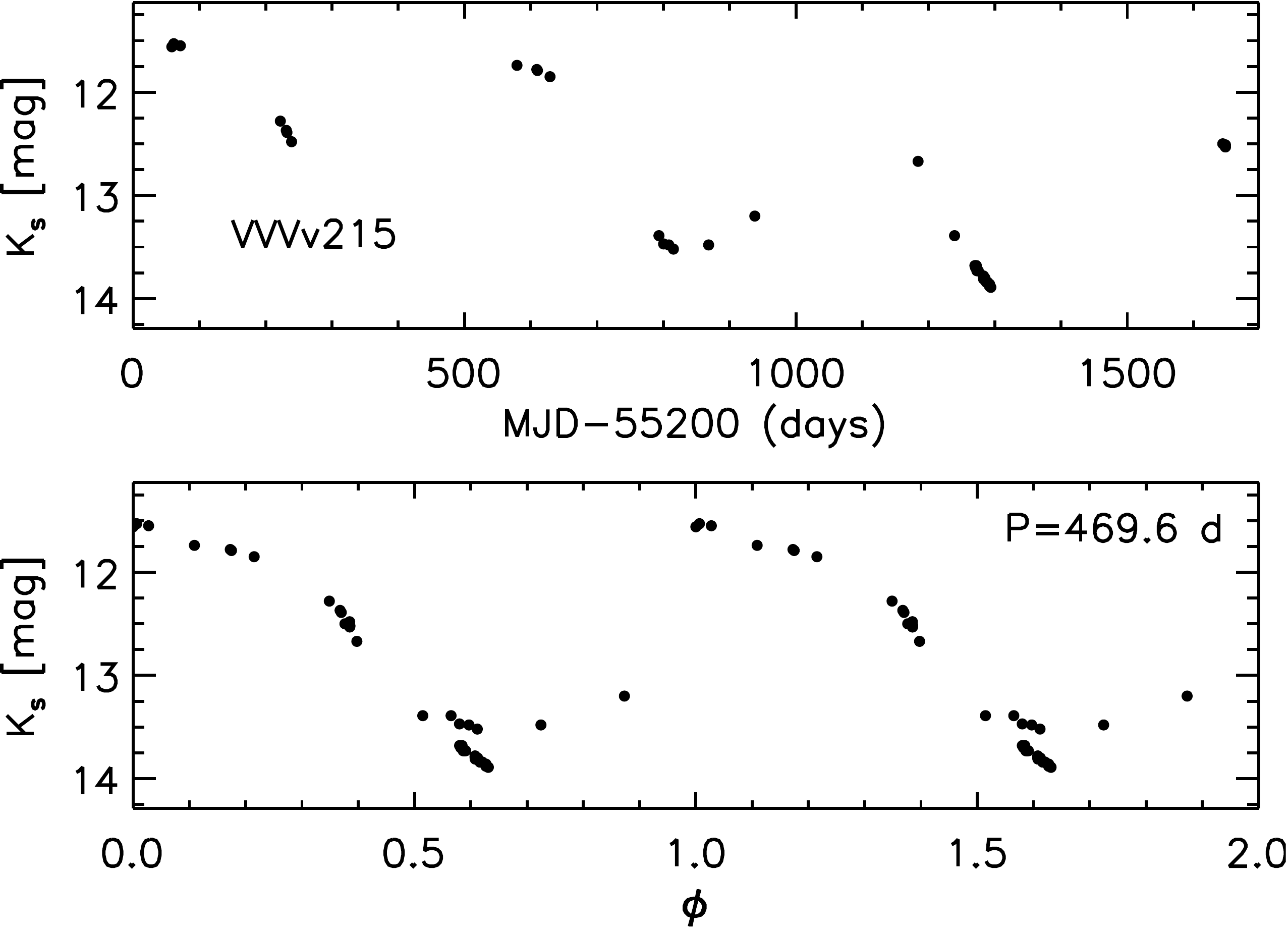}}\\
\resizebox{\columnwidth}{!}{\includegraphics{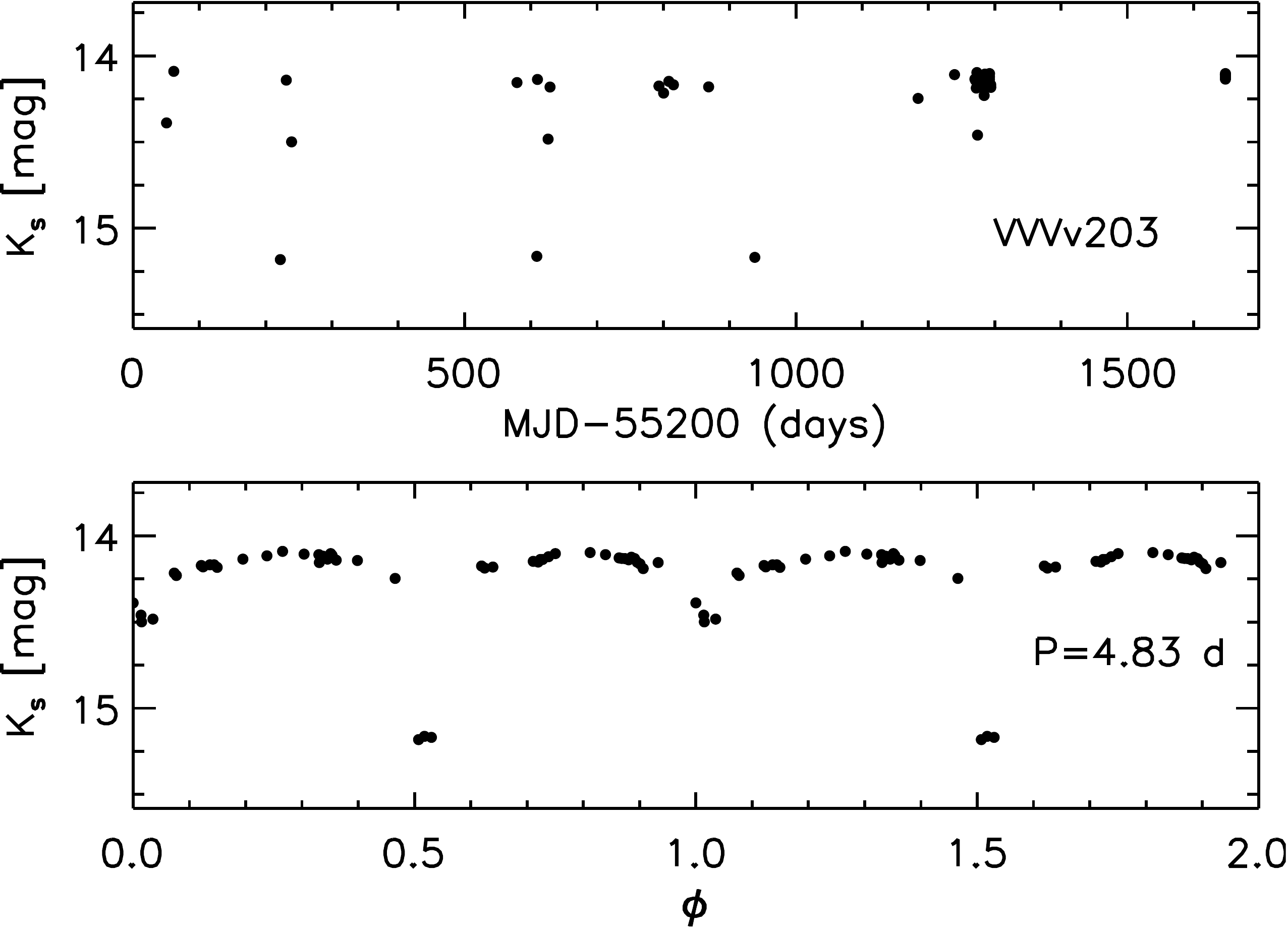}}
\caption{Examples of $K_{\rm s}$ light curves for objects not associated with SFRs. (top) the long period variable VVVv215. (bottom) the eclipsing binary VVVv203. }
\label{vvv:nonSFRs_examples}
\end{figure}

Out of the 286 stars in this subsample, 5 objects correspond to known objects from the literature (novae, EBs and a high mass X-ray binary), 45$\%$ of them are LPVs and 17$\%$ are EBs where we are able to measure a period. In addition, 30$\%$ of the sample is comprised of objects in which variability seems to occur on short timescales. The light curves of many objects in the latter group resemble those of EBs with measured periods, but with only 1 or 2 dips sampled in the dataset. We suspect that most of these could also be EBs but we are not able to establish the periods. Finally, we also find 18 objects (6$\%$) that do not appear to belong to any of the former classes.  In Fig. \ref{vvv:nonSFRs_examples} we show two examples of objects belonging to these different subclasses where a period could be derived.
The LPV VVVv215 is typical of many of the dusty Mira variables in the dataset that show long term trends caused by changes in 
the extinction of the circumstellar dust shell. These trends are superimposed on the pulsational, approximately sinusoidal 
variations, with the result that the $K_{\rm s}$ magnitude as a given point in the phase curve can differ between cycles.



\begin{figure}
\centering
\resizebox{\columnwidth}{!}{\includegraphics{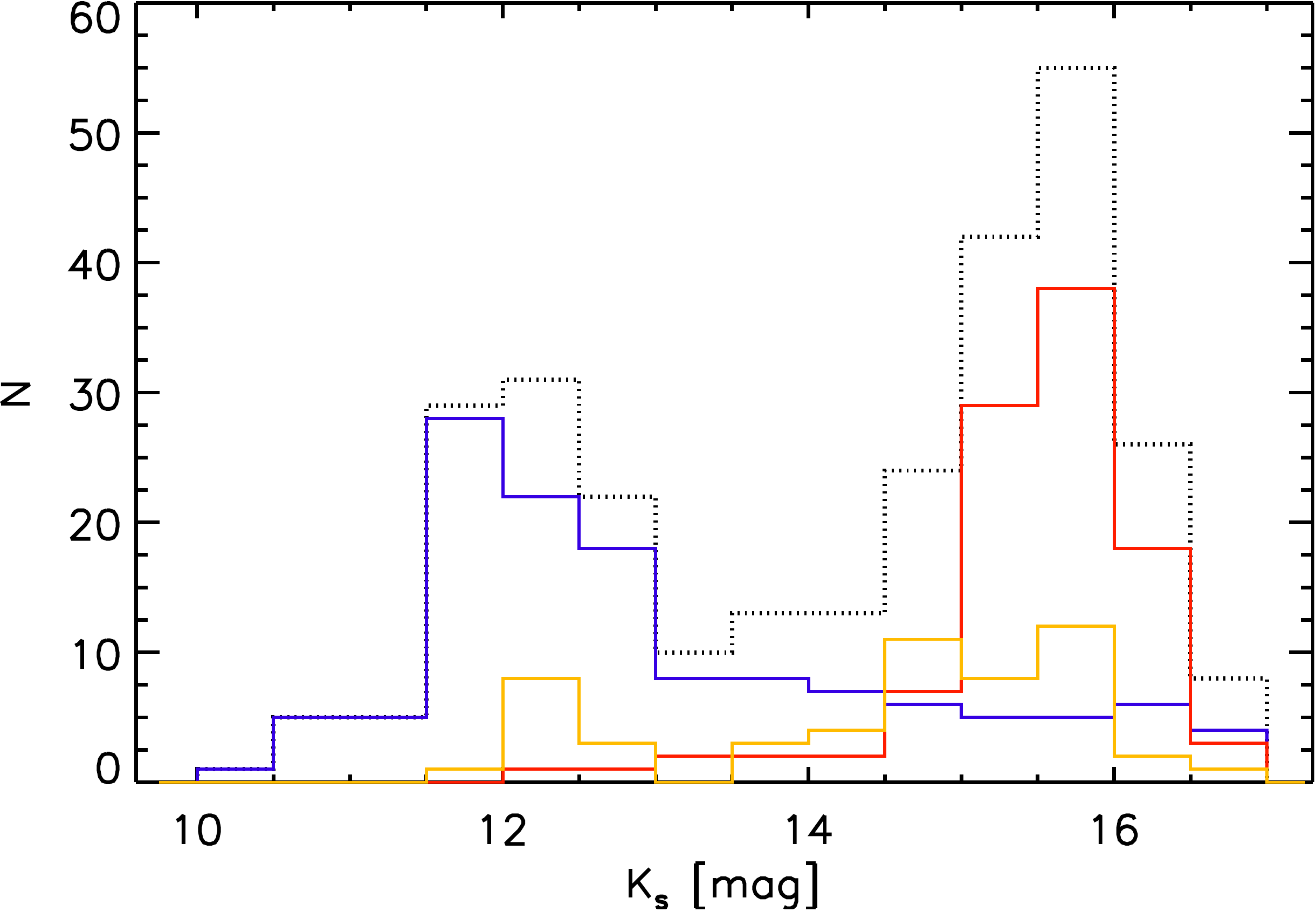}}\\
\resizebox{\columnwidth}{!}{\includegraphics{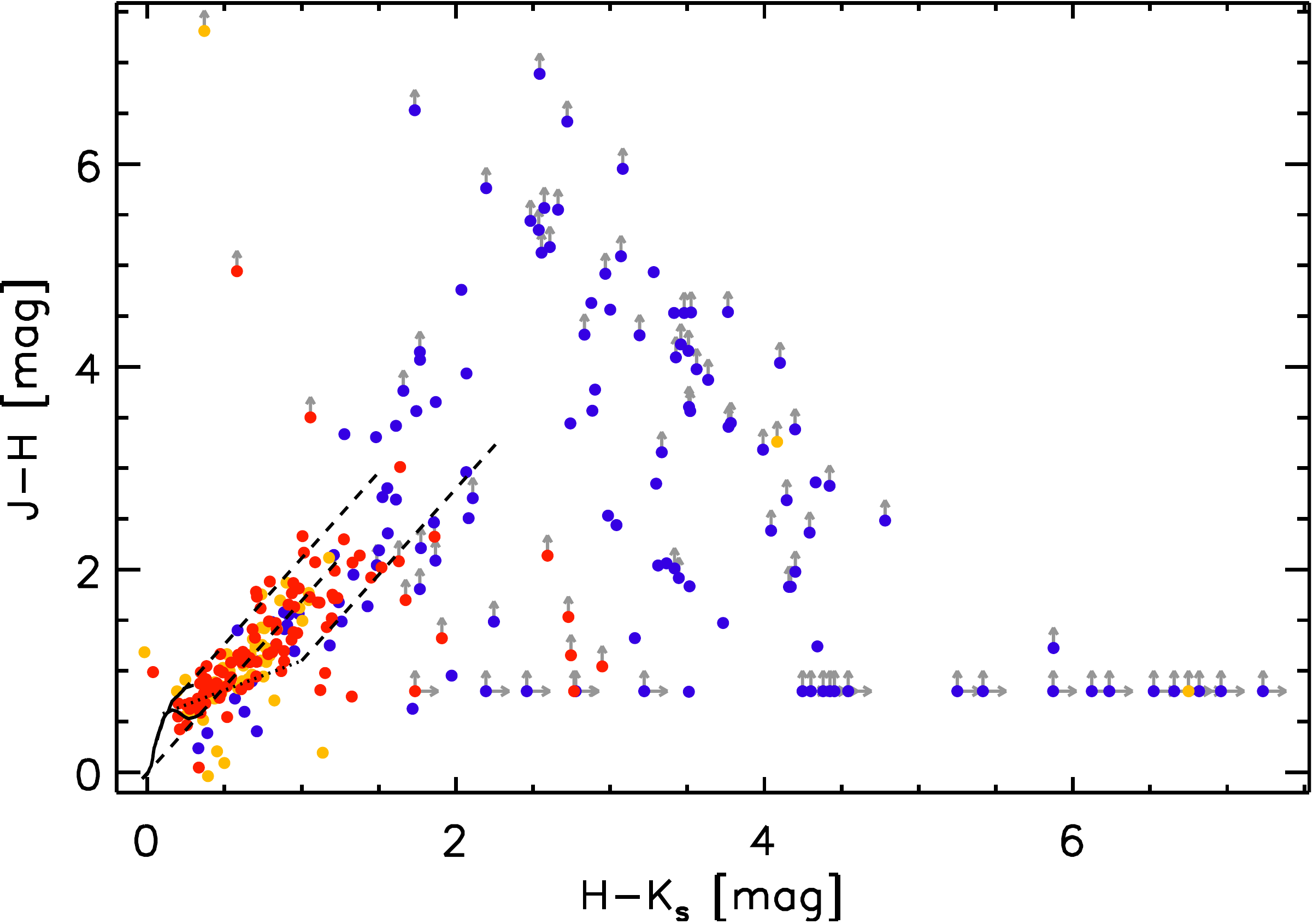}}
\caption{(top) Overall $K_{\rm s}$  distribution (from 2010 data) of objects not associated with SFRs (dotted line), and the same distribution separated for LPVs (solid blue line), EBs and known objects (solid orange line), and other classes of variable stars (solid red line). (bottom) $(J-H)$, $(H-K_{\rm s})$ colour-colour diagram for LPVs (blue circles), EBs and known objects (orange circles), and other classes of variable stars (red circles). In the diagram, arrows mark lower limits. The solid curve at the lower left indicates the colours of main sequence stars. The short-dashed line is the CTTS locus of \citet{1997Meyer} and the long-dashed 
lines are reddening vectors.}
\label{vvv:nonsfrs_prop}
\end{figure}

The objects belonging to different classes show very different properties. Figure \ref{vvv:nonsfrs_prop} shows the $K_{\rm s}$ distribution for these objects, where it can be seen that the LPVs dominate the bright end of the distribution with a peak at $K_{\rm s} \sim 11.8$ magnitudes, and showing a sharp drop at brighter magnitudes, probably due to the effects of saturation. EBs and other classes are usually found at fainter magnitudes. The near-infrared colours of the two samples (bottom plot of Fig. \ref{vvv:nonsfrs_prop}) show that LPVs tend to be highly reddened objects or have larger $(H-K_{\rm s})$ colours than EBs and other classes, which usually have the colours of lightly reddened main sequence and giant stars. We will see in 
Sect. \ref{vvv:sec_physmec} that this low reddening and lack of K band excess (in most cases) distinguishes the EBs and other shorter period 
variables from the sample spatially associated with SFRs, so contamination of the SFR sample by these shorter-period 
variables should not be very significant. 

The colour-colour diagram of Fig. \ref{vvv:nonsfrs_prop} also supports the idea that this sample might contain some 
bona fide YSOs that are not revealed by our searches of SIMBAD and the WISE images, as mentioned in our discussion of contamination. In the figure we observe objects (red circles) that show $(H-K_{\rm s})$ colour excesses and are neither known variables nor classified as LPVs or EBs. By simply selecting red circles located to the right of the reddening vector passing through 
the reddest main sequence stars we estimate that 44 objects have colours consistent with a YSO nature. This would represent 15$\%$ of objects outside SFRs and 5\% of the full sample of 816 variables. A more detailed study would of course be needed 
to confirm their nature as YSOs. We also note that the lower left part of the classical T Tauri stars (CTTS) locus plotted in the figure extends into the region
occupied by lightly reddened main sequence stars, so in principle some of these individual red circles can potentially be YSOs.


The bright LPVs are very likely pulsating Asymptotic Giant Branch (AGB) stars. These type of stars are usually divided into Mira variables, which are characterized by displaying variability of $\Delta K > 0.4$~magnitudes and  with periods in the range $100 <$~P$< 400$~d, and dust-enshrouded AGB stars, which are heavily obscured in the optical due to the thick circumstellar envelopes (CSE) developed by heavy mass loss ($\sim 10^{-4}$~M$_{\odot}$~yr$^{-1}$). The latter group, comprised of carbon-rich and oxygen-rich stars (the latter often referred to as OH/IR stars if they display OH maser emission), show larger amplitudes in the K-band (up to 4 magnitudes) and have periods between 400 $<$~P$< 2000$~d \citep[for the above, see e.g,][]{2006Jimenez2,2008Whitelock}.

AGB stars are bright objects and should be saturated at the magnitudes covered in VVV. However, due to the large extinctions towards the Galactic mid-plane we are more likely to observe these type of objects in VVV compared to our previous UKIDSS study. We can estimate the  apparent magnitude of Mira variables at the different Galactic longitudes covered in VVV, and by assuming that these objects are located at the Galactic disc edge \citep[R$_{GC}=14$~kpc][]{2011Minniti} and then considering other Galactocentric radii. At a given longitude, $l$, we derive A$_{V}$ as the mean value of the interstellar extinction found at latitudes $b$ between $-1^{\circ}$ and $1^{\circ}$. The interstellar extinction is taken from the \citet{1998Schlegel} reddening maps, and corrected following \citet{2011Schlafly}, i.e. $E(B-V)=0.86E(B-V)_{Schlegel}$. We then assume that extinction increases linearly with distance, at a rate A$_{V}/D_{edge}$ (mag kpc$^{-1}$), with $D_{edge}$ the distance to the Galactic disc edge at the corresponding $l$. We finally take the absolute magnitude as M$_{K}=-7.25$ \citep{2008Whitelock}. 

\begin{figure}
\centering
\resizebox{\columnwidth}{!}{\includegraphics{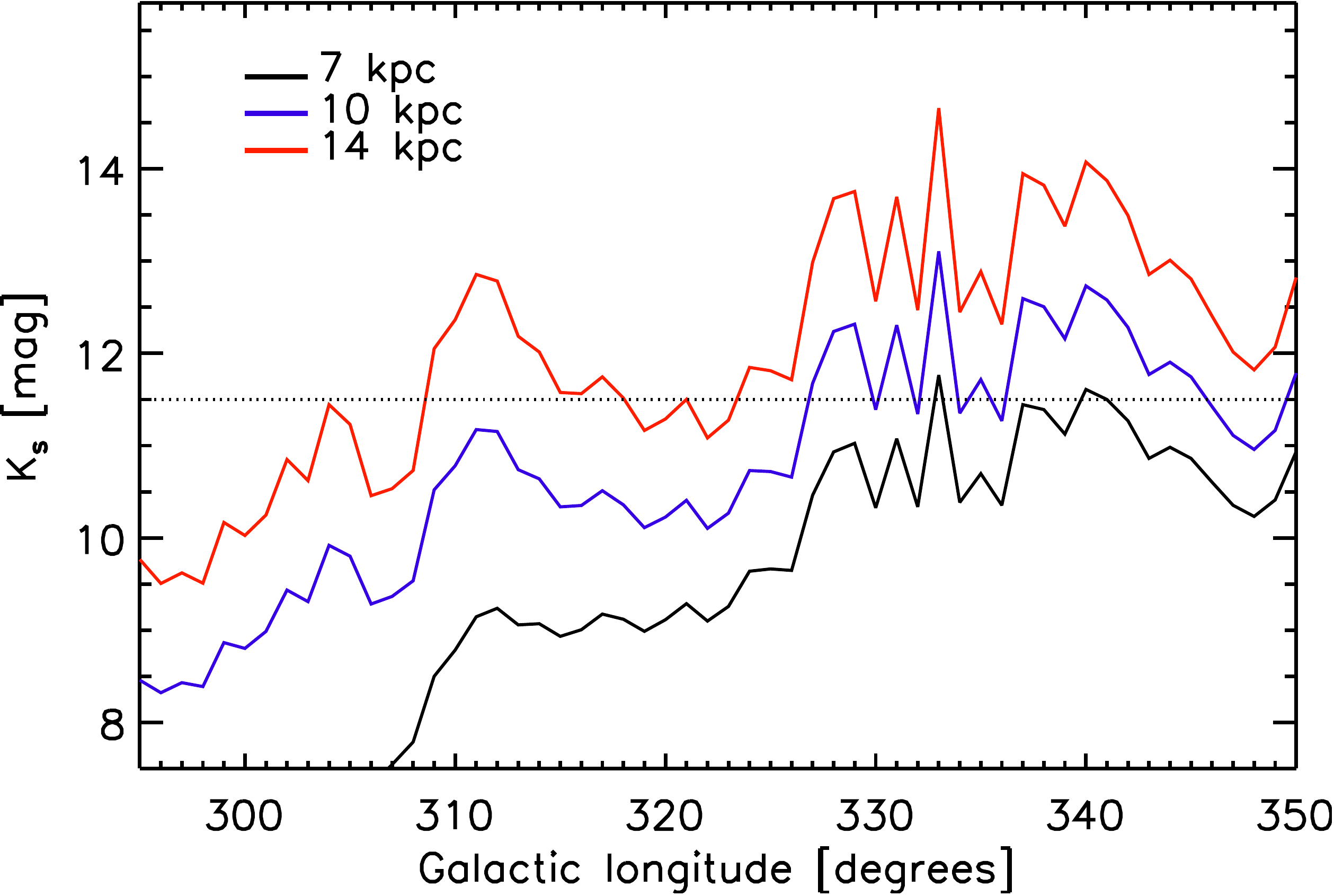}}\\
\resizebox{\columnwidth}{!}{\includegraphics{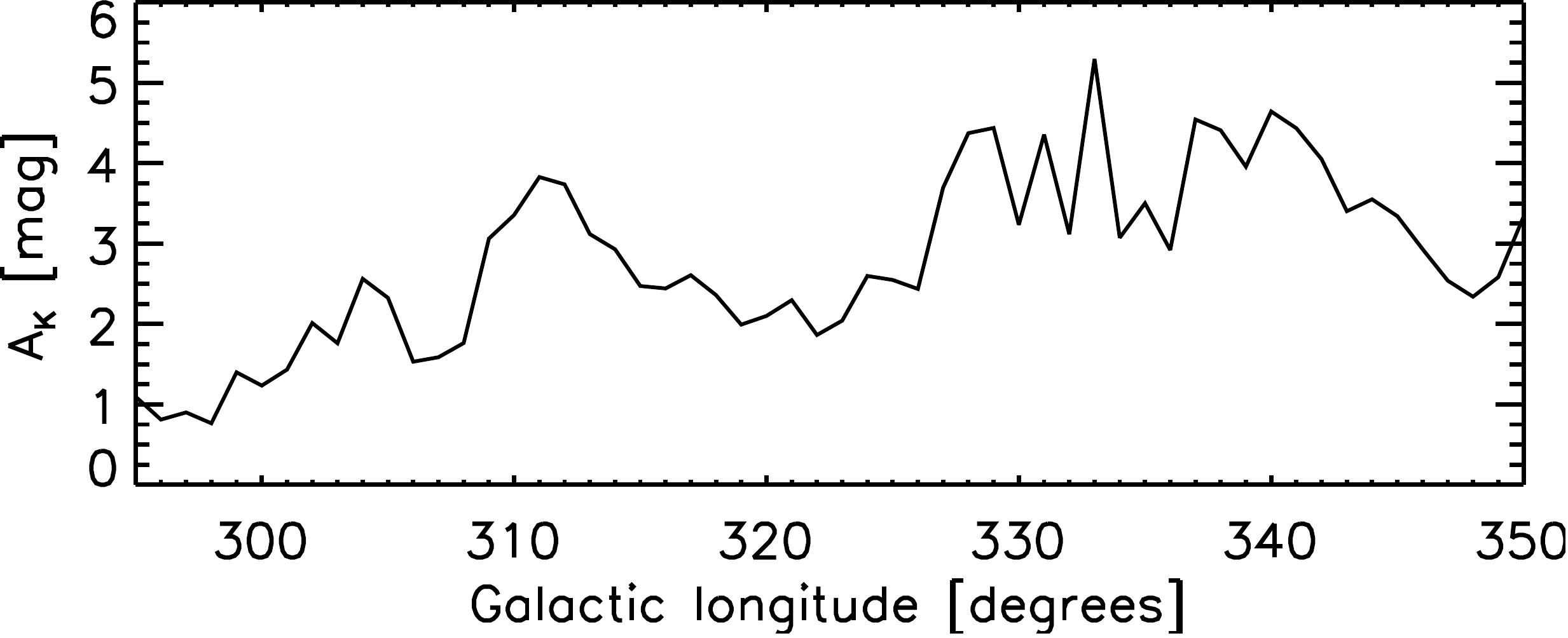}}
\caption{(top) Apparent $K_{\rm s}$ magnitude, derived as explained in the text, for a Mira variable located at the Galactic disc edge (solid red line). The same value is shown for a Mira located at lower Galactic radii R$_{GC}=10$~kpc (blue line) and R$_{GC}=7$~kpc (black line). The magnitude where the number of detections for LPVs drops ($K_{\rm s}=11.5$ magnitudes) is marked by a dotted line. (bottom) K-band Galactic extinction column as a function of Galactic longitude.}
\label{vvv:miras_apmag}
\end{figure}

Figure \ref{vvv:miras_apmag} shows the estimated apparent magnitudes of Mira variables at different $l$ and at varying R$_{GC}$. In the figure we also show the magnitude, $K_{\rm s}=11.5$, that marks the drop in the number of the detection of these objects, as observed in the histograms of the $K_{\rm s}$ distribution (Fig. \ref{vvv:nonsfrs_prop} ). We can see as we move away from the Galactic center, a Mira variable would most likely saturate in VVV, especially at $l < 310^{\circ}$. This occurs due to the effects of having relatively larger extinctions towards the Galactic center (see bottom plot of Fig.\ref{vvv:miras_apmag}) and that a star at R$_{GC}=14 $~kpc is located farther away from the observer as $l$ approaches $l=0^{\circ}$. We note that \citet{2011Ishihara} finds that most AGB stars are found at $R_{GC}< 10$ kpc so if we place a Mira variable at smaller Galactic radii (R$_{GC}=$7, 10 kpc), we see that it is less likely for such a star to show up in our sample. However, variable dust-enshrouded AGB stars which undergo heavy mass loss, suffer heavy extinction due to their thick circumstellar envelope, and thus are fainter than Mira variables \citep[AGB stars with optically thick envelopes are found to be $\sim$5 $K_{\rm s}$ magnitudes fainter than objects with optically thin envelopes in the work of][]{2006Jimenez} and thus less likely to saturate in VVV, even at large distances. Then most AGB stars in our sample are probably dust-enshrouded objects.

\begin{figure}
\centering
\resizebox{\columnwidth}{!}{\includegraphics{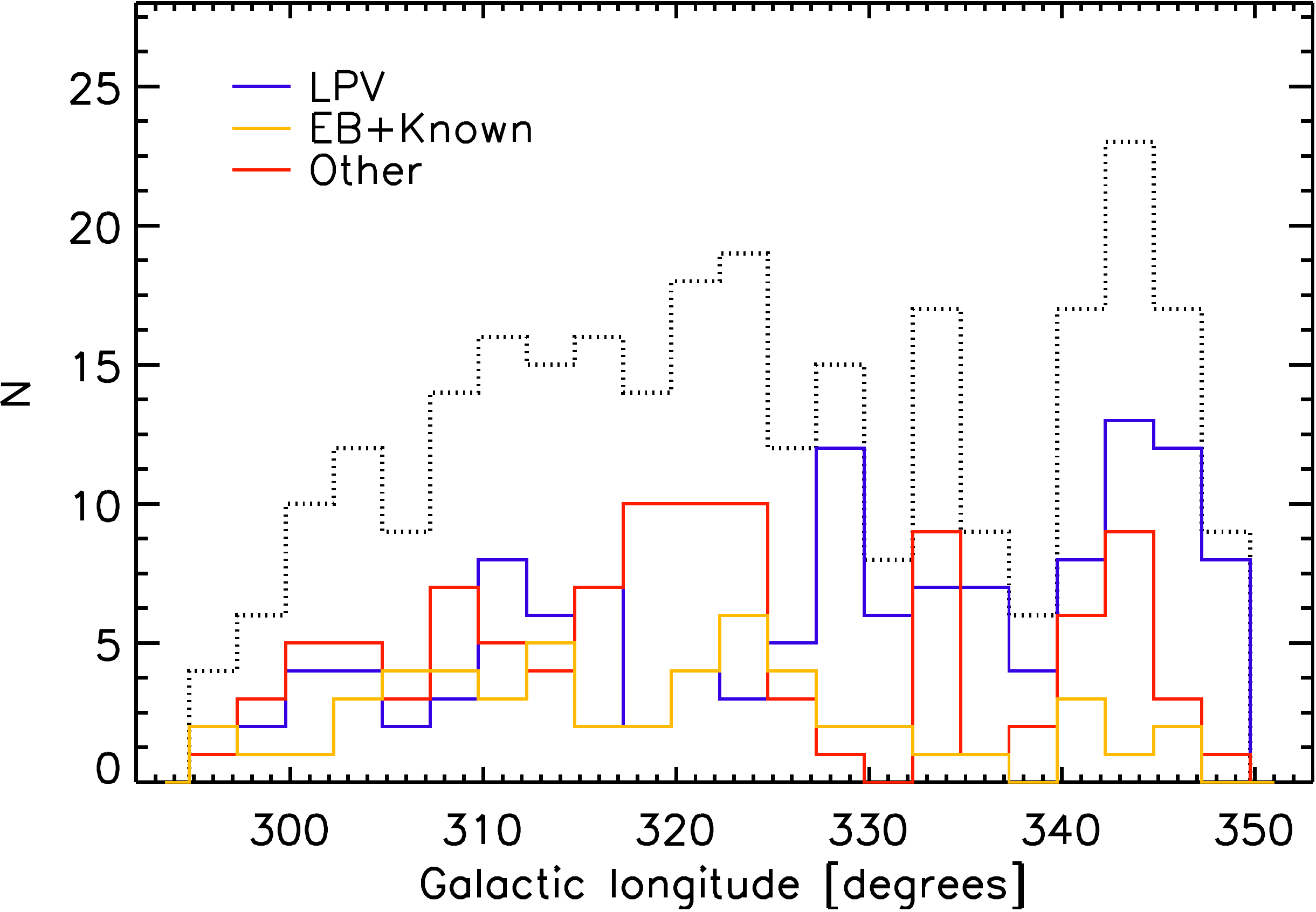}}\\
\resizebox{\columnwidth}{!}{\includegraphics{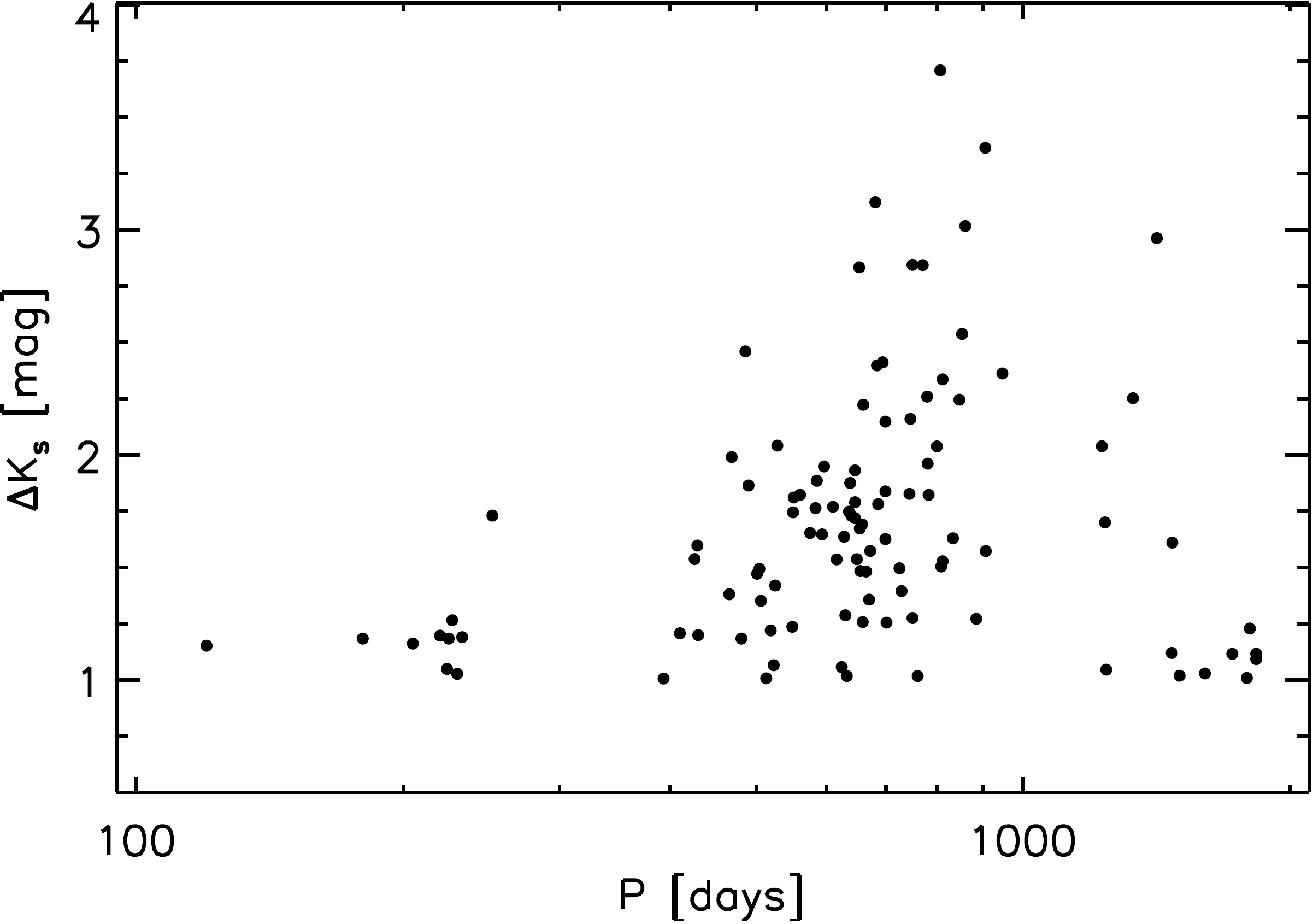}}
\caption{(top) Overall Galactic longitude distribution of objects outside SFRs (black line). We also show the same distribution divided into LPVs (blue line), EBs and known objects (orange line), and other classes of variable stars (red line). (bottom) Period vs $K_{\rm s}$ amplitude for LPVs with a measured period.}
\label{vvv:miras_periodamp}
\end{figure}

The observations confirm the trend expected from the analysis above. Figure \ref{vvv:miras_periodamp} shows that the number of LPVs increases as we come closer to the Galactic center. In addition, when taking into account AGB stars with measured periods, we confirm that the majority of AGB stars show periods longer than 400 days and large amplitudes (see lower panel in Fig. \ref{vvv:miras_periodamp}), as expected in heavily obscured AGB stars. It is interesting to see in the same figure, that variable objects with periods longer than 1500 days show lower amplitudes than expected for their long periods. This is similar to the observed trend in the variable OH/IR stars of \citet{2006Jimenez2}. According to the authors, these objects correspond to stars at the end of the AGB. We note that the apparent lack of high amplitude objects at longer periods could relate to the fact that more luminous (longer period) objects have smaller amplitudes expressed in magnitudes \citep[red supergiants often display $\Delta K<1$~mag, see e.g.][]{2008VanLoon}.

\begin{figure}
\centering
\resizebox{\columnwidth}{!}{\includegraphics{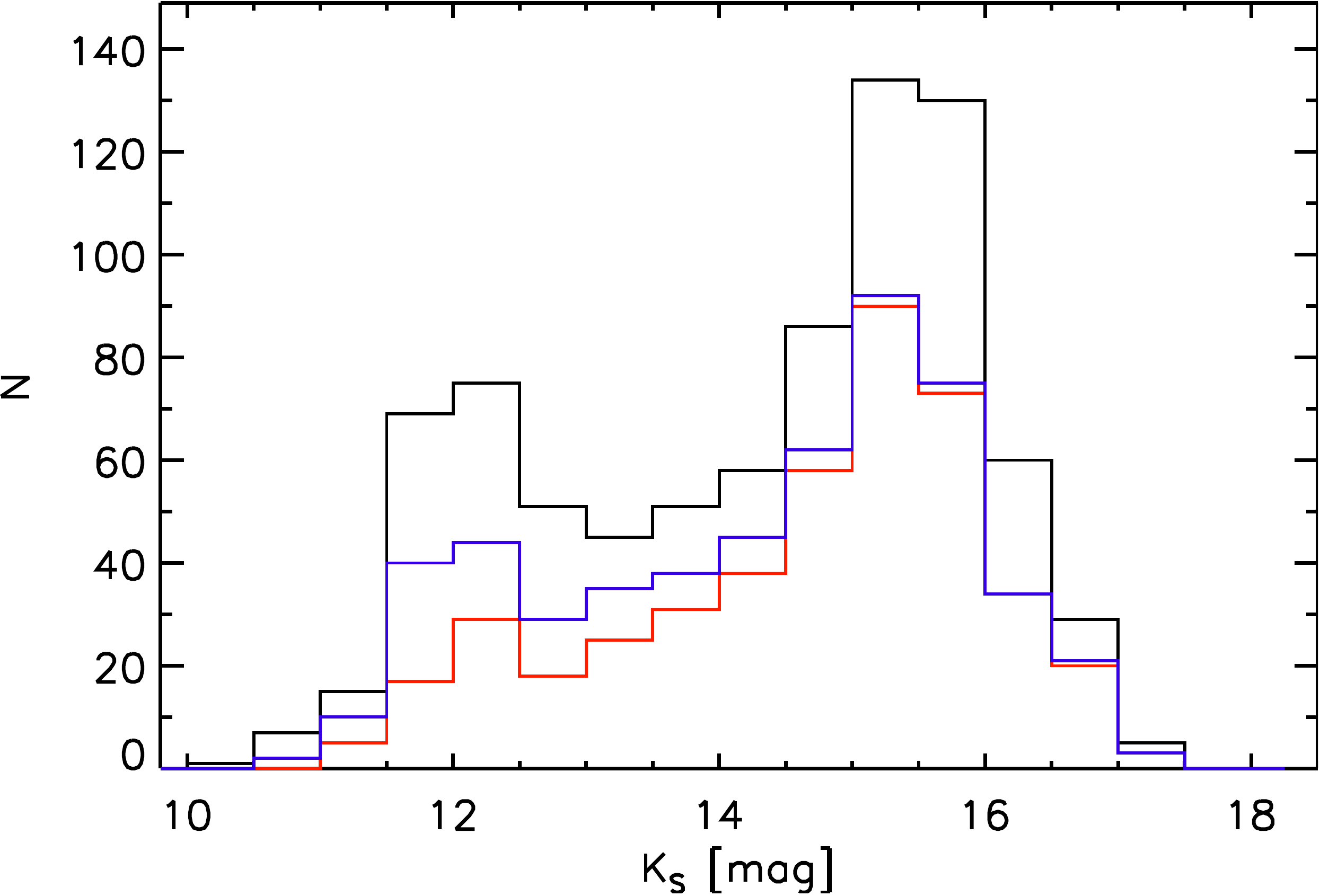}}\\
\caption{$K_{\rm s}$ distribution (from 2010 data) of the 816 VVV selected variable stars (solid black line). We show the same distribution for the overall sample of stars found to be likely associated with areas of star formation (blue line) and for the same sample but removing objects that show Mira-like light curves in Sect. \ref{vvv:sec_lcmorp} (red line). }
\label{vvv:k_dist}
\end{figure}

This population of bright pulsating AGB stars can also explain the observed bimodality of the $K_{\rm s}$ distribution for the full sample of VVV high amplitude variable stars (see Fig. \ref{vvv:k_dist}). The peaks of the distribution occur at $K_{\rm s}\sim 11.8$ and $K_{\rm s}\sim 15.8$. The peak at the bright end is at the same magnitude as the peak for LPVs. 

When we only plot objects which are found to be likely associated with areas of star formation, the peak at the bright end becomes less evident (see blue histogram in Fig. \ref{vvv:k_dist}).  When we plot only SFR-associated sources that do not have LPV-like
light curves (see Sect. \ref{vvv:sec_lcmorp}) the bimodality almost disappears, as shown by the red histogram in Fig. \ref{vvv:k_dist}. 
AGB stars are probably the main source of contamination in our search for eruptive YSOs in SFRs, especially dust-enshrouded AGB stars which can have infrared colours resembling those of YSOs. Hence it is fortunate that we can remove most of this 
contamination by selecting against LPV-like light curves.

\subsection{YSO source density}

Our finding in Sect. \ref{vvv:sec_sfrass} that YSOs constitute about half of the detected population of high amplitude variables in VVV disc fields indicates that they represent the largest single population of high-amplitude infrared variables in the Galactic mid-plane, at least in the range $11 < K < 17$. We note that extragalactic studies of high-amplitude stellar variables are dominated by the more luminous AGB star population \citep[see e.g.][]{2015Javadi}. Our analysis only considered the VVV disc tiles 
with $\left|b\right|<1^{\circ}$, this amounts to 76 tiles, covering each 1.636 deg$^{2}$ of the sky. After
allowing for the small overlaps between adjacent tiles, the total area covered in this 
part of the survey is 119.21 deg$^{2}$.  Adopting a 50\% YSO fraction for the full sample of 816 variables
implies a source density of 3.4 deg$^{-2}$. As noted in Sect. \ref{sec:vvvselec}, the stringent data quality cuts in our selection procedure excluded $\sim 50\%$ of high amplitude 
variables down to $K_{\rm s}$=15.5, and a high fraction at fainter magnitudes where completeness falls (see Fig. \ref{vvv:k_dist}). The corrected source density is 
therefore $\sim$7 deg$^{-2}$.

When considering the source density of high amplitude infrared variables in the UGPS, \citet{2014Contreras} argued that the observed source density under-estimates 
the actual source density due to three main effects: (1) with only two epochs of K-band data most high amplitude 
variables will be missed; (2) the source density rises towards the magnitude
cut of K$<16$, indicating that many low luminosity PMS variables that are detected at distances of 
1.4 to 2 kpc would be missed at larger distances; and (3) the dataset used in the UGPS search of \citeauthor{2014Contreras} excludes the mid-plane and
is therefore strongly biased against SFRs. The UGPS YSO source density is estimated to reach 12.7 deg$^{-2}$ when correcting for these 3 factors. In the case of VVV, given the higher number of epochs obtained from this survey, and that this analysis is not biased against areas of star formation, the source density is only likely to be affected by item (2) of the UGPS analysis. Figure \ref{vvv:k_dist} shows the magnitude distribution of the VVV variables associated with SFRs, where we can see a similar behaviour to the UGPS results, with the density of sources rising steeply towards faint magnitudes. Contrary to the UGPS search, we do not have a hard magnitude cut in the VVV sample, which includes sources as faint as $K_{\rm s} \sim 17$. However, the number of sources decreases at $K_{\rm s} > 16$, so we estimate an effective magnitude detection limit of $K_{\rm s}=16.25$~magnitudes. This implies that if typical sources from VVV have similar characteristics to UGPS objects in Cygnus and Serpens ($K=14.8, d=1.4-2$~kpc), then we would not detect them at distances $d>3.32$ kpc. The complete sample of star forming complexes from \citet{2003Russeil} shows that $83\%$ of them are located beyond these distances. Correcting for this factor we then estimate a true source density of 41 $\deg^{-2}$, though this figure does not include YSOs with low mass and luminosity that are too faint to be sampled by VVV due to the absence of nearby SFRs in the survey area. This figure of 41 $\deg^{-2}$ is three times larger than the one estimated from the UGPS analysis of \citet{2014Contreras} (12.7 deg$^{-2}$).


Two effects can account for the larger source density in VVV than UGPS. (1) High luminosity YSOs are less common, but they can be observed at larger distances. The UGPS study would not find such objects at large distances because the available dataset did not cover the mid-plane of the Galactic disc, in which all distant SFRs are located due to their small scale 
height. Since VVV does 
cover the midplane we are able to detect these rare higher luminosity YSOs. This seems to be supported by the slightly larger distances established for members
of the spectroscopic subsample in paper II. (2) In the UGPS study, most (23/29) of the variables in SFRs were located in just 2 large SFRs: the Serpens OB2 association and 
portions of Cygnus X. The much smaller size of the UGPS sample (in number of SFRs and number of variables) meant that there was considerable statistical uncertainty
in the area-averaged source density. Moreover, the incidence of high amplitude variability is greater at the earlier stages of YSO evolution (see Sect. \ref{vvv:sec_alphaclass}) so the 
numbers in the UGPS study may have been reduced by a relative lack of YSOs at these stages in the two large SFRs surveyed.



The estimated highly variable YSO source density remains much larger than that estimated for Mira variables in \citet{2014Contreras}, 
indicating a higher average space density for the variable YSOs. The observed variables in SFRs also outnumber the EBs and unclassified 
variables in the magnitude range of this study. However, we are likely to miss a large part of the population of high amplitude EBs due to 
the sparse time sampling of VVV.


In Appendix \ref{appen:EBs} we attempt to calculate the source density and space density 
of high amplitude EBs from the OGLE-III Galactic disc sample of \citet{2013Pietrukowicz}. In this we are aided by a recent analysis of the physical 
properties of the large sample of {\it Kepler} eclipsing binaries 
\citep{2014Armstrong}, which indicates that EBs with high amplitudes in the 
VVV $K_{\rm s}$ and OGLE $I$ passbands are dominated by systems with 
F to G-type primaries. We use simple calculations to show that while EBs can 
have very high amplitudes at optical wavelengths, the eclipse depth should 
not exceed 1.6 mag in $K_{\rm s}$. Similarly, we find that eclipse depths 
should not exceed 3 mag in $I$. These results are supported by the VVV and 
OGLE-III datasets \citep{2013Pietrukowicz} in which the distribution of EB
amplitudes falls to zero by these limits. YSOs with $\Delta K_{\rm s} > 1.6$~mag
are very numerous in our sample so we conclude that high amplitude YSOs 
greatly outnumber EBs above this limit. Below $\Delta K_{\rm s} = 1.6$~mag it 
is harder to reach a firm conclusion (see Appendix \ref{appen:EBs}). The space densities of 
EBs and those YSOs massive enough to be sampled by VVV may be comparable at 
$1< \Delta K_{\rm s} < 1.6$~mag. High amplitude YSOs are likely to be more 
numerous if the variability extends down to the peak of the 
Initial Mass Function at low masses, given that high 
amplitude EB systems contain a giant with mass of order 1~M$_{\odot}$.

\section{Analysis of variables in star formation regions}\label{vvv:sec_physmec}

The results presented here  concern YSO variability in the K$_{\rm s}$ bandpass. At these wavelengths, variability in typical YSOs is produced by physical mechanisms (or a combination of them) affecting the stellar photosphere, the star-disc interface, the inner edge of the dust disc as well as spatial scales beyond 1 au \citep[see e.g.][]{2015Rice}. These mechanisms include cold or hot spots on the stellar photosphere \citep[e.g.][]{2012Scholz}, changes in disc parameters such as the location of the inner disc boundary, variable disc inclination and changes in the accretion rate \citep[as shown by][]{1997Meyer}. Variable extinction along the line of sight can also be responsible for the observed changes in the brightness of YSOs. Dust clumps that screen the stellar light have been invoked to explain the variability observed in Herbig Ae/Be stars and early-type Classical T Tauri stars \citep[group also known as UX Ori stars, see e.g.][]{1999Herbst,2002Eiroa}. In other scenarios variable extinction can be produced by a warped inner disc, dust that is being uplifted at larger radii by a centrifugally driven wind, azimuthal disc asymmetry produced by the interaction of a planetary mass companion embedded within the disc or by occultations in a binary system with a circumbinary disc \citep[see e.g.,][and references therein]{2013Romanova,2012Bans,2013Bouvier, 2014Windemuth}. Finally, sudden and abrupt increases in the accretion rate (of up to 3 orders of magnitude) explain the large changes observed in eruptive variable YSOs. The variability in these systems traces processes occurring at the inner disc \citep[in EXors, see][]{2012Loren} or at larger spatial scales beyond 1 au, such as instabilities leading to outbursts events \citep[in FUors, see e.g.][]{2014Audard}.

The amplitude of the variability induced by most of these mechanisms is not expected to be larger than $\Delta K\sim1$~magnitude. Table 6 of \citet{2013Wolk} shows the expected  amplitude of the $K$-band variability that would be produced by these different mechanisms. Cold and hot spots, and changes in the size of the inner disc hole are not expected to show $\Delta K$ larger than 0.75 magnitudes. We do note that the variability produced by hot spots from accretion depends on the temperature of the spot and the percentage of the photosphere that is covered by such spots, thus sufficiently hot spots can produce larger changes in the magnitude of the system. The range in $\Delta$K from variable extinction is effectively limitless as it depends on the amount of dust that obscures the star. Large changes ($\Delta$K $>$~1 mag) have been observed from variable extinction in YSOs, e.g. AA Tau, V582 Mon \citep[][]{2013Bouvier, 2014Windemuth}. Nevertheless, variable extinction can be inferred from colour variability (see e.g. Sect. \ref{vvv:sec_nirchange}). \citet{2013Wolk} also estimate that a change in the accretion rate of a class II object of $\log \dot{\mathrm{M}}($M$_{\odot}$~yr$^{-1})$ from $-8.5$~to $-7$ yields $\Delta$K$\sim$0.75 magnitudes. Thus, larger changes as observed in eruptive variables will produce large amplitudes.  

Given all of the above, it is reasonable to expect variability in our YSO sample to be dominated by accretion-related variability and/or events of obscuration by circumstellar dust.

\subsection{Light curve morphologies}\label{vvv:sec_lcmorp}

We have visually inspected the light curves of our 530 SFR-associated variables in order to gain insight into the physical
mechanism causing the brightness variations. In addition, we used {\sc pdm} in {\sc IRAF} and
{\sc LombScargleFast} in {\sc GATSPY} to search for periodicity in the light curves of our objects. We stress that this is 
a simple and preliminary classification that is highly influenced by the sparse sampling of VVV. A more detailed study is planned in future, with improved precision by applying the differential photometry method of \citet{2014Huckvale} to the VVV images. We have divided the morphologies in the following classifications. 

\begin{figure}
\centering
\resizebox{\columnwidth}{!}{\includegraphics{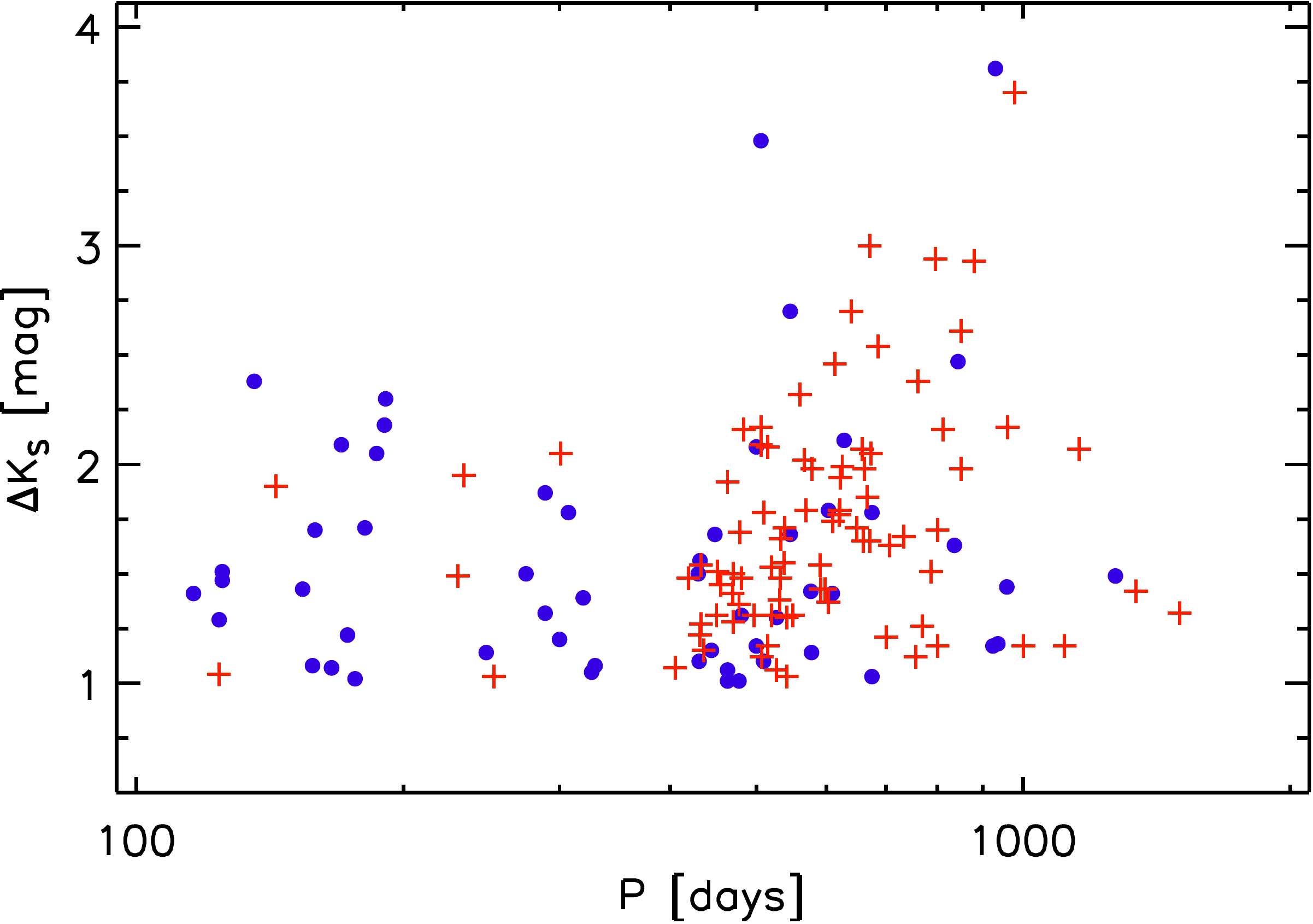}}\\
\resizebox{\columnwidth}{!}{\includegraphics{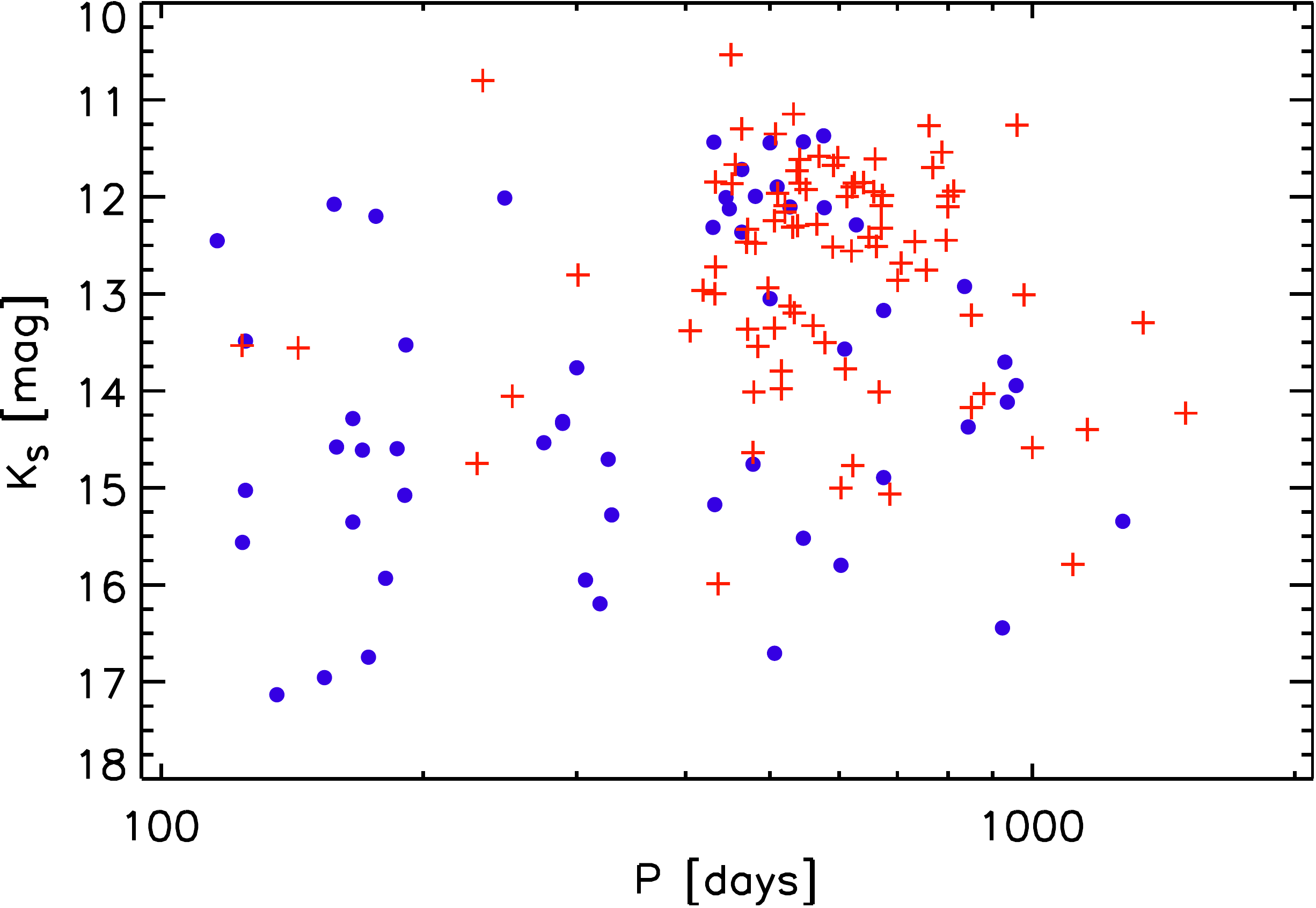}}
\caption{(top) $K_{\rm s}$ amplitude vs period for stars in SFRs with long-term periodic variability. 
LPVs that show Mira-like light curves are shown in red plus signs while other sources are 
shown as blue circles. (bottom) $K_{\rm s}$ magnitude (2010) vs period for the same sample of stars.}
\label{vvv:ysos_lpv_periodamp}
\end{figure}

\begin{itemize}

\item{{\bf Long-term Periodic Variables.}} Defined as objects showing periodic variability with $P>100$~days. This limit is adopted for consistency with the limit used in the analysis of objects outside of SFRs, with the benefit that contamination by 
long period AGB stars will be confined to this group. We measure periods for most of these objects, albeit with some difficulty in phase-folding the data in many of them. In this subsample we find 154 stars, representing 29$\%$ of objects spatially associated with SFRs. In Sect. \ref{sec:cont} we contended that field high-amplitude infrared variables with periodic light curves ($P>100$~days) are very likely dust-enshrouded AGB stars, these being identifiable
by their smooth, approximately sinusoidal light curves. We estimated that $\sim$27\% of the 530 SFR-associated variables 
would be non-YSOs and up to 45\% of these would be LPVs, implying that this subsample may contain $\sim$64 dusty Mira 
variables. Visual inspection of the 154 light curves indicates that while some have a smooth sinusoidal morphology (after 
allowing for long term trends due to variable extinction in the expanding circumstellar dust shell) others display short
timescale scatter superimposed on the high amplitude long term periodicity. In Fig. \ref{vvv:lpvmorph}, we show the 
examples of objects VVVv309 and VVVv411. The short timescale variability in their light curves is definitely not consistent 
with the typical light curves of Mira variables and their periods of 143.95 and 190.6 days, respectively, are shorter than
those of the dusty Miras detected outside SFRs (see Fig. \ref{vvv:miras_periodamp}). This short timescale scatter is typical of the scatter
observed in normal YSOs due to a combination of hot spots, cold spots, and small variations in accretion rate or extinction,
so it is reasonable to think that most of the long period variables with short timescale scatter are in fact YSOs.

\begin{figure*}
\centering
\resizebox{0.44\textwidth}{!}{\includegraphics{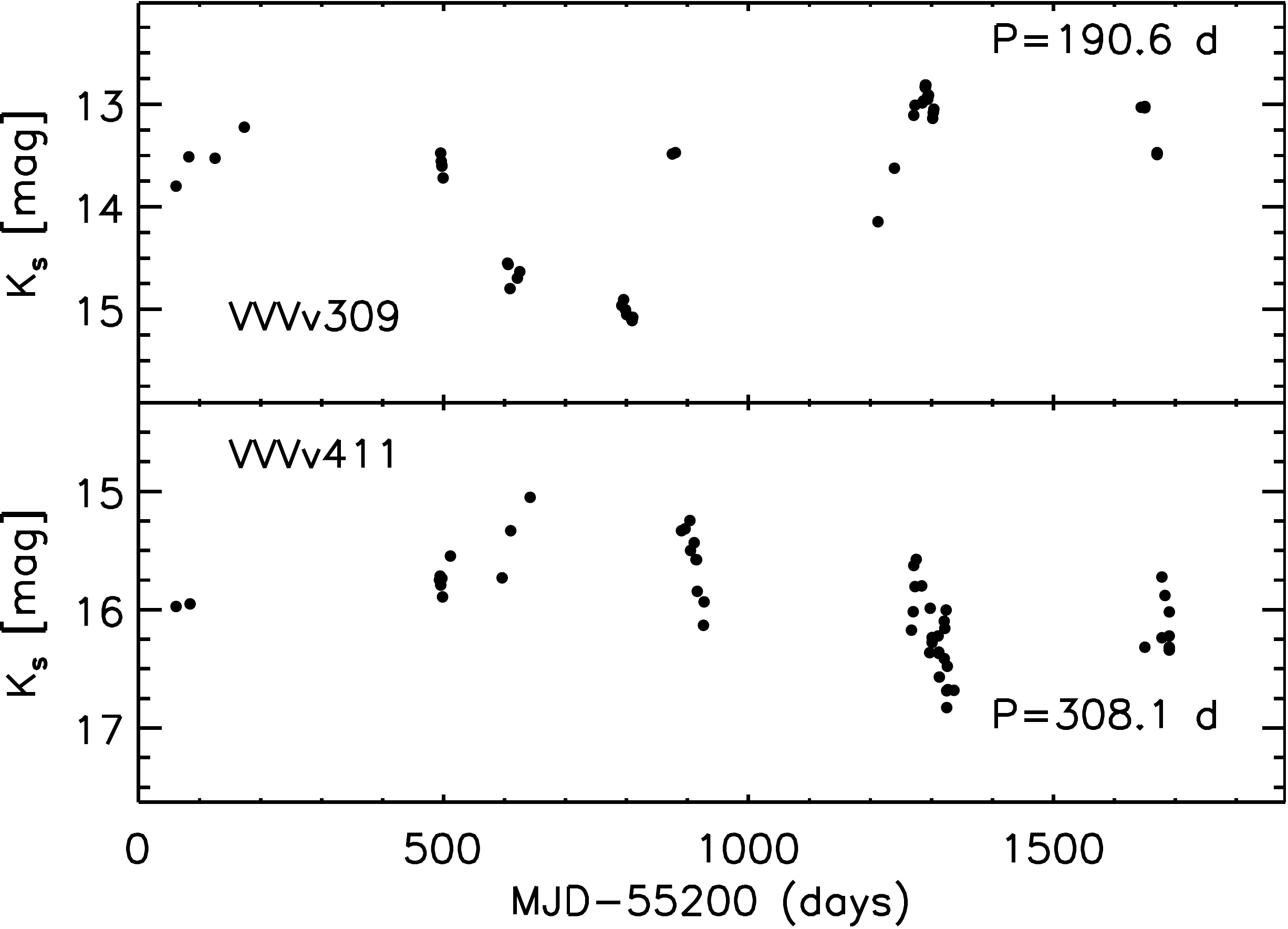}}
\resizebox{0.45\textwidth}{!}{\includegraphics{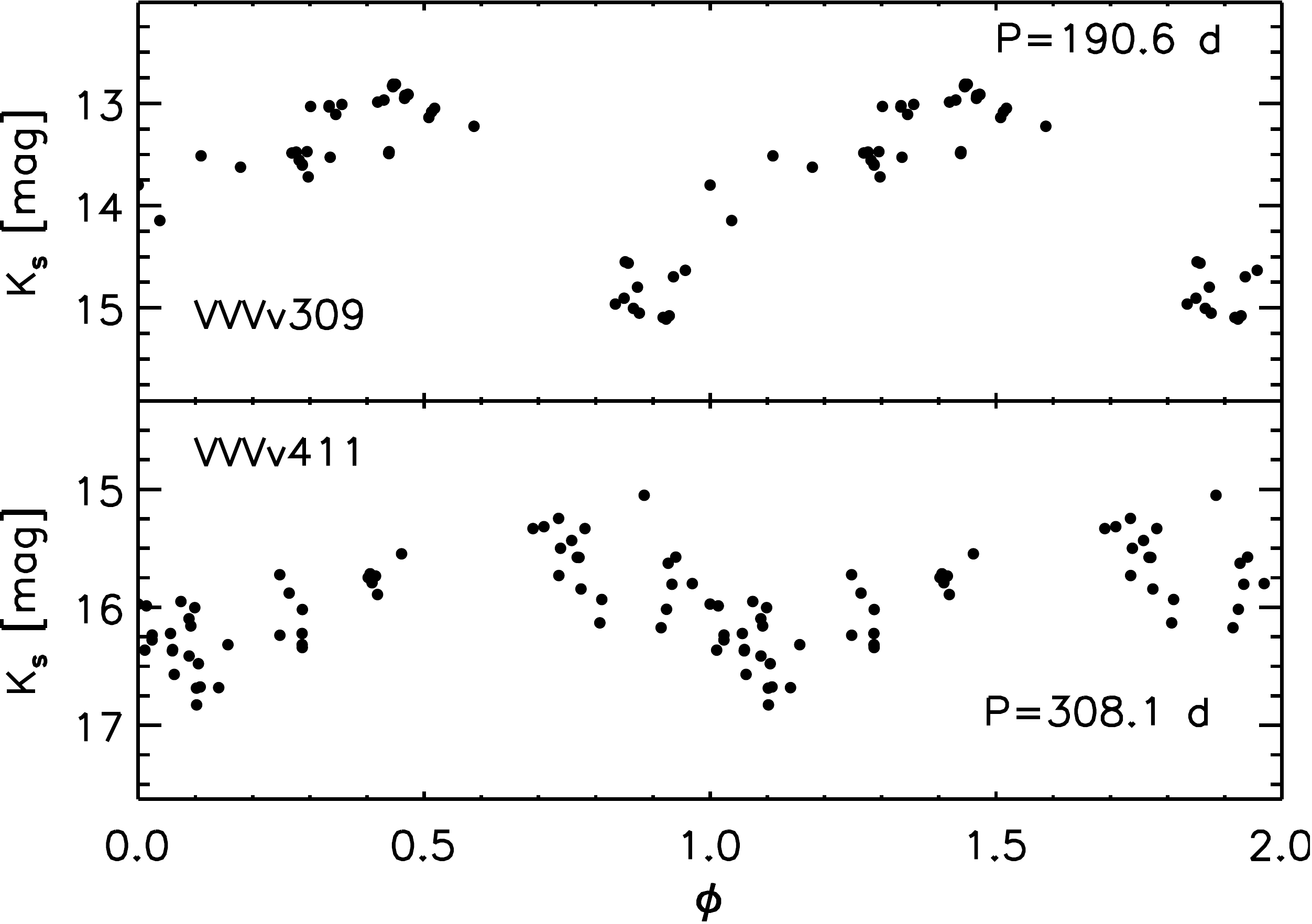}}\\
\resizebox{0.45\textwidth}{!}{\includegraphics{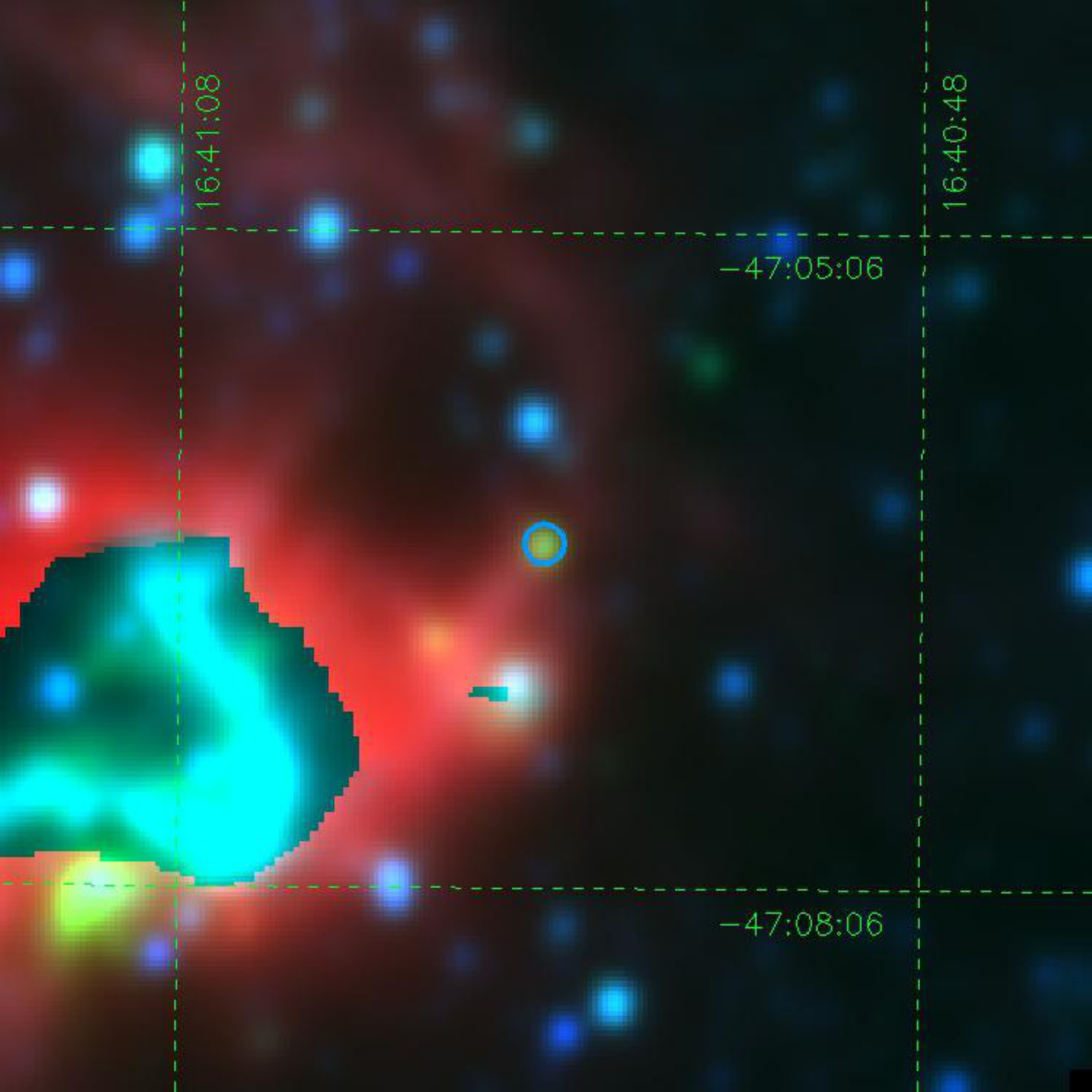}}
\resizebox{0.45\textwidth}{!}{\includegraphics{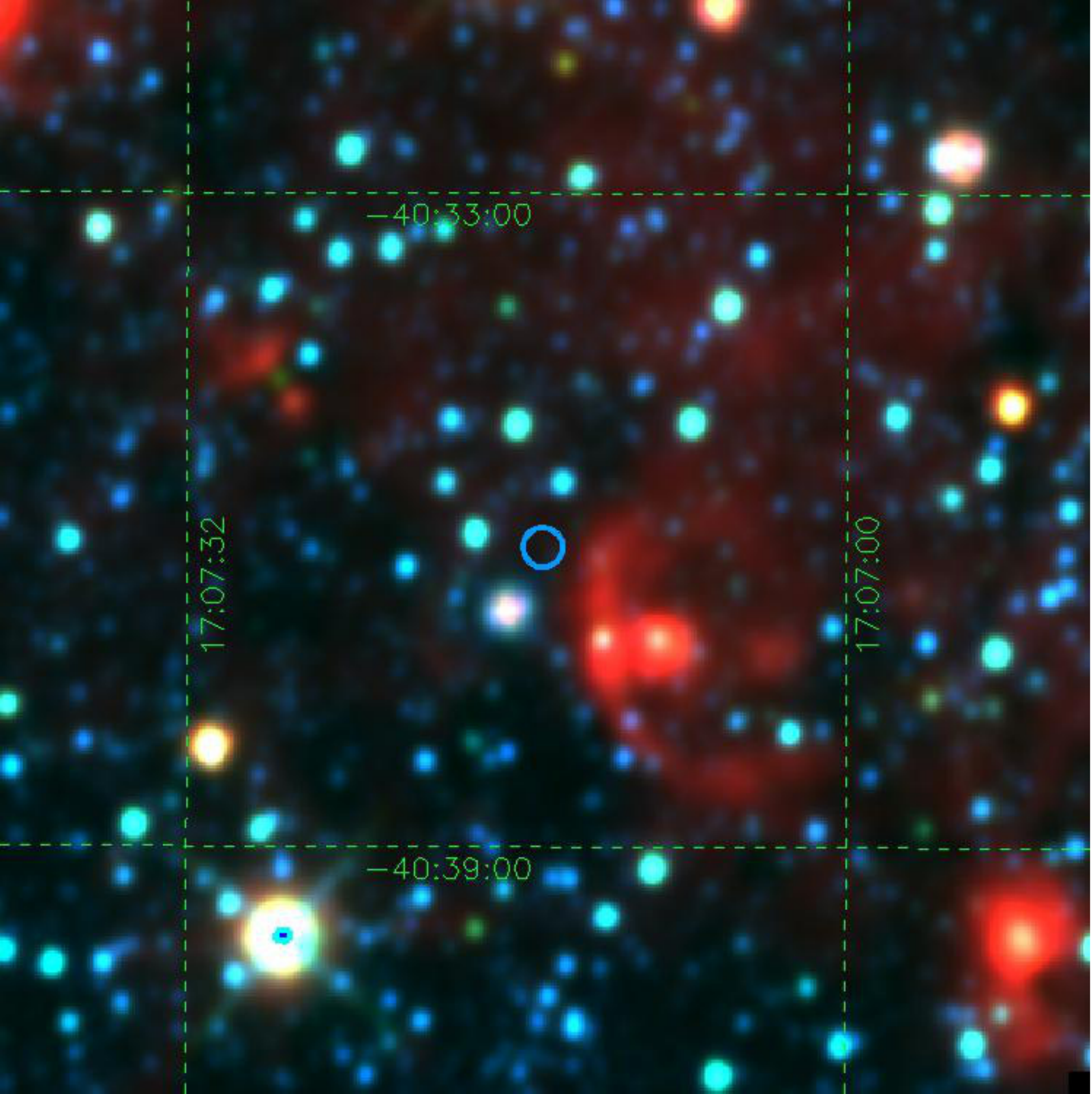}}
\caption{(top left) Examples of $K_{\rm s}$ light curves of the long-period variables VVVv309 and VVVv411, which are found in areas of star formation. (top right) Phased light curves for the same objects. (bottom) 10\arcmin$\times$10\arcmin\ WISE false colour images  (blue=3.5 $\mu$m, green=4.6 $\mu$m, red=12 $\mu$m) centred on VVVv309 (left) and VVVv411 (right).  In both images the location of the variable star is marked by a ring around the location of the object. VVVv309 is 114\arcsec\ from HII region GRS G337.90 -00.50 \citep[see e.g.][]{2011Culverhouse}. The 12$\mu$m WISE image of VVVv309 saturates at the centre of the HII region creating the blue/green ``inset'' in the false colour image. VVVv411 is located 104\arcsec\ from the infrared bubble [CPA2006] S10 \citep{2012Simpson} as well as other indicators of ongoing star formation.} 
\label{vvv:lpvmorph}
\end{figure*}

To support this interpretation, Fig. \ref{vvv:ysos_lpv_periodamp} shows the period vs $\Delta K_{\rm s}$ and 
period vs $K_{\rm s}$ distributions, for objects where we are able to measure a period. The period vs $\Delta K_{\rm s}$ 
distribution is similar to that observed in LPVs outside SFRs (Fig. \ref{vvv:miras_periodamp}) except that 
there is a larger number of ``long-term'' periodic variables found with periods, $100<P< 350$ days. The 65 blue points are 
the objects with short timescale scatter and the red points are the remaining 89, categorised by careful inspection of 
the light curves of the 154 long-term periodic variables in SFRs. The blue points clearly dominate the group with $P <350$~days
and they also have a distinctly fainter distribution of $K_{\rm s}$ magnitudes, similar to that shown in the red histogram
in Fig.\ref{vvv:k_dist}, which represents all objects in SFRs except those with Mira-like light curves.
As expected, the red points with Mira-like light curves typically have $K_{\rm s}\sim 12$, similar to the LPV 
distribution plotted in Fig. \ref{vvv:nonsfrs_prop}. We conclude that inspection of the light curves can separate the
evolved star population of LPVs from the YSOs in SFRs with fair success, though we caution that this is an imperfect 
and somewhat subjective process that can be influenced by outlying data points and our limited knowledge of the time domain
behaviour of circumstellar extinction in dusty Mira systems. The limitations are demonstrated by the presence of a number 
of blue points with $K_{\rm s}\sim 12$ and $P>350$~days in the lower panel of Fig. \ref{vvv:ysos_lpv_periodamp} and a 
hint of bimodality even in the ``decontaminated'' magnitude distribution in Fig. \ref{vvv:k_dist} (red histogram).
In the subsequent discussion of YSOs from our sample we only include the 65 objects with short timescale scatter (called LPV-yso)
and assume the other 89 sources are dusty AGB stars or other types of evolved star (or LPV-Mira). 

This decontamination of AGB stars reduces the SFR-associated sample to 441 objects. The long term periodic YSOs represent 15$\%$ of this sample.

Periods, $P>$ 15 days are longer than the stellar rotation period of YSOs, or the orbital period of their inner discs \citep{2015Rice}. Some YSOs have been observed to show variability with periods longer than even 100 days.  WL~4 in $\rho$ Oph shows periodic variability with $P=130.87$ days \citep{2008Plavchan}, which can be explained by obscuration of the components of a binary system by a circumbinary disc. The $K$-band amplitude of the variability in that system is somewhat less than 1 magnitude. However, it is possible to think that a similar mechanism might be responsible for the variations in some of our objects. \citet{2012Hodapp} show that variable star V371 Ser, a class I object driving an H$_{2}$ outflow, has a periodic light curve with $P=543$~days. The authors suggest that variability arises from variable accretion modulated by a binary companion. In view of this, the variability in some of the long term periodic variables might be driven by accretion and we discuss this in paper II, based on spectroscopic evidence for a sub-sample of them.

\begin{figure}
\centering
\resizebox{\columnwidth}{!}{\includegraphics{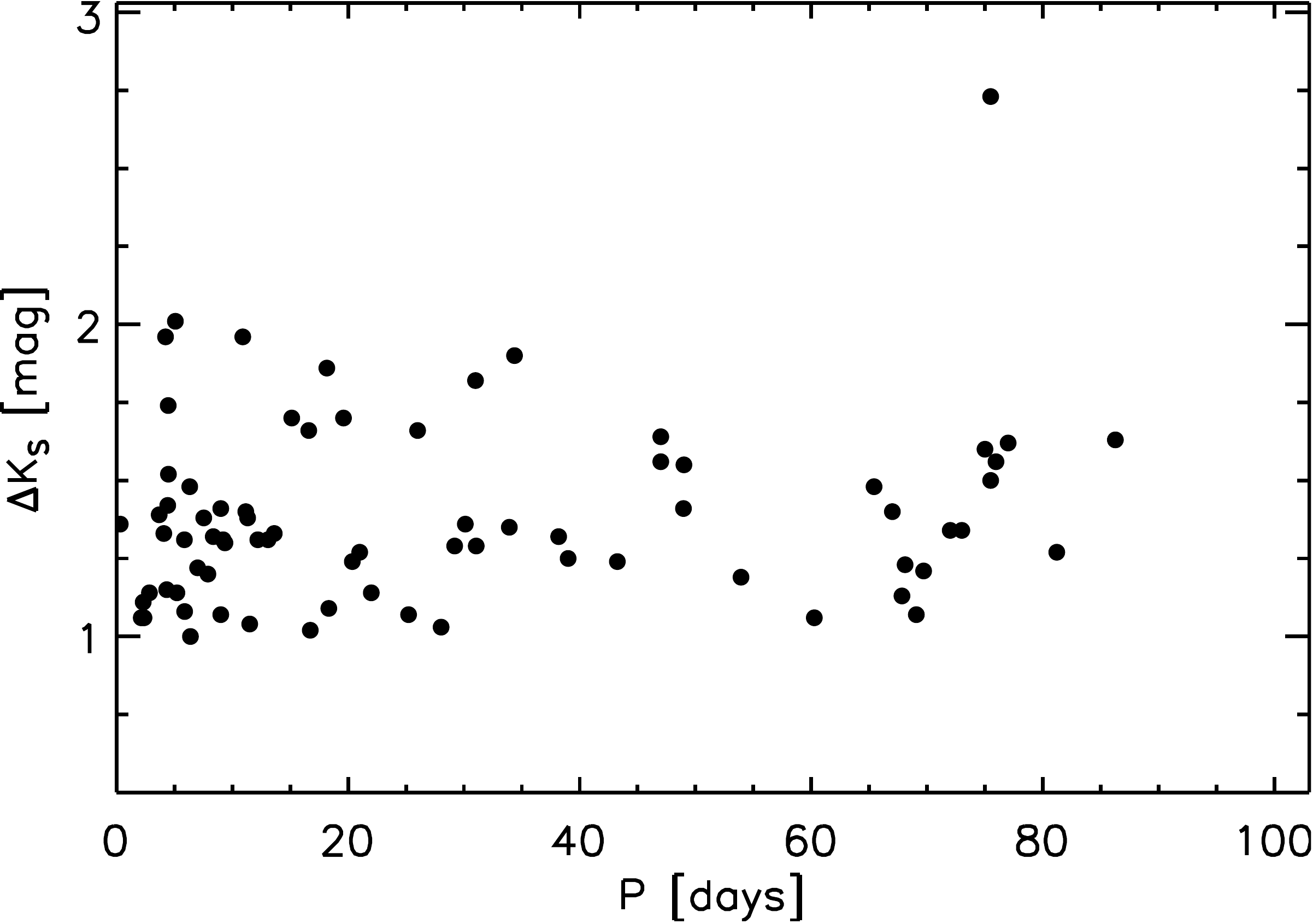}}
\caption{$K_{\rm s}$ amplitude vs period for stars with short-term variability and with a measured period.}
\label{vvv:ysos_spv_periodamp}
\end{figure}


\item{{\bf Short-term Variability.}} This group comprises objects that either have periodic variability and measured periods, $P<100$ days, (75 objects) or else have light curves that appear to vary continuously over short timescales ($t<100$~days) but not with an apparent period (87 objects). Their
light curves do not resemble those of detached EBs because they vary continuously and cannot be contact binaries 
(W UMa variables) because their periods are typically longer than 1 day. For objects in this classification that have measured periods, we observe a broad
distribution from 1 to 100 days and the amplitudes are in the range $\Delta K_{\rm s}$=1 to 2 magnitudes (see Fig. \ref{vvv:ysos_spv_periodamp}). If we join together the long-term periodic variables and the short-term variables (STVs) with measured periods, we find that sources with periods, $P>100$~days show higher amplitudes, on 
average, and sources with $P>600$~days have redder SEDs (larger values of the spectral index $\alpha$). There are no clear
gaps in the period distribution, so the 100~day division between the two groups that we adopted to aid decontamination is arbitrary.
We find 162 stars in the STV group, which represents 37$\%$ of the decontaminated SFR-associated sample. 

High-amplitude periodic variability has been observed in YSOs over a wide range of periods. RWA1 and RWA26 in Cygnus OB7 \citep[][]{2013Wolk} vary with periods of 9.11 and 5.8 days respectively. The variability has been explained as arising from extinction and inner disc changes. As mentioned before variability with $P>15$ days is not expected to arise from the stellar photosphere or changes in the inner disc of YSOs. This instead could be related to obscuration events from a circumbinary disc, such as in V582 Mon \citep{2014Windemuth}, and YSOs ONCvar 149 and 479 in \citet{2015Rice}. Variable accretion has been invoked to explain the observed periodic variability ($P\sim30$~d) of L1634 IRS7 \citep{2015Hodapp}. The shorter periods within this group
may indicate rotational modulation by spots in objects with amplitudes not far above 1 magnitude, see below.

\begin{figure}
\centering
\resizebox{0.45\textwidth}{!}{\includegraphics{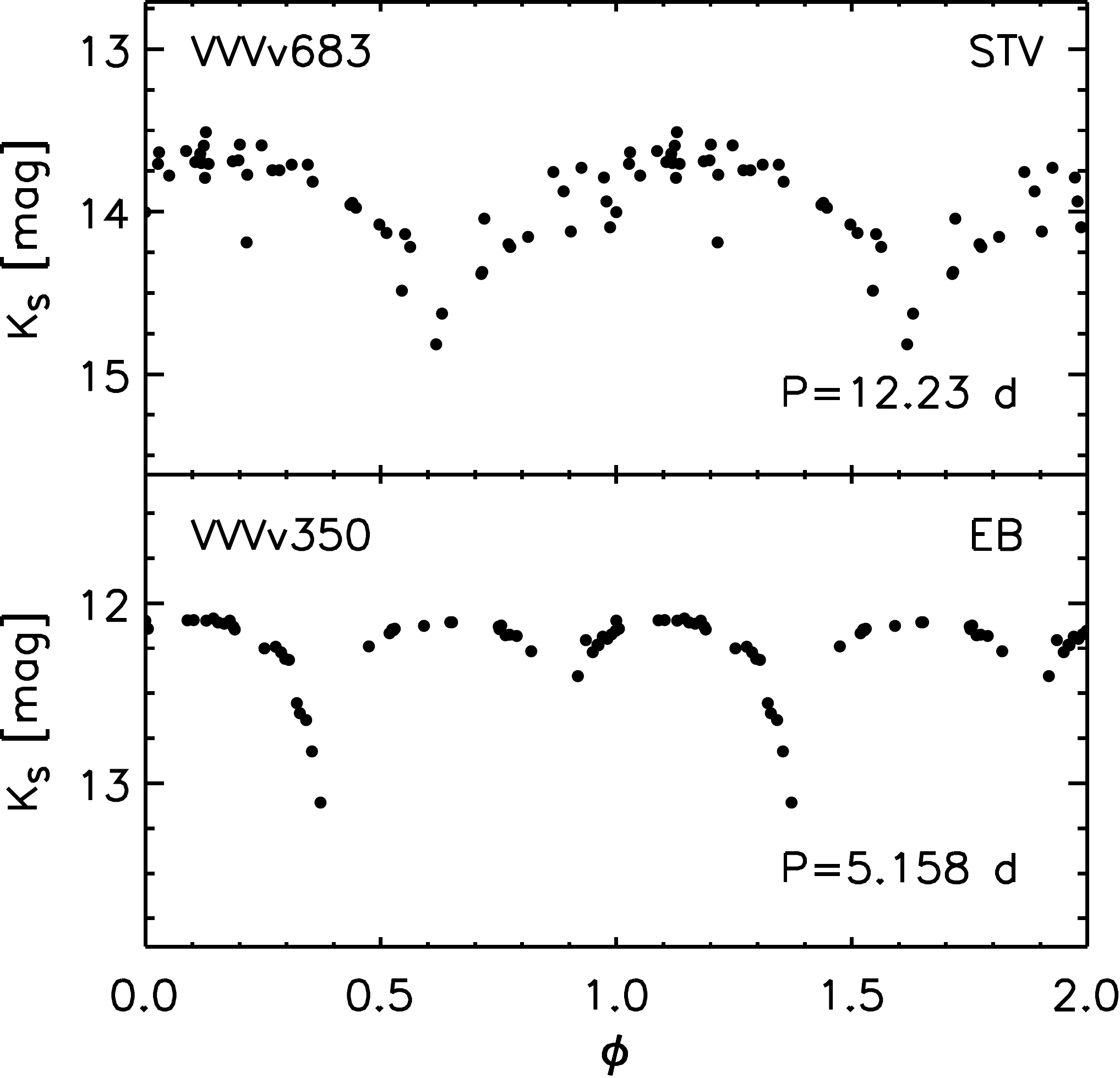}}\\
\resizebox{0.45\textwidth}{!}{\includegraphics{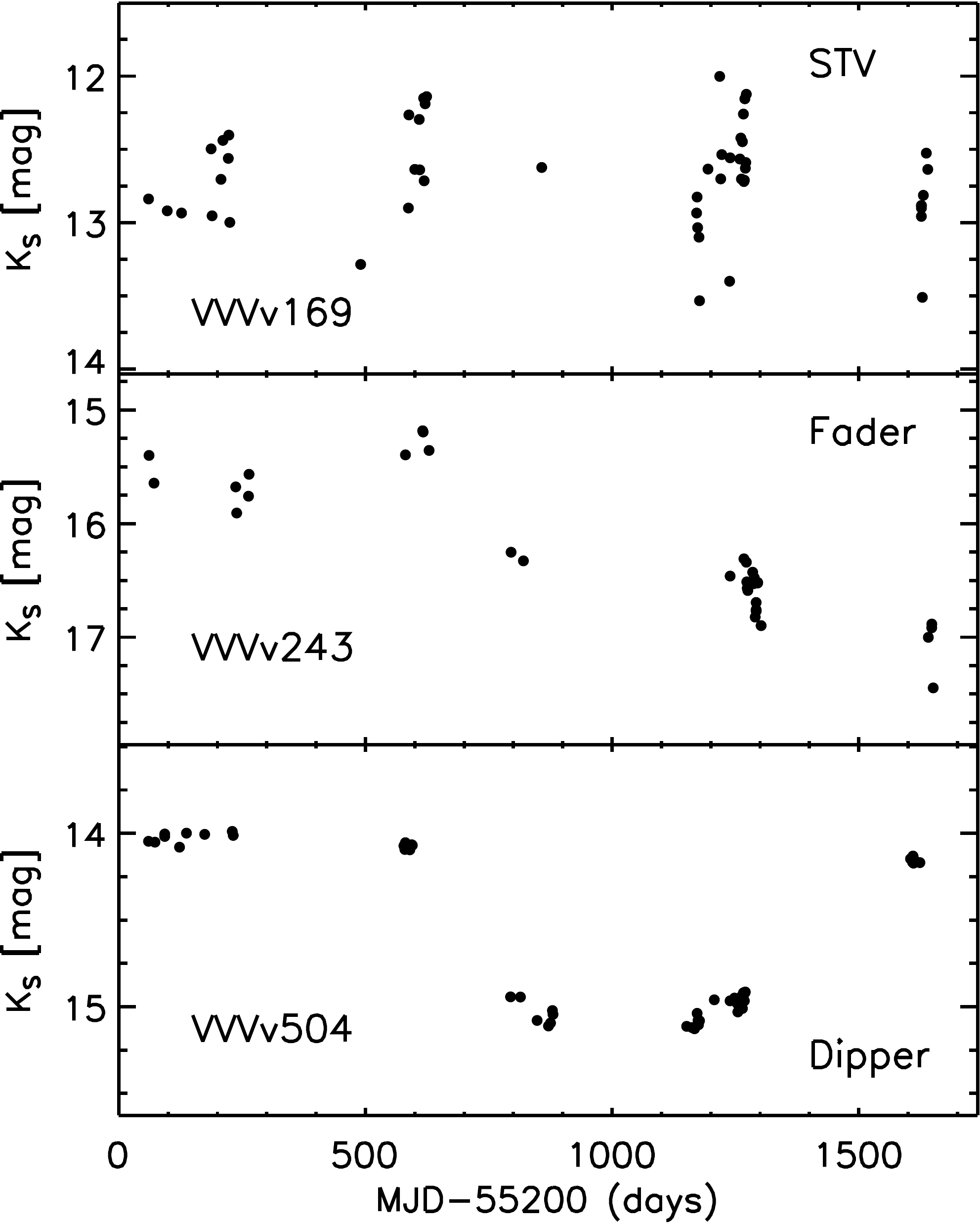}}
\caption{Examples of $K_{\rm s}$ light curves for the different classifications as explained in the text. (top) Phased light curves of short-term variable star with a measured period, VVVv683 and eclipsing binary VVVv350. (bottom) Light curves of short-term variable star, without a measured period, VVVv169, the fader VVVv243 and the dipper VVVv504.}
\label{vvv:lpvmorph2}
\end{figure}


\begin{figure}
\centering
\resizebox{0.48\textwidth}{!}{\includegraphics{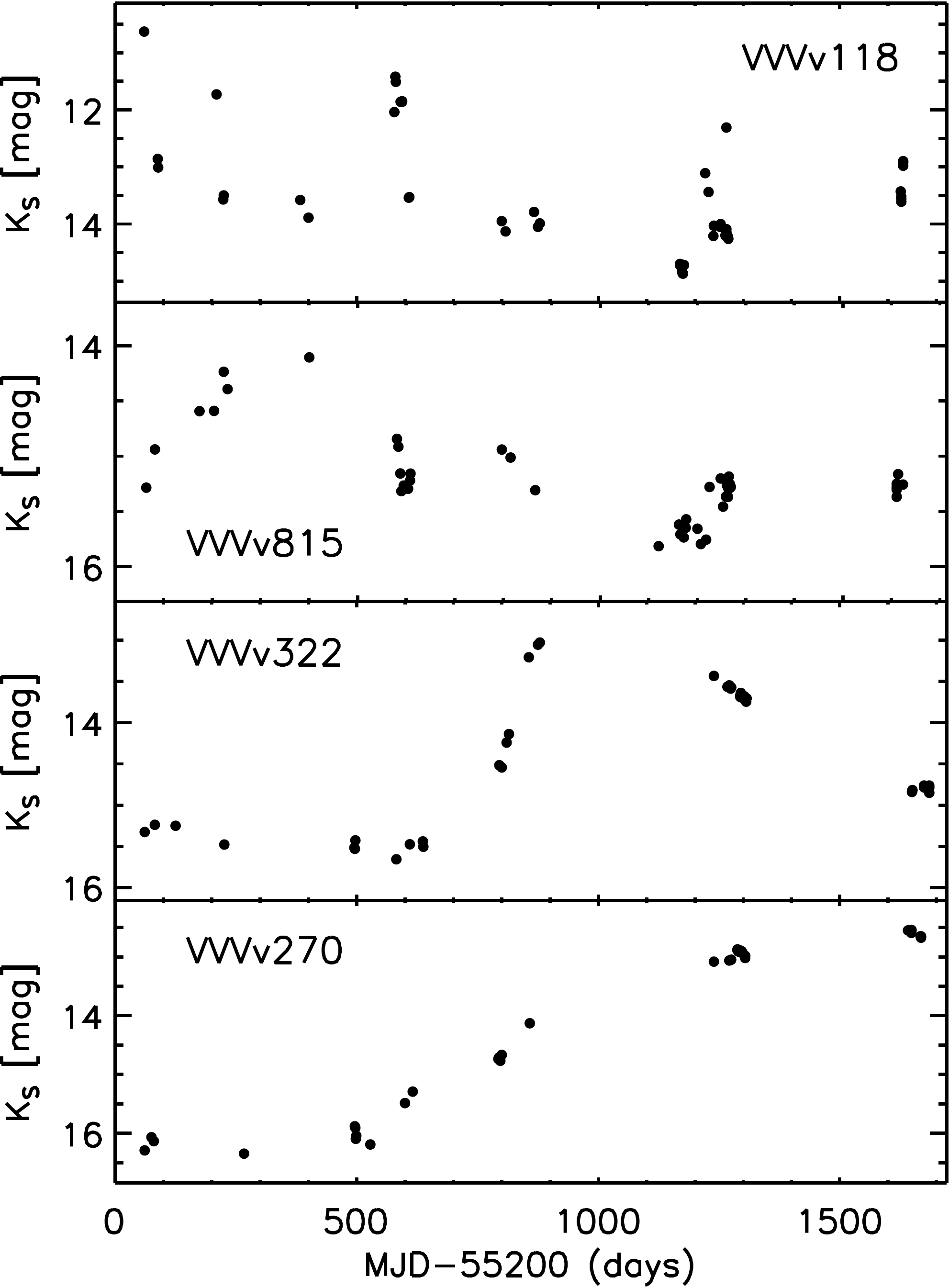}}
\caption{Examples of $K_{\rm s}$ light curves for  different objects in the eruptive classification as explained in the text. From top to bottom we show objects VVVv118, VVVv815, VVVv322 and VVVv270.}
\label{vvv:erupmorph}
\end{figure}

\item{{\bf Aperiodic Long-term variability.}} This category can be divided into three different subclasses: a) Faders. Here the light curves show a continuous decline in brightness or show a constant magnitude for the first epochs followed by a sudden drop in brightness that lasts for a long time ($\ge 1$~year), continuing until the end of the time series in 2014. This type of object might be related to either stars going back to quiescent states after an outburst or objects dominated by long-term extinction events similar to the long-lasting fading event in AA Tau \citep{2013Bouvier}, or some of the faders in \citet{2013Findeisen}. b) Objects that show long-lasting fading events and then return to their normal brightness (such as VVVv504 in Fig. \ref{vvv:lpvmorph2}), which we refer to as dippers. These might also be related to extinction events. Examples of objects in groups (a) and (b) can be seen in Fig. \ref{vvv:lpvmorph2}. 

Group (c) contains sources with outbursts, typically of long duration ($\geq 1$~yr). In a very small number of objects the outburst duration appears to be much shorter, on the order of weeks. The increases in brightness are also unique or happen no more than twice during the light curve of the object, thus not resembling the light curves of objects in the STV category. An exception is VVVv118, which shows four brief rises on timescales of weeks.

The light curves in this category typically have a monotonic rise of 1 magnitude or more, though sometimes a lower level scatter is present atop the rising trend. In a small number of cases the rise in the light curve is poorly sampled, starting at or before the beginning of the time series, but the subsequent drop exceeds 1 magnitude. Figure \ref{vvv:erupmorph} shows four examples of objects falling in the eruptive classification. The examples have been selected in order to illustrate the different temporal behaviour observed in objects belonging to this class. As we have already mentioned, VVVv118 shows multiple short, high-amplitude rises. In general objects show outbursts which last between 1-4 years (see VVVv815 and VVVv322). We also detect a few cases where the outburst duration cannot be measured as it extends beyond 2014 data (e.g. VVVv270).

When comparing to the behaviour of known classes of eruptive variables, VVVv118 would resemble that of EXors and VVVv270 could potentially be an FUor object (based only on photometric data). However most of the objects have outburst durations that are in between the expected duration for EXors and FUors. 

Considering the outburst duration of the known subclasses of young eruptive variables, we are likely to miss detection of FUor outbursts if they went into outburst prior to 2010. In the case of EXors, which have outbursts that last from few weeks to several months, we would expect to detect more of these objects given the time baseline of VVV. However, our results show a lack of classical EXors, which could be a real feature or it could be related to the sparse VVV sampling. Thus, we need to test our sensitivity to short, EXor-like eruptions.


We simulate outbursts with timescales from ~2 months to ~3 yr. First we generate a very rough approximation of an eruptive light curve with outburst duration, $T_{o}$. The light curve consists of: 1) A quiescent phase of constant magnitude that lasts until the beginning of the outburst, which is set randomly at a point within the 2010-2012 period (between 0 and 1000 d). 2) A rise which is set arbitrarily to have a rate of 0.15 mag/day, lasting $T_{rise}=$10 d until reaching an outburst amplitude of 1.5 mag, which is a little below the
median for the VVV eruptive variable candidates. 3) Plateau phase with a constant magnitude set to the peak of the outburst. This phase lasts for $T_{o}-T_{rise}-T_{decline}=T_{o}-20$~d. 4) The decline, which lasts 10 days and has a rate of 0.15 mag/day. Finally, 5) A second quiescence phase. To every point in the light curve we add a randomly generated scatter of $\pm$0.2~mag. Once the light curve is generated, we measure the magnitude of the synthetic object at the observation dates of a particular VVV tile. If the synthetic object shows $\Delta K_{\rm s}\geq1$~mag then it is marked as a detection. This procedure is repeated 1000 times for each outburst duration (which is set to be between 30 and 900 days). We also repeat this procedure for four different VVV tiles.


The simulation shows that the number of detections is very similar ($\sim 80\%$) for $T_{o} >$~ 7 months and declines slowly as $T_{o}$ is
reduced, falling by a factor of 2 for $T_{o}=2$ months.
However, this is not enough to cause the apparent lack of eruptive 
variables with EXor-like outbursts in our sample. We conclude 
that the longer (1-4 yr) durations that we observe are 
typical values for infrared eruptive variables, rather than a
a sampling effect.

The characteristics of our eruptive sample (see Paper II) agree with recent discoveries of eruptive variables that show a mixture of characteristics between the known subclasses of eruptive variables \citep[see e.g.][]{2009Aspin}. We note that classification of our sample into the known subclasses becomes even more problematic when taking spectroscopic characteristics into account, as e.g. VVVv270, the potential FUor from its light curve, shows an emission line spectrum, or VVVv322 shows a classical FUor near-infrared spectrum. In Paper II we propose a new class of eruptive variable to describe these intermediate eruptive YSOs.


Sources classified as eruptive are very likely to be eruptive variables, where the changes are explained by an increase of the accretion rate onto the star due to instabilities in the disc of YSOs \citep[see e.g.][]{2014Audard}.
We find 39 objects in subgroup (a), 45 in (b) and 106 in (c). The whole class of faders/bursts represent 43$\%$ of the likely YSO sample.

\item{{\bf Eclipsing Binaries.}} We find 24 objects with this light curve morphology, representing 5$\%$ of the sample. We are able to measure a possible period in 15 of them. The remaining 9 objects are left with this classification given the resemblance of their light curves to the objects with measured periods. We expect that a number of them will be field EBs
contaminating our YSO sample. However, inspection of the 2 to 23~$\mu$m spectral index for each object, $\alpha$,
(see Fig. \ref{vvv:gc2}), indicates that 12 objects are classified as either class II or flat-spectrum sources. If these are in fact YSOs, they would represent a significant discovery as YSO EBs are invaluable anchors for stellar evolutionary models, which generally lack empirical data on stellar radii. Figure \ref{vvv:ebyso} shows the light curve and location of one candidate YSO EB, VVVv317, with P$=6.85$~d and $\alpha=-1.57$. The spectral index places it at the edge of the classification of class II YSOs.

\begin{figure}
\centering
\resizebox{0.95\columnwidth}{!}{\includegraphics{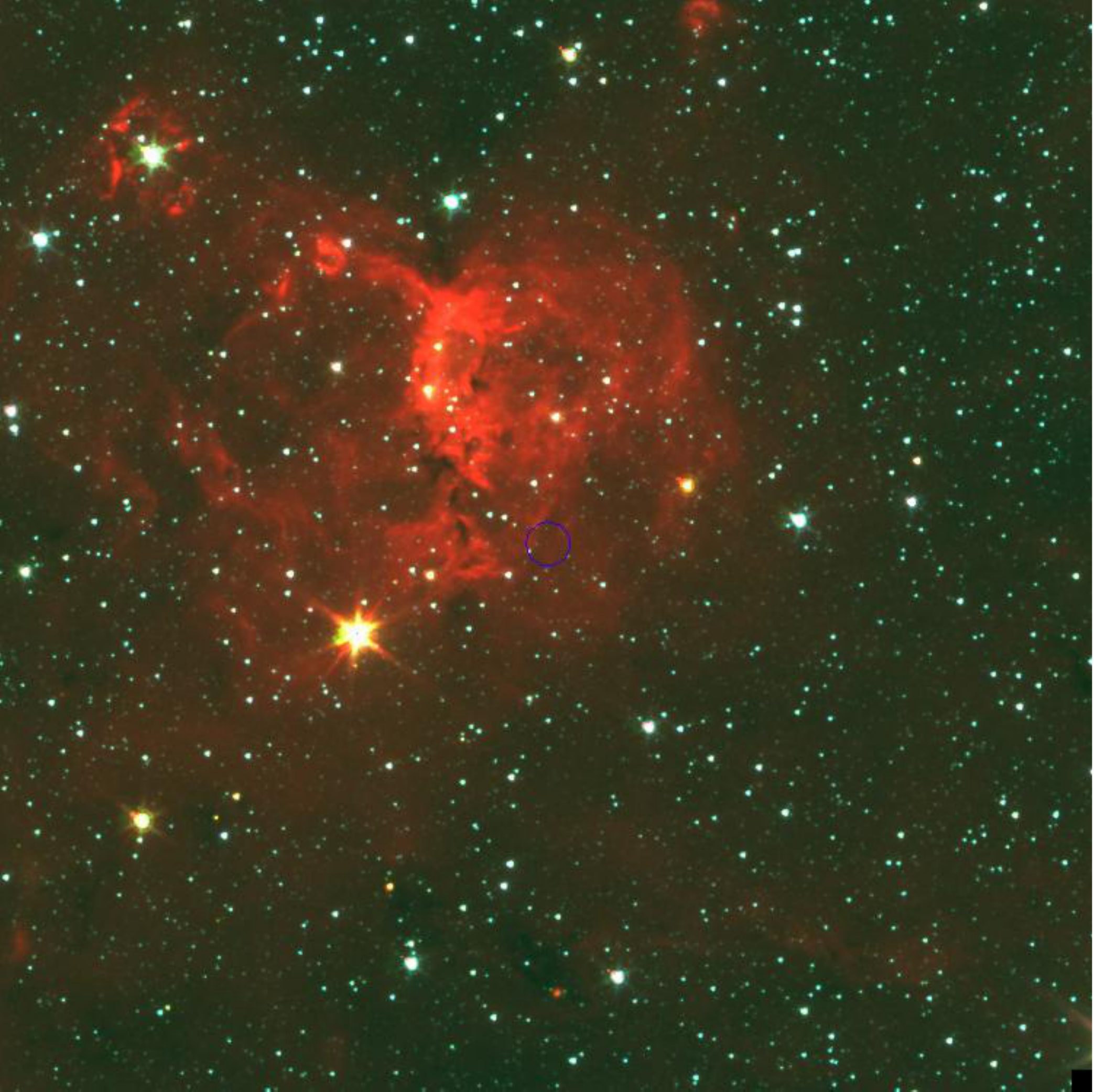}}\\
\resizebox{\columnwidth}{!}{\includegraphics{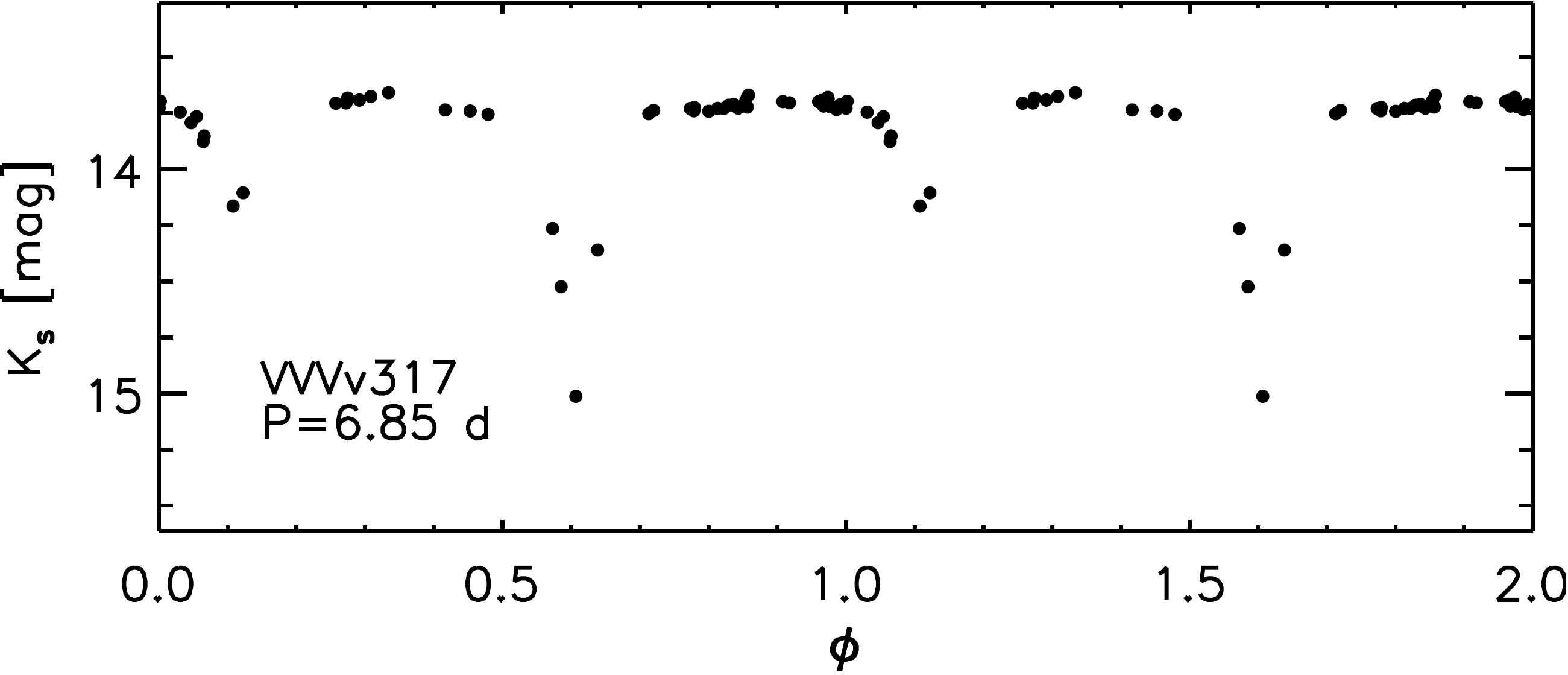}}
\caption{(top) False colour image (blue=3.6 $\mu$m, green=4.5 $\mu$m and red=8 $\mu$m) from GLIMPSE, showing the location of candidate YSO EB VVVv317 (blue ring). (bottom) Phased light curve of the object.}
\label{vvv:ebyso}
\end{figure}


\end{itemize}

\begin{figure}
\centering
\resizebox{0.95\columnwidth}{!}{\includegraphics{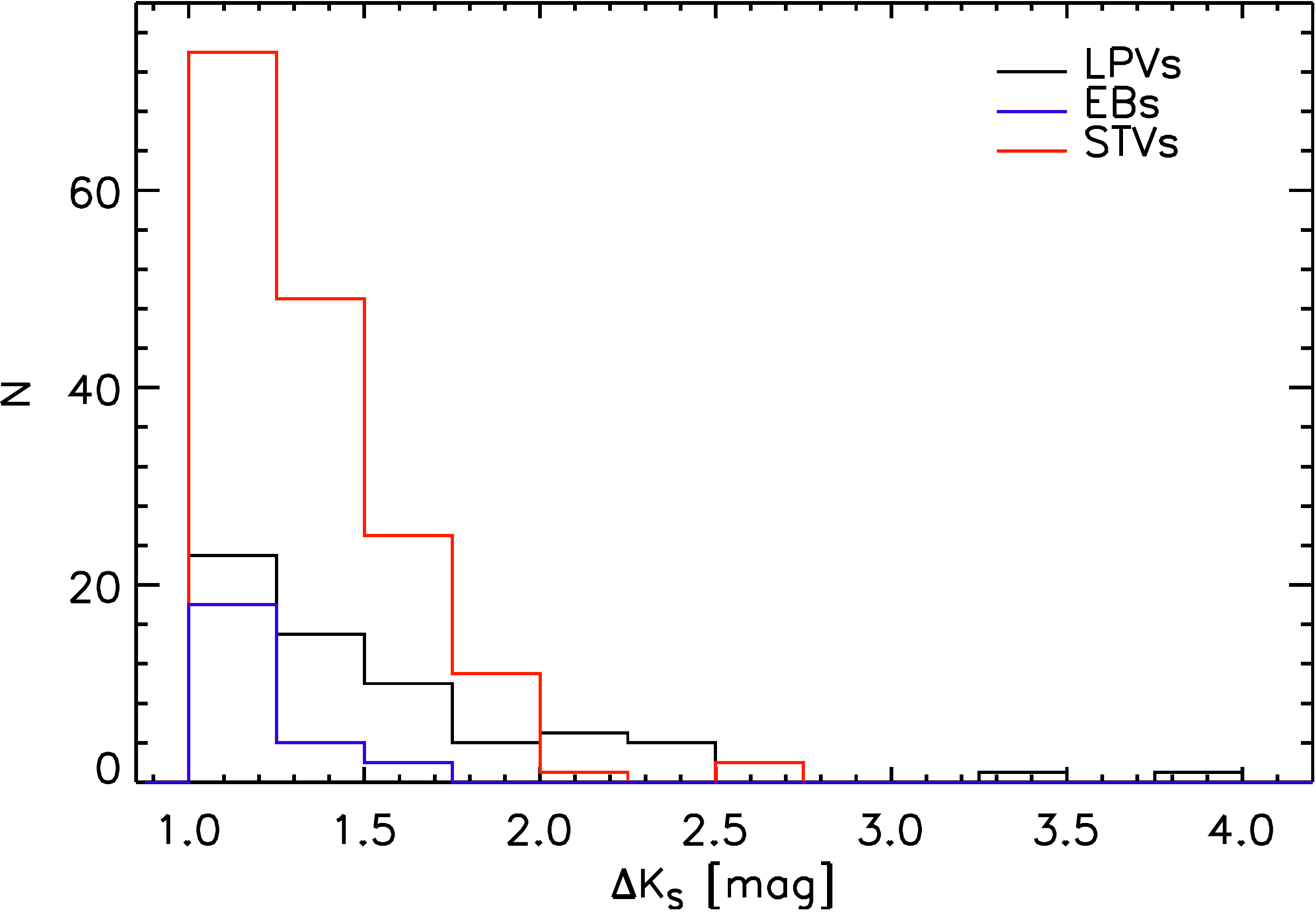}}\\
\resizebox{0.95\columnwidth}{!}{\includegraphics{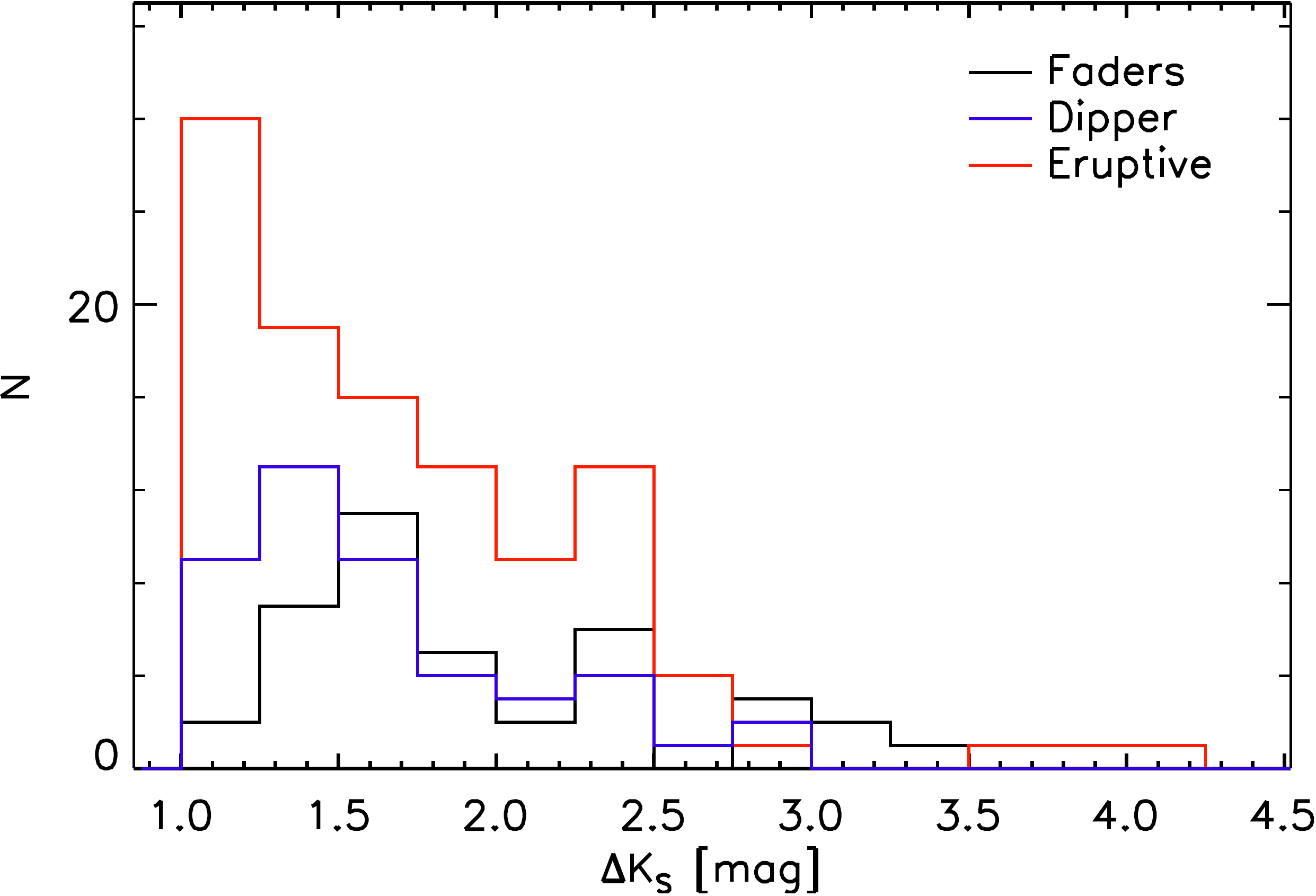}}
\caption{$\Delta K_{\rm s}$ distribution for the different light curve morphology YSO classes. (top) Distribution for LPVs (black line), STVs (red line) and EBs (blue line). (bottom) Distribution for aperiodic variables, faders (black line), dippers (blue line) and eruptive objects (red line).}
\label{vvv:classlc_dkdist}
\end{figure}

In Fig. \ref{vvv:classlc_dkdist} we compare the amplitude distributions ($K_{\rm s,max}-K_{\rm s,min}$) of the different categories of 
variable YSO. We can see that
the EBs and STVs typically have the smallest amplitudes (means of 1.18 mag and 1.33 mag, respectively), whereas
the faders and eruptive variables have the highest amplitudes (means of 1.95 mag and 1.72 mag, and medians of 1.75 mag and 1.61 mag respectively). The
dippers and long-term periodic variables have mean amplitudes of 1.64 mag and 1.57 mag respectively, which are similar
to the mean amplitude of 1.56 mag for the full sample. The substantial number of STVs with amplitudes only a little
over 1 magnitude is consistent with our suggestion that in the shorter period variables in this category variability may be explained by 
rotational modulation of dark or bright spots on the photosphere. The relatively high amplitudes of the eruptive 
variables are not unexpected, whilst the high amplitudes of the faders could be explained if some of these objects are eruptive variables returning to quiescent states.

\subsection{Near infrared colour variability}\label{vvv:sec_nirchange}

Considering the various classes of light curve defined above for
SFR-associated variables, it is reasonable to expect that extinction causes the variability in dippers and perhaps
some of the STVs. Extinction variability can be observed in eruptive objects. However, we do not expect the main cause of variability in this class to be due to this mechanism. It is less clear what to expect for faders and 
long-term periodic variables. We can test this by looking at colour variability data.

The VVV survey was initially designed with only 1 epoch of 
contemporaneous JHK$_{\rm s}$ colour data but a 2nd such epoch was added to the programme for observation in 2015, both 
to benefit the YSO variability science and to  help to understand VVV variables of unknown nature. We note that many objects in the SFR sample are not detected in J- and H-band (see Sect. \ref{vvv:sec_alphaclass}), making a colour comparison impossible. Also, in many objects the two epochs do not span a large fraction of the full range of magnitudes in the light 
curve. E.g. in many eruptive objects we do not have direct comparison between quiescent and outburst states. Nevertheless, comparison of the change in colour vs magnitude still provides some valuable information on the mechanisms driving variability, particularly for those sources in which source magnitudes differ substantially at the 2 epochs.



\begin{figure*}
\centering
\resizebox{0.8\textwidth}{!}{\includegraphics{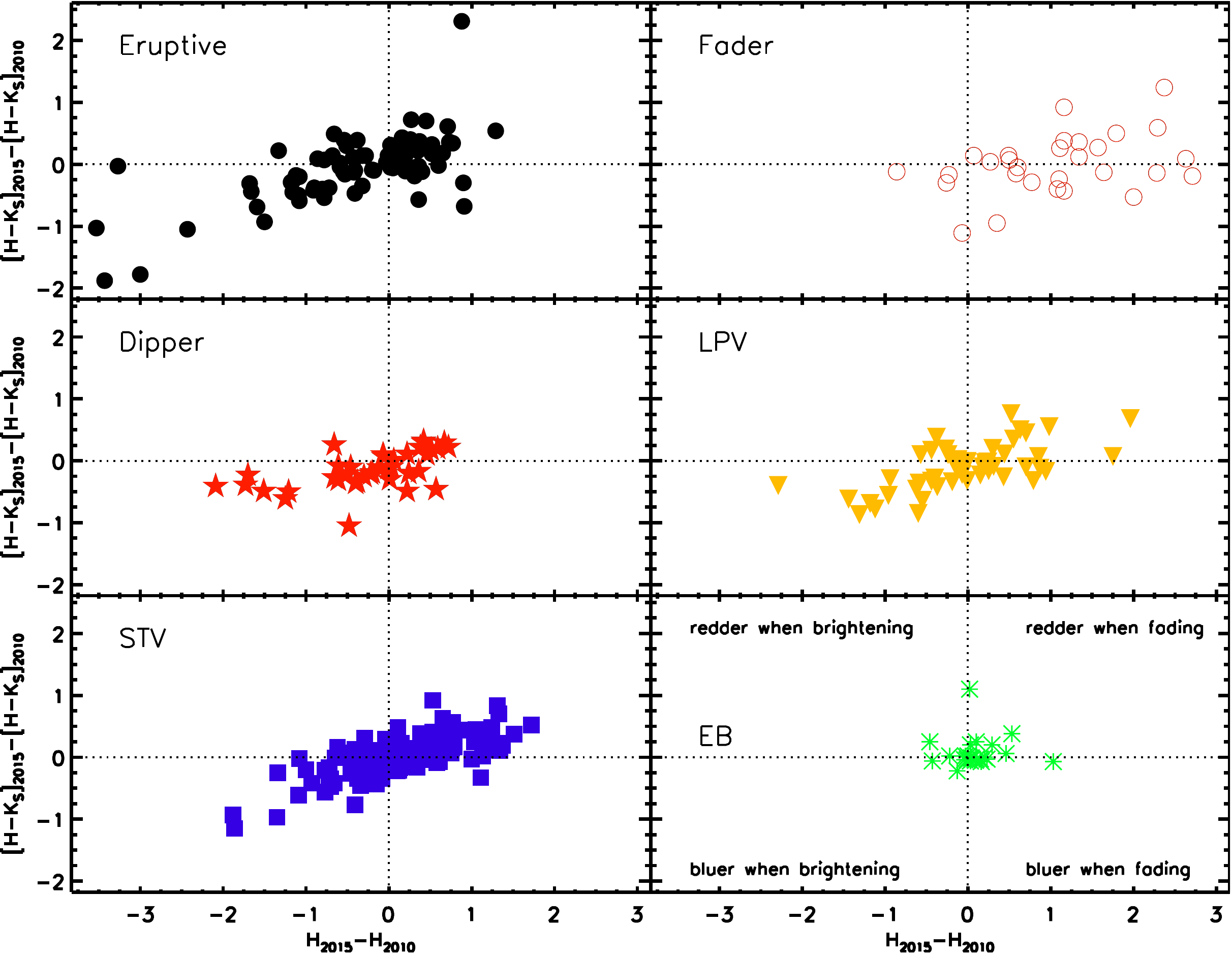}}
\resizebox{0.8\textwidth}{!}{\includegraphics{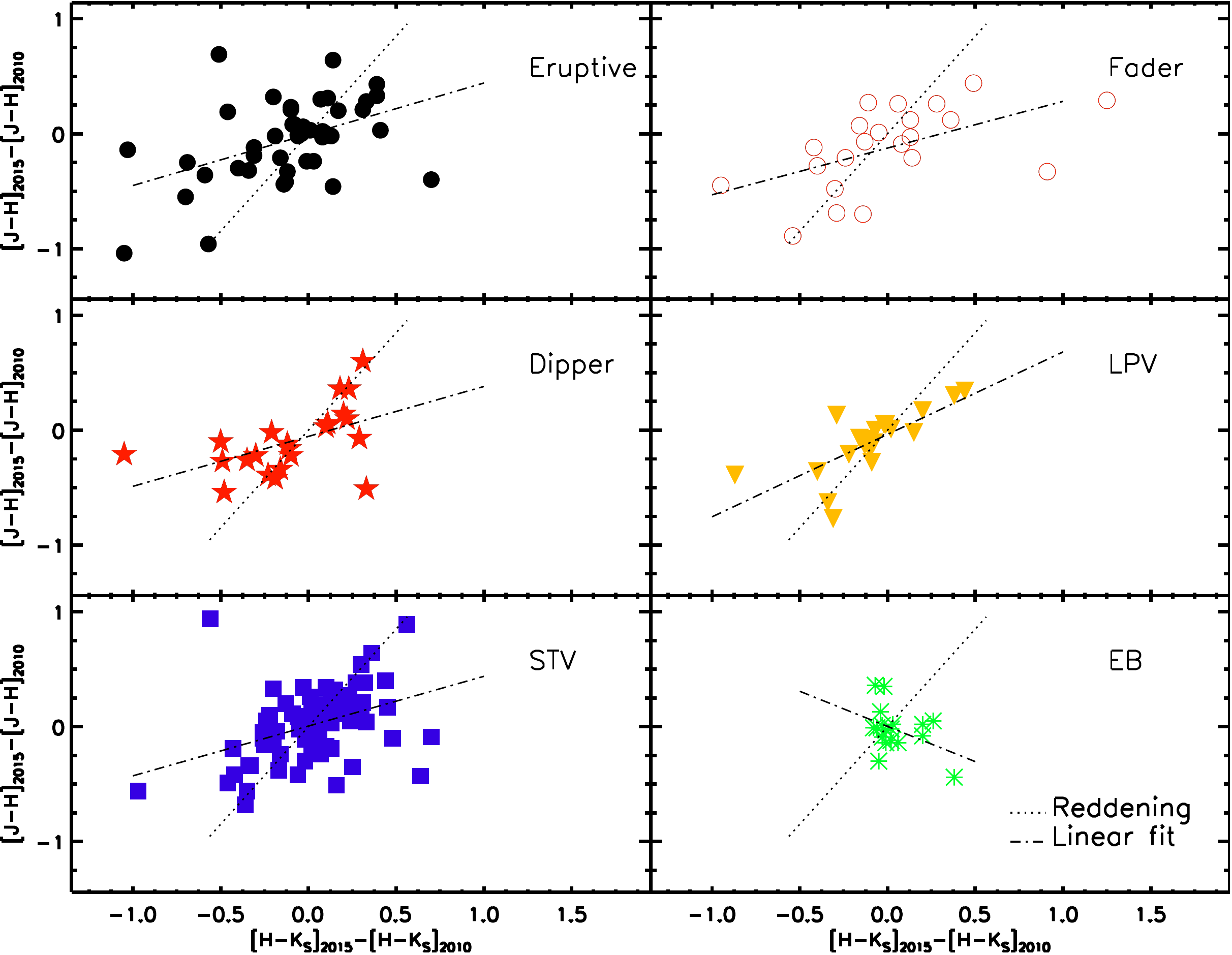}}
\caption{$\Delta (H-K_{\rm s})$ vs $\Delta H$ (top) and $\Delta (J-H)$ vs $\Delta (H-K_{\rm s})$ (bottom) for YSOs with an available second $JHK_{\rm s}$ epoch from VVV. In the plots we mark the different classes from light curve morphology. In the bottom plot we mark the expected changes which occur parallel to the reddening vector (dotted line) as well as the best fit to the observed change (dot-dashed line). In the bottom right of the upper panel we mark four distinct regions as explained in the text.}
\label{vvv:ysoext}
\end{figure*}

Following a similar procedure to \cite{2012Loren} and \citet{2014Antoniucci} we compare the change in colour, $(H-K_{\rm s})$, vs magnitude, $H$, between 2010 and 2015 for the different classes of YSOs (see Fig. \ref{vvv:ysoext}). In the figure we see that in most cases the changes in both colour and magnitude are small, thus objects cluster around the origin. This is especially true in EBs. The overall distribution of eruptive, dippers, LPVs and STVs on the other hand, seems to be elongated along an axis passing through the ``bluer when brighter'' and ``redder when fainter'' quadrants.
This agrees with the behaviour expected from changes due to accretion or extinction in YSOs and resembles the near-infrared variability observed in EXors \citep{2009Loren,2012Loren}, the classical T Tauri sample of \citet{2012Loren}, and the mid-infrared variability of candidate EXors from \citet{2014Antoniucci}. It is interesting to see that many objects classified as faders fall in the bluer when fading quadrant. This behaviour is still consistent with YSO variability due to changes in disc parameters described by \citet{1997Meyer} and observed in some  Cygnus OB7 YSOs \citep[see e.g.][]{2013Wolk}. The different behaviour might also be caused by the different geometry (inclination) of the system with respect to the observer. Scattering of light by the circumsteller disc or envelope may also be contributing to the sources that are bluer when fainter.


Figure \ref{vvv:ysoext} also shows the observed change in both $(J-H)$ and $(H-K_{\rm s})$ colours for YSOs detected in the three filters in both epochs. In there we also plot the expected change if the variability occurs parallel to the reddening line (independent of the direction of the change) as well as a linear fit to the different YSO classes. We would expect that variability similar to that observed in EXors \citep[see e.g. fig. 1 in][]{2012Loren} would show a behaviour that is not consistent with reddening. It is hard to say much from EBs as they do not show much variability. The overall change in STVs, faders and dippers appears to be different from the reddening path, although it appears to depend on the selection of objects from those samples. The path followed by LPVs is also different from the reddening line, but it does suggest a more similar behaviour to extinction compared to the other classes. Eruptive variables appear to follow a very different path from reddening. We would expect such behaviour if the variability is similar to that observed in EXors \citep[see e.g. fig. 1 in][]{2012Loren}. Given that variable extinction and eruptive variability are the only known mechanisms to produce variability well in excess of 1~mag, the fact that colour variability disfavours the former in eruptive systems suggests that the latter is more likely.

We have checked the individual $(J-H)$ vs $H$, $(H-K_{\rm s})$ vs $K_{\rm s}$ and $(J-H)$ vs $(H-K_{\rm s})$ colour-magnitude (CMD) and colour-colour diagrams for the 15 eruptive variables that showed $\Delta K_{\rm s}>0.75$ mag between the two multi-wavelength epochs (this representing a significant fraction of the total amplitude in most systems). 
 From 15 objects, 10 do not show changes consistent with extinction (e.g. they are bluer when fainter or show negligible colour change) and 5 were found to show variability approximately following the reddening vector. However, the colour behaviour in these 5 objects does not contradict the idea that accretion is the mechanism driving variability because: 1) we are not directly comparing quiescent vs outburst states, as the two near-infrared epochs cover random points in the light curve; 2) As previously mentioned, extinction does play a role in outburst variability. E.g. the near infrared colour variation of V1647 Ori follows the reddening path in fig. 13 of \citet{2008Aspin}. Extinction might also be involved in the observed variability or the recent eruptive object V899 Mon \citep{2015Ninan}.

The same analysis of individual CMD and colour-colour diagrams for the remaining classes shows that 3/7 LPV-YSOs, 5/9 STVs, 
1/4 dippers and 3/11 faders have colour changes consistent with variable extinction (again considering only systems
with $\Delta K_{\rm s}>0.75$ between the 2 multi-colour epochs). No results can be derived from EBs as none of them show changes larger than 0.75 magnitudes between the two epochs. It is interesting to see that the changes in the majority of faders are not consistent with extinction. This supports the idea that variability in many of the objects in this class could be related to accretion changes.   

In Appendix \ref{apenn:mid-ir} we briefly summarise the colour and magnitude changes detected 
by the multi-epoch photometry from the WISE satellite. We note that this adds 
little to the preceding discussion of near-infrared colour changes, though 
large mid-infrared variability is observed in a minority of sources where 
the satellite happened to sample both a peak and a trough in the light curve.     



\subsection{Variability trends with SED class}\label{vvv:sec_alphaclass}

In order to study the possible evolutionary stage of the variable stars in SFRs, we use the slope of the SEDs of the stars between $2 < \lambda < 24~\mu$m. Following \citet{1987Lada}, we define the parameter $\alpha$ as $\alpha=d(\log (\lambda F_{\lambda}))/d(log (\lambda))$.  The value of $\alpha$ is determined from a linear fit to SED points between $2 < \lambda < 24~\mu$m. Objects are then classified according to their value of $\alpha$ following \citet{1994Greene}, also shown in Table \ref{table:vvvclass}. We note that this class definition might not necessarily relate to the actual evolutionary stage of the object. As stated in e.g. \citet{2006Robitaille}, parameters such as inclination or stellar temperature can affect the shape of the SED at the wavelengths used to classify YSOs. We have removed numerous objects that showed Mira-like characteristics from their light 
curves but there may still be some contamination by non-YSOs amongst the remaining 
objects projected close to an SFR.


We derive $\alpha$ using the photometry arising from VVV and WISE, given that these were taken in the same year (2010). When the objects are not detected in WISE, we use {\it Spitzer}/GLIMPSE photometry. The use of the latter is more likely to cause errors in the estimation of $\alpha$ due to the time difference between {\it Spitzer}/GLIMPSE and VVV measurements, but {\it Spitzer}/GLIMPSE benefits from higher spatial resolution. We find that if we use {\it Spitzer} instead of WISE the difference in $\alpha$ shows a random offset of 0.1-0.2 for the majority of the sample detected in both surveys.



\begin{figure*}
\centering
\resizebox{\columnwidth}{!}{\includegraphics{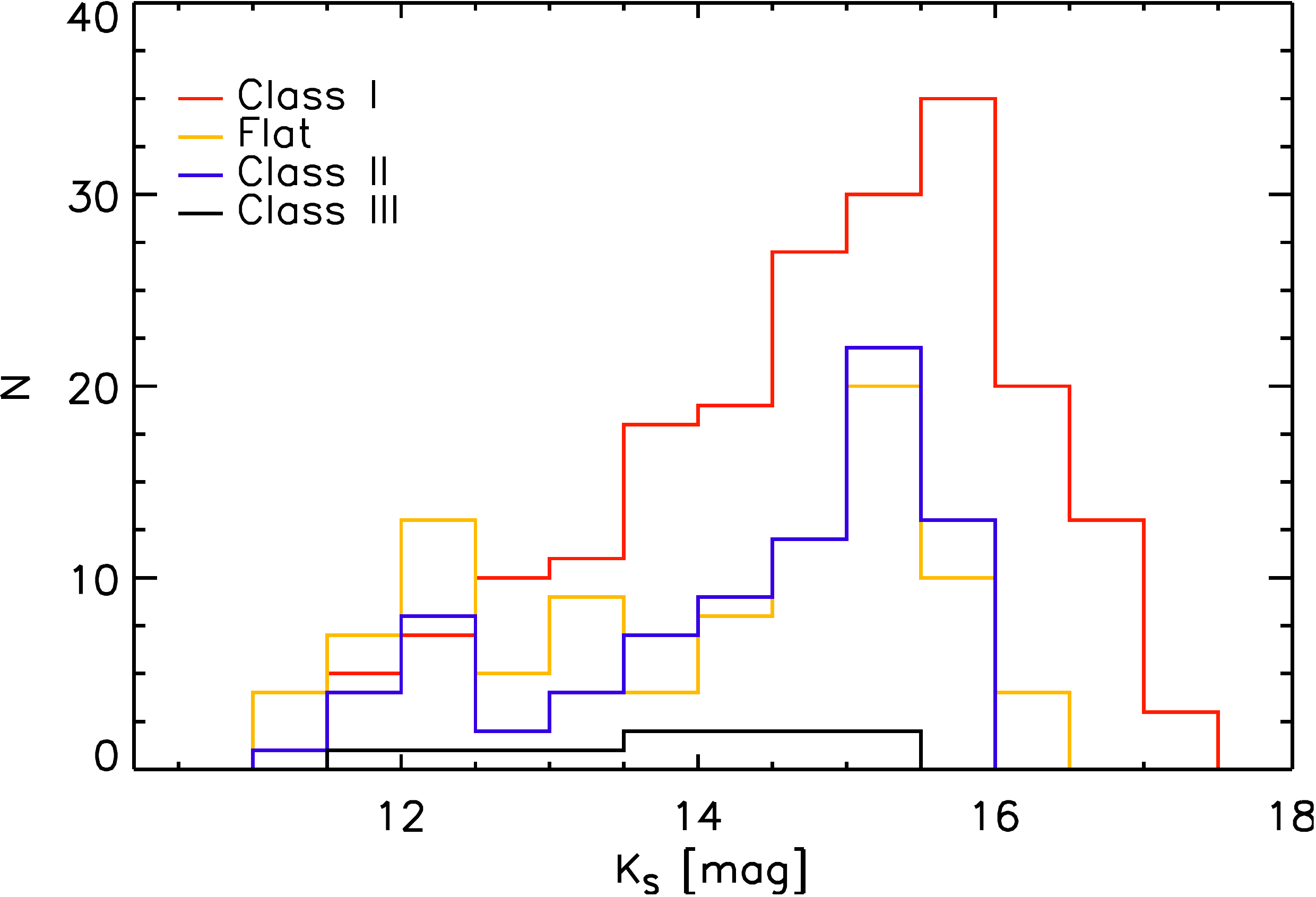}}\\
\resizebox{\columnwidth}{!}{\includegraphics{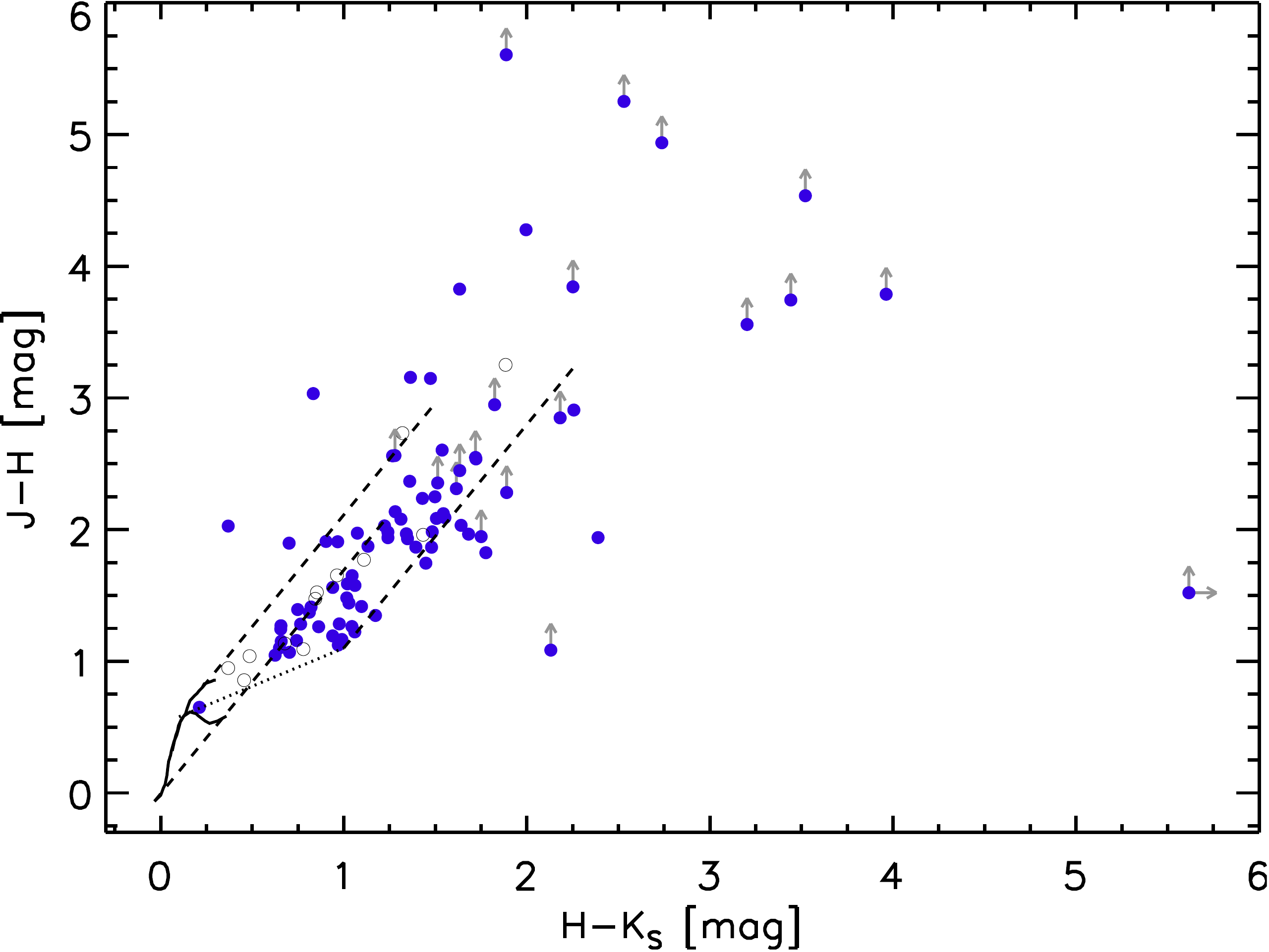}}
\resizebox{\columnwidth}{!}{\includegraphics{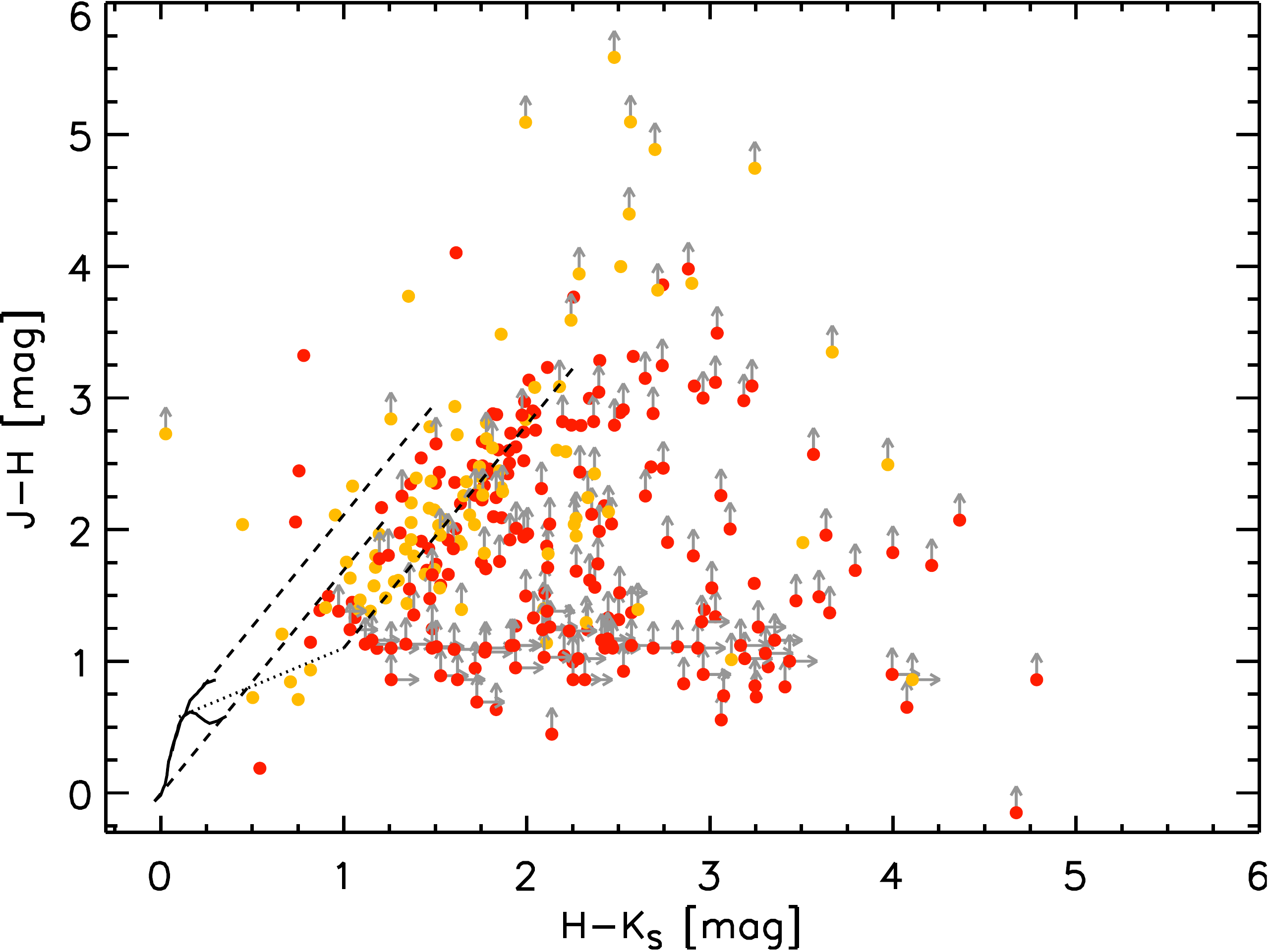}}
\caption{(top) $K_{\rm s}$ distribution (from 2010 data) of class I (red), flat-spectrum (orange), class II (blue) and class III (black) YSOs. (bottom left) Colour-colour diagram for class II (blue filled circles) and class III YSOs (black open circles) from VVV. In the figure, lower limits in colour are marked by arrows. The classical T Tauri locus of \citet{1997Meyer} is presented (long-dashed line) along with intrinsic colours of dwarfs and giants (solid lines) from \citet{1988Bessell}. Reddening vectors of $A_{V} = 20$~mag are shown as dotted lines. (bottom right) Colour-colour diagram for class I (red) and flat-spectrum (orange) YSOs. }
\label{vvv:gc}
\end{figure*}

The number of objects belonging to different classes are shown in Table \ref{table:vvvclass}. We find that the majority (67\%) of objects in our sample are either class I or flat spectrum sources (45\% and 22\%, respectively). Objects belonging to different classes show some differences in their global properties. Figure \ref{vvv:gc} shows the near-infrared colours of objects in SFRs. As expected the vast majority of objects show colours consistent with them being YSOs. The $H-K_{\rm s}$ 
colour tends to be redder for stars belonging to younger evolutionary stages. 
The fraction of objects detected at J and H  bands also decreases with objects belonging to younger stages, as would be expected for typical deeply embedded class I objects. Table \ref{table:vvvclass} shows that 66$\%$ of class I objects are not detected in the J band, whilst $23\%$ of them are not detected in either the J nor H bands. These near IR colour trends
confirm that, as with typical YSOs, highly variable YSOs whose spectral index indicates an earlier evolutionary stage 
also have higher extinction by circumstellar matter, along with more infrared emission from circumstellar matter.

It might be thought that the relatively high reddening of most of the YSOs in Fig. \ref{vvv:gc} is due to foreground extinction,
given that in Paper II we derive typical distances of a few kpc for these sources. We measure foreground extinction for a sub-sample of VVV objects (the 28 variable YSOs in Paper II) by estimating the extinction of red clump giants found at distances similar to those of our objects. The red clump giants are identified in the local $K_{\rm s}$ vs $(J-K_{\rm s})$  colour-magnitude diagrams of the VVV objects (6\arcmin$\times$6\arcmin~ fields), and distances are estimated from their observed magnitudes $K_{\rm s}$ and mean ($J-K_{\rm s}$) colours using equation 1 in \citet{2011Minniti}. The excess of the mean $(J-K_{\rm s})$ with respect to the intrinsic colour of red clump giants \citep[$(J-K_{\rm s})_{0}$=0.70 mag][]{2011Minniti}, gives us a measure of the foreground extinction to the front of the
molecular cloud containing each YSO. In most cases the YSO itself is redder than the red giant branch stars, due to extinction by matter within 
the cloud and by circumstellar matter. This method yields values of A$_{K_{\rm s}}\sim0.8$ to 1.4 mag (i.e. A$_{V} \sim$7 to 12 mag) which is higher than the typically very low diffuse interstellar extinction between
the sun and nearby molecular clouds. We infer that the optical faintness and steeply rising near-infrared SEDs of these objects are due not only to their early evolutionary stage but also to foreground extinction. Correcting our values of $\alpha$ accounting for A$_{V}=11$ magnitudes, produces changes of 0.2-0.4 in this parameter. These changes would alter the classification of 1/3 of the sample where we can estimate $\alpha$, mostly in objects where $\alpha$ has a value that is close to the limits set by \citet{1996Greene}. Despite this correction flat-spectrum and class I sources still dominate the sample of YSOs with a measured value of $\alpha$: with no correction these represent 76$\%$ of the sample, whereas
with a correction to $\alpha$ of 0.3 the proportion remains high, at 59$\%$.

We choose not to apply a correction in Table \ref{table:vvvclass} and our subsequent analysis
because the extinctions and distances
are uncertain (e.g. the red giant branch is not well 
defined in the CMDs for about a third of sources) and we cannot be sure this
method is entirely correct. Our derived extinctions are a factor of $\sim$2
higher than indicated by the 3D extinction map of \citet{2006Marshall}, which
was also based on red clump giants but used 2MASS data on a coarser angular
scale. Moreover, in Paper II we compare the ratio of 70~$\mu$m flux to 24~$\mu$m flux in the spectroscopic subsample with that found in a nearby sample of YSOs. At these far-infrared wavelengths, where extinction is very
low, we find similar flux ratios for class I systems in both datasets, indicating that most VVV class I YSOs have
been correctly classified. In the following discussion, we directly compare embedded (class I and flat 
spectrum) YSOs with sources in nearby SFRs. We simply ask the reader to note 
firstly that $\alpha$ may be slightly inflated by interstellar extinction and 
secondly that this more distant sample will be biased towards more luminous 
YSOs of intermediate mass (see Paper II).



\begin{table}
\begin{center}
\begin{tabular}{@{}l@{\hspace{0.3cm}}c@{\hspace{0.25cm}}c@{\hspace{0.2cm}}c@{\hspace{0.2cm}}c@{\hspace{0.2cm}}c@{\hspace{0.2cm}}}
\hline
Class & $\alpha$ & N  & N$_{Jdrop}$ & N$_{JHdrop}$ & N$_{\Delta K_{\rm s} \geq 2}$\\
\hline
class I & $\alpha>0.3$ & 198 & 130 & 45 & 49\\
flat & $-0.3 \leq \alpha \leq 0.3$ & 95  & 35 & 1 & 12\\
class II & $-1.6 < \alpha < -0.3$ & 83 & 19 & 1 & 6\\
class III &  $ \alpha \leq -1.6 $ & 12  & 0 & 0 & 0\\
Undefined & n/a & 53 & 10 & 2 & 3\\
\hline
\end{tabular}
\caption{Number of VVV variable stars belonging to the different evolutionary classes of YSOs, as determined from their SEDs.}\label{table:vvvclass}
\end{center}
\end{table}

\begin{figure*}
\resizebox{\columnwidth}{!}{\includegraphics{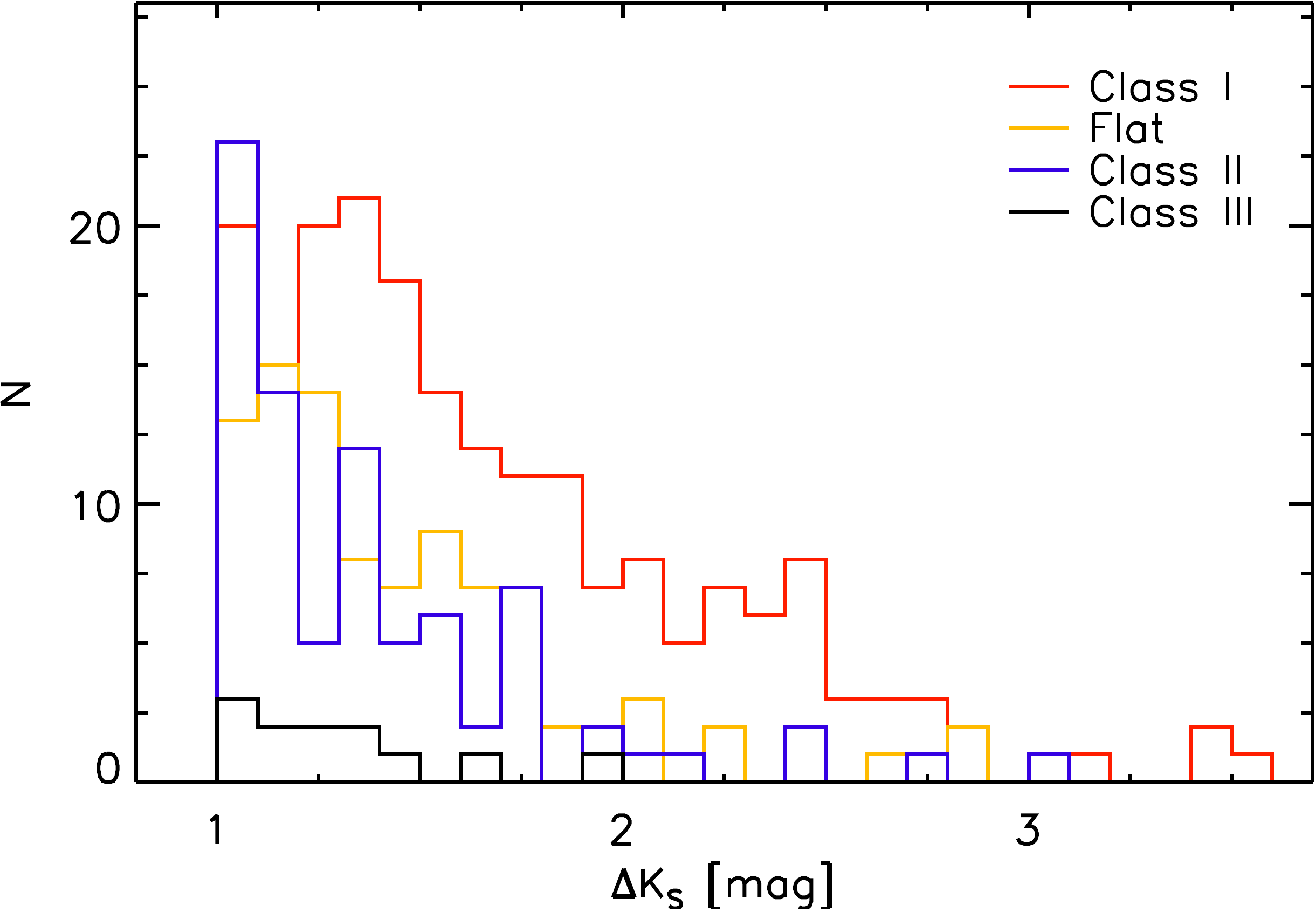}}
\resizebox{\columnwidth}{!}{\includegraphics{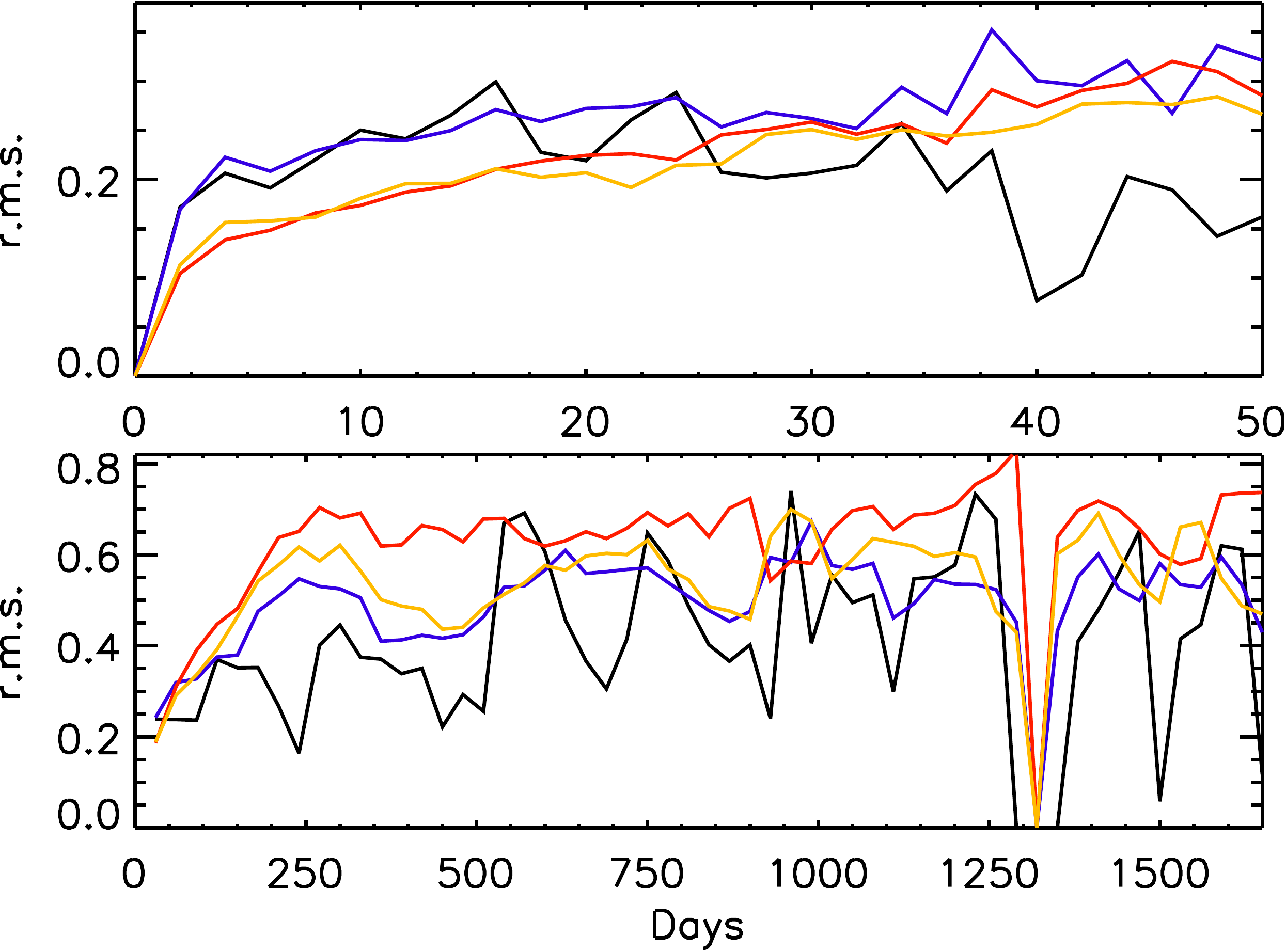}}\\
\resizebox{\columnwidth}{!}{\includegraphics{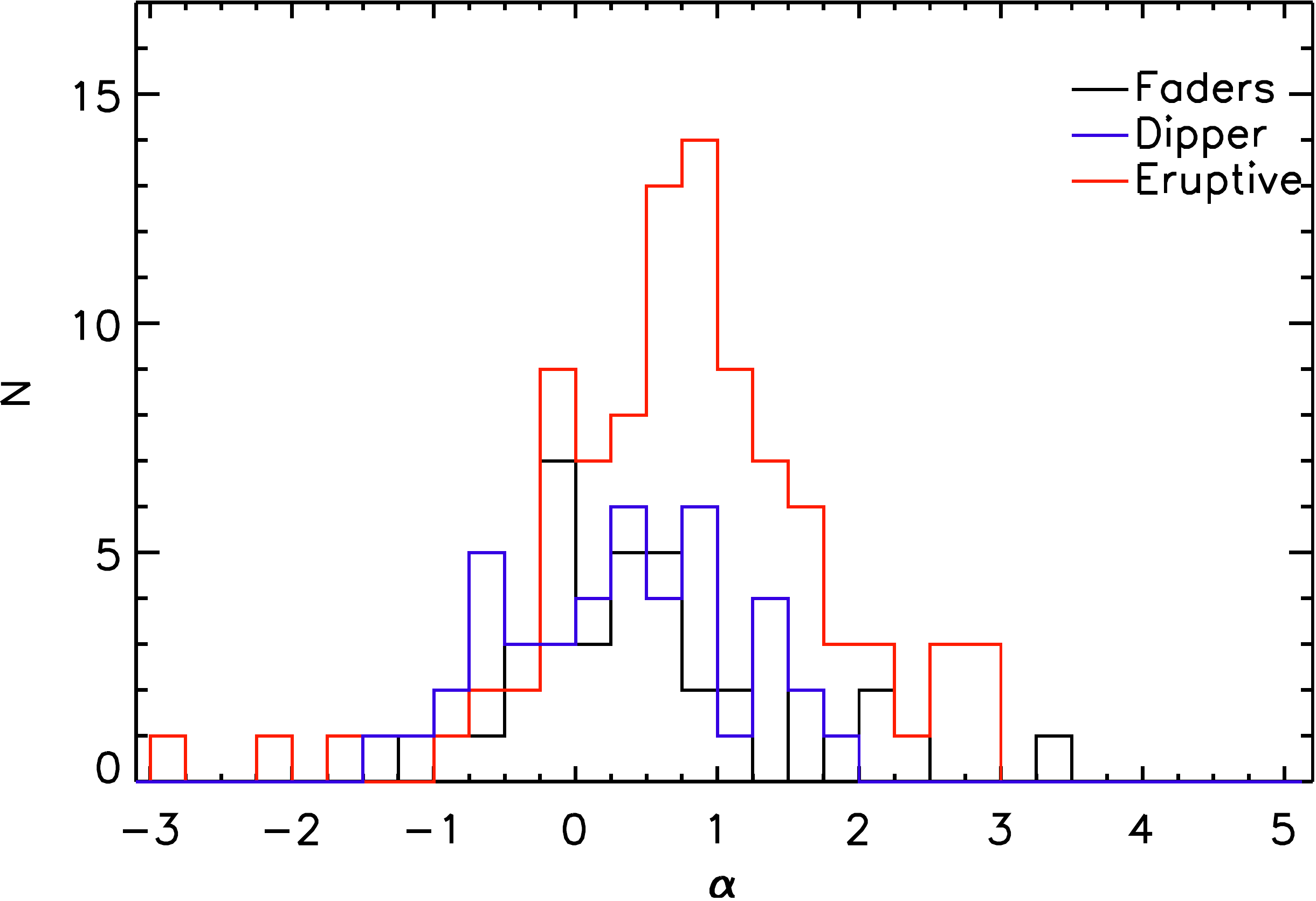}}
\resizebox{\columnwidth}{!}{\includegraphics{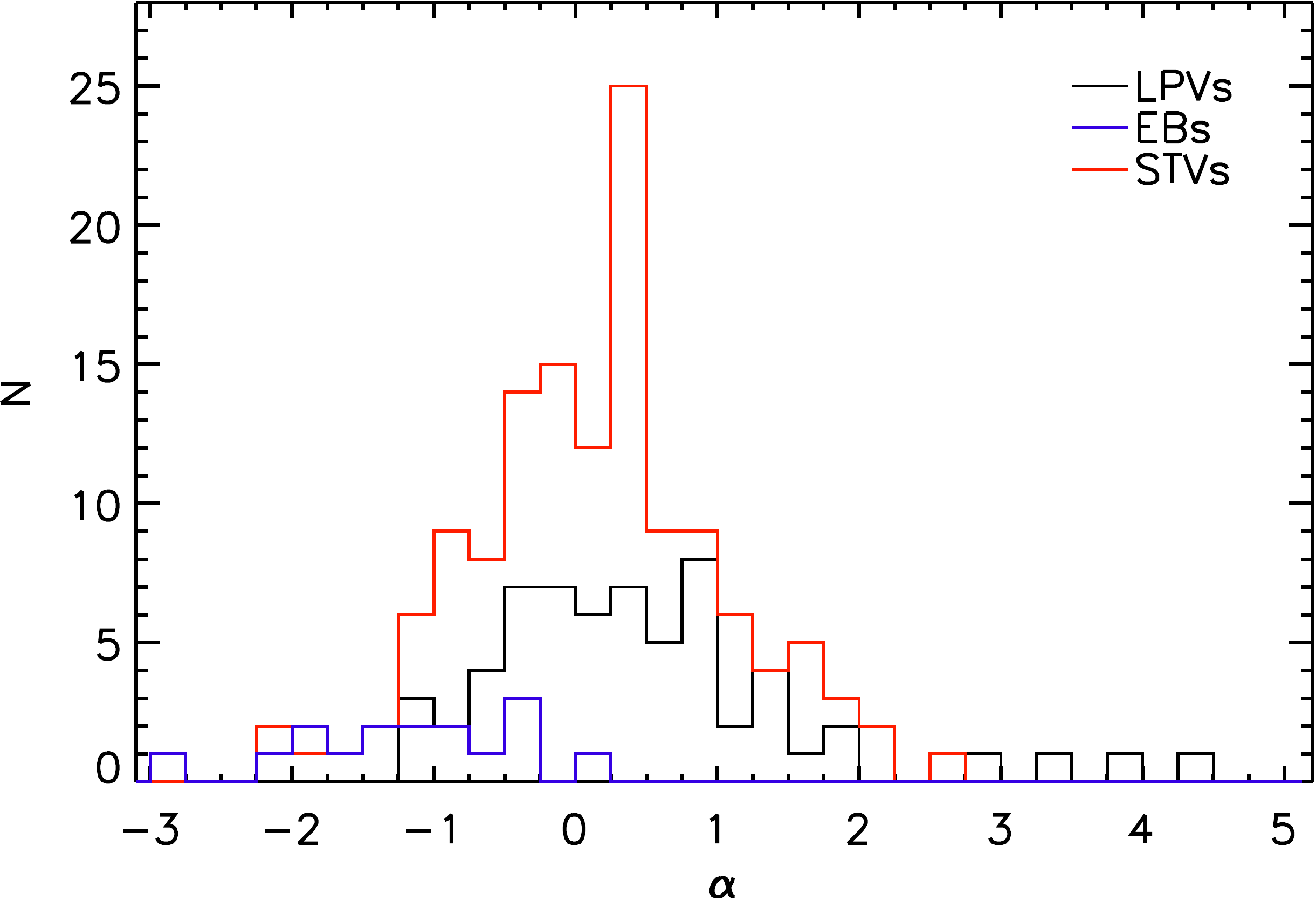}}\\
\resizebox{\columnwidth}{!}{\includegraphics{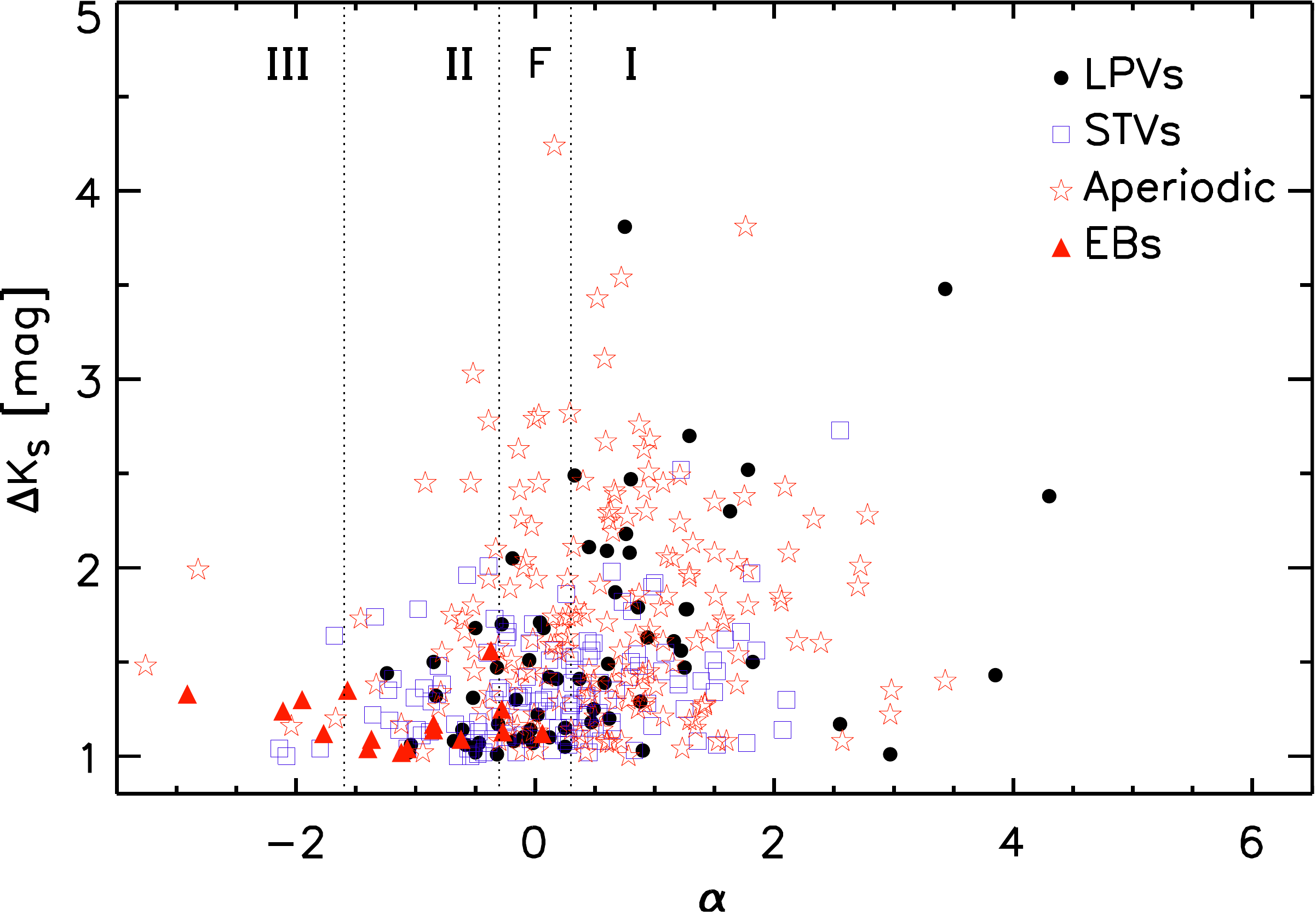}}
\caption{(top, left) $\Delta K_{\rm s}$ distribution of the different YSO classes. (top, right) Mean r.m.s. variability for the different YSO classes in intervals up to 50 days (calculated with time bins of 1 day), and for intervals up to the full 1600 day baseline of the dataset (using time bins of 30 days). The colour coding in the top (left and right) figures is the same as in Fig.\ref{vvv:gc}. (middle, left) $\alpha$ distribution for fader (black line), dipper (blue line) and eruptive (red lines) objects. (middle, right) Same distribution but for long-term periodic objects (black lines), stars with short-term variability (red line) and EBs (blue line). (bottom) $\alpha$ vs $\Delta K_{\rm s}$  for objects associated with SFRs. The limits for the different YSO classes are marked by dashed lines. Objects in this figure are divided according to the morphological light curve classification (Sect. \ref{vvv:sec_lcmorp}). The different classifications are marked in the plot.}
\label{vvv:gc2}
\end{figure*}

The distribution of $\Delta K_{\rm s}$ (Fig. \ref{vvv:gc2}, upper left panel) shows that peak of the distribution is found at larger $\Delta K_{\rm s}$ for younger objects. We also observe a higher fraction of objects with $\Delta K_{\rm s} > 2$~mag for class I objects ($25\%$) than for flat-spectrum ($13\%$) and class II ($7\%$) objects.  The comparison of $\alpha_{class}$ vs $\Delta K_{\rm s}$ in the bottom panel of the same figure further illustrates the increase in amplitude at younger evolutionary stages, as well as the higher incidence of $\Delta K_{\rm s}>1$ variability at the younger stages. The class I and flat 
spectrum YSOs constitute 87\% of  the $\Delta K_{\rm s} > 2$ subsample, dominating it even more than the full 
SFR-associated sample.

We note that 70 objects with $\Delta K_{\rm s} > 2$~mag also have redder near infrared colours than 
the full sample: the proportions of $J$ band non-detections and $JH$ non-detections rises from 44\% and 11\% in the full SFR-associated sample to 56\% and 24\%  in the $\Delta K_{\rm s} > 2$ subsample. This further highlights the simple 
fact that efficient detection of the majority of YSOs with the most extreme variations requires observation at wavelengths $\lambda \ge 2~\mu$m. Most importantly, this further emphasises that younger objects have higher 
accretion variations. Eruptive variables are the largest component of the
$\Delta K_{\rm s}>$2 sample, comprising 30/70 objects.



In Fig. \ref{vvv:gc2} we also show the mean variability of YSOs belonging to different evolutionary classes as a function of time baseline. This is
calculated by averaging the values of r.m.s. variability vs time interval computed using every possible pairing of two points within the light curve of each star. In the figure we show the variability over intervals up to 50 days (calculated with time bins of 1 day), and for intervals up to the full 1600 day baseline of the dataset (using time bins of 30 days). The r.m.s. variability over short timescales appears to be larger for more evolved objects than flat and class I sources. The variability increases with time for every YSO class and becomes flat at $t\sim250-350$ days, although this is less clear for class III sources because of noise due to the lower number of objects in this class. Class I and flat sources have higher r.m.s. variability on these longer timescales. 

The higher r.m.s variability in class II and III systems on timescales $<$25 days can be explained by the fact that in these more evolved YSOs the stellar photosphere contributes a greater proportion of the K-band luminosity of the system, whereas in less evolved YSOs the luminosity is more dominated by the accretion disc.
Consequently we may expect a greater contribution to variability from cold and hot spots in the photosphere of class II and class III YSOs, which manifests itself on the 
timescale of stellar rotation. We note that variability on rotational timescales of a few days also contributes to the measured mean r.m.s. variability shown in Fig. \ref{vvv:gc2} on all longer timescales, which is why the variation increases rapidly on baselines from zero to 3 days \citep[a typical rotation timescale in YSOs, e.g][]{2010Alencar} and then increases more slowly thereafter.


It is also very interesting to see that the 250-350 day timescale at which the maximum of mean r.m.s. variability is reached in all classes of YSO corresponds to variablity on spatial scales of 1-2 au, assuming the timescale is determined by Keplerian rotation about low to intermediate mass YSOs.  \citet{2014Connelley} found from a spectroscopic variability study that mass accretion tracers in their sample of class I YSOs, such as Br$\gamma$ and CO emission, are highly variable over timescales of 1-3 yrs and they proposed the above explanation of the timescale.  



Studies of the optical, near infrared and mid-infrared temporal behaviour of YSOs have shown that  the great majority are variable at these wavelengths, with their light curves showing a diversity of amplitudes, timescales and morphologies \citep[see e.g.][]{2013Findeisen, 2014Cody, 2014Rebull, 2015Rice, 2015Wolk}. These studies, and the earlier large scale study of \citet{2012Megeath} showed that the amplitude of the variability increases for younger embedded objects, though these works contained few if 
any YSOs with $\Delta K_{\rm s}>1$. \citet{2012Rice} found indications that amplitudes $\Delta K > 1$~mag are more common
amongst class I systems ($13\pm7\%$, based on 2 high amplitude objects in a sample of 30 in the Braid Nebula within Cygnus OB7), 
whereas such high amplitudes are found to be less common in more evolved YSOs \citep[see e.g.][]{2001Carpenter}.


\subsection{Eruptive variability}\label{vvv:sec_erupvars}

We have found from the morphological classification that 106 of our SFR-associated high amplitude variables have light curves that show sudden and large increases in brightness. Our near infrared colour variability data, though limited,
appears to verify that variability does not arise from changes in the extinction along the line of sight in most cases. As we have discussed previously, large magnitude changes in our sample are more likely explained by either changes in the accretion rate or in the extinction along the line of sight. Since we appear to discard the latter effect in our eruptive variables, we infer that large changes in the accretion rate are the most likely explanation for the observed variability in these objects.

In Fig. \ref{vvv:gc2} (middle panels) we show histograms of the spectral index of the YSOs of each light curve type. 
We also see that the YSOs classified as eruptive have larger values of $\alpha$ (i.e. redder SEDs) than the other categories
of YSO variables. The redder SEDs of the eruptive objects supports the idea that fluctuations in the accretion rates are larger and much more common at early stages of PMS evolution than in the class II stage.  This is especially true given that class II YSOs typically outnumber class I YSOs in nearby SFRs by a factor of $\sim 3.7$-4.8 \citep[see e.g. table 1 in][]{2014Dunham} due to the greater duration of the class II stage. We have 70
eruptive variables classified as class I YSOs and 5 classified as class II YSOs (the remainder being flat spectrum or class III
systems). If we assume that the YSOs classified as eruptive are mainly genuine eruptive variables (which is supported by our
spectroscopic follow up in paper II) this tells us that the incidence of eruptive variability is $\sim$50-70 times higher in 
class I YSOs than class II YSOs. If we consider the possible correction to $\alpha$ due to foreground extinction (see \ref{vvv:sec_alphaclass}), then we have
52 class I YSOs and 15 class II YSOs in the eruptive category, so eruptive variability is still 13 to 17
times more common in class I YSOs. We conclude that the
difference is at least an order of magnitude.

It is interesting to see that some of the theoretical models that explain the outbursts observed in young stars, predict that luminosity bursts are more common in the class I stage compared to later stages of stellar evolution. These models mainly involve gravitational instability (GI) in the outer disc, which is most likely to occur during the embedded phase. GI can cause the outer disc to fragment, forming bound fragments that later migrate into the inner disc. The infall of these fragments can lead to mass accretion bursts. GI can also produce a persistent spiral structure which efficiently transfers mass to small radii. The continuous pile up of mass at lower radii can trigger magneto-rotational instabilities (MRI), which lead to sudden disc outbursts \citep[see e.g.,][ and references therein]{2009Zhu,2014Audard, 2015Vorobyov}. Thus invoking these mechanisms might explain the higher occurrence of eruptive variables at younger stages. However, we note that most mechanisms that explain eruptive variability, such as bursts due to binary interaction \citep{1992Bonnell}, MRI activated by layered accretion \citep{2009Zhu} or thermal instabilities \citep{1996Hartmann}, predict outbursts during class I through class II stages of YSO evolution.



The classification as eruptive variables only comes from $K_{\rm s}$ light curves, and spectroscopic follow up is needed to confirm their YSO nature. However, we note that potentially adding 106 more objects to the YSO eruptive variable class would increase the known members by a factor of five. Moreover, our survey covers just a portion of the Galactic plane. Therefore eruptive variability in YSOs might be more common than previously thought, especially for the most embedded and young objects. Spectroscopic follow up of a subsample of the objects shows that a large fraction of them are indeed eruptive YSOs (see Paper II).





\section{Incidence of eruptive variability}

To determine whether episodic accretion plays an important role in the assembly of stars it is essential to have an estimate of the incidence of the phenomenon. \citet{1996Hartmann} estimate that stars must spend $\sim 5\%$ of their lifetime in high states of accretion to gain their final mass during the infall phase. \citet{2009Enoch} and \citet{2009Evans} find that 5-7$\%$ of class I stars in their sample are at high accretion states. In this section we attempt to derive an initial estimate of this number using our sample of eruptive variables from VVV and the intrinsically red {\it Spitzer} sources of \citet{2008Robitaille}.

We investigated the intrinsically red {\it Spitzer} sources of \citet{2008Robitaille}, specifically those objects classified by the authors as likely YSOs. The use of GLIMPSE and MIPSGAL \citep{2009Carey} photometry allows us to study the SEDs of this sample, where we find 2059 class I YSOs in the Galactic disc area studied in our work. From these objects, 51 are found in our list of high amplitude variables in SFRs, 26 of them having the eruptive variable classification. These numbers would imply that approximately 2.5$\%$ of class I YSOs show large amplitude variability, and that eruptive variability is observed in 1.3$\%$ of YSOs at this evolutionary stage.

Once again we need to take into account the completeness of our sample due to our selection criteria. In Sect. \ref{sec:vvvselec} we showed that our strict selection criteria caused our sample of high amplitude variables to be only $\sim50\%$ complete, down to $K_{\rm s}=15.5$~mag. In Sect. \ref{sec:vvvselec} we noted that at magnitudes fainter than $K_{\rm s}=15.5$~mag the completeness of our selection falls more steeply. Thus the incompleteness factor in the \citeauthor{2008Robitaille} sample is likely to be higher.

We studied the completeness of the \citeauthor{2008Robitaille} sample in four widely separated VVV tiles (d052, d065, d072 and d109) using 2010-2014 $K_{\rm s}$ data, where we found 138 counterparts in the VVV catalogues. If we only consider 2010-2012 data, 22 objects show high amplitude variability \footnote{The true variability of these objects was confirmed via visual inspection of 1\arcmin$\times$1\arcmin images.}, are located in SFRs, and have light curves that do not resemble those of AGB stars. From these, 8 are part of the list of high amplitude variables presented in this work, which implies an incompleteness factor of 22/8 or 2.75, caused by the quality cuts that were required in Sect. \ref{sec:vvvselec} to reduce the number of false positives to a manageable level. This figure is higher than the factor of $\sim$2 (50\% completeness) quoted above because many of the class I YSO variables from the Robitaille et al.(2008) list have mean $K_{\rm s}>$15.5 magnitudes. If we consider the whole period of 2010-2014, then the number of high amplitude variables increases to 29 raising the incidence by a further factor of 1.32.

We should also consider whether there is a bias towards erupting systems in the
\citet{2008Robitaille} sample, which could in general be detected at greater distances,
or lower down the mass function, than quiescent systems. Such a bias would only affect our
calculation if erupting YSOs detected by GLIMPSE in $\sim$2003 also erupted
a second time in 2010-2014. We tested for this by comparing the GLIMPSE $I2$ (4.5~$\mu$m)
and WISE $W2$ (4.6~$\mu$m) magnitudes of the 26 systems from the \citet{2008Robitaille} sample
that are included in our list of eruptive YSOs. $W2$ and $I2$ magnitudes for YSOs are
typically similar \citep[e.g.][]{2014Antoniucci}. Considering the 20/26 of these
systems with $W2$ detections in the WISE AllSky catalogue (data from 2010, typically
before the eruption) the median $I2-W2$ is 0.02 mag. If we transform the {\it Spitzer} $I2$ magnitudes
to the $W2$ passband using the equation given in \citet{2014Antoniucci} then the median
difference between the {\it Spitzer} and WISE data is $-$0.14 mag. This small number indicates that
any luminosity bias toward eruptive systems can be neglected in this initial estimate of their
incidence.

After correcting for incompleteness, the incidence of high amplitude variability among class I YSOs rises to 6.8$\%$, including only stars that varied in the 2010-2012 time period. This rises to 9\% if we extend the time period to 2010-2014. More importantly the incidence of eruptive variables amongst class I YSOs reaches 3.4$\%$(or 4.6$\%$ in the 2010-2014 data). The 4.6$\%$ figure has a statistical uncertainty of 40$\%$, not including any biases arising from our use of the sample of \citet{2008Robitaille}, so we should express the incidence of eruptive variability as about 3 to 6$\%$ over a 4 year timescale. These figures happen to agree with the incidence of outbursts inferred by \citet{2009Enoch} and \citet{2009Evans} from the observation of class I YSOs with high bolometric luminosities. However, their observations are perhaps more likely to trace long-duration, FUor-like outbursts than those detected by VVV.

If we assume that all class I YSOs go through episodes of enhanced accretion at the same average rate, then our estimate that $\sim4\%$ of them burst over a 4 year period, would imply that roughly every source suffers at least one burst over 100 yr. \citet{2016Froebrich} study the distribution of the separation between large H$_{2}$ knots in jets. The knots likely trace the accretion burst history of a particular source. \citeauthor{2016Froebrich} find that the bright knots have typical separations that correspond to about 1000 yr, which they conclude is too large to be EXor driven and too small to be FUor driven. This number is also larger than the 100 yr burst frequency of the VVV sample. However, further study needs to be done to obtain a more robust estimate, and the assumption of similar behaviour for all class I YSOs is of course questionable.

\section{Summary and Conclusions}

We have searched for high amplitude infrared variables in a 119 deg$^2$ area of the Galactic midplane covered by the VVV survey, using a method that tends to exclude transients and eruptive variables that saturated during outburst or very faint in 
quiescence, owing to our requirement for a high quality detection at every epoch. We discovered 816 bona fide variables in the 
2010-2012 data with $\Delta K_{\rm s}>1$ in that time interval. Nearly all of these were previously unknown as variable stars, though 
a significant minority had been identified as embedded YSO candidates in the Robitaille et al.(2008) catalogue of stars with 
very red [4.5]-[8.0] {\it Spitzer}/GLIMPSE colours.

We have presented a fairly simple analysis of the sample using the 2010-2014 VVV light curves, supplemented by a recently 
obtained 2nd epoch of multi-filter $JHK_{\rm s}$ data and photometry from WISE, and {\it Spitzer}. Our 
main conclusions are as follows.

\begin{itemize}
\item In agreement with the previous results from searches in the UKIDSS GPS \citep{2014Contreras}, we observe a strong 
concentration of high-amplitude infrared variables towards areas of star formation. The two point correlation function and nearest neighbour 
distribution of VVV objects show evidence for clustering on angular scales typical of distant Galactic clusters and SFRs.
The variable stars found in SFRs are characterized by having near-infrared colours and SEDs of YSOs. 

\item The most common types of variable outside SFRs are LPVs (typically dust-obscured Mira variables because other types are 
saturated in VVV) and EBs. By visual inspection of the light curves of the variables in SFRs we were able to identify and remove 
most of these contaminating systems and provide a reasonably clean sample of high amplitude YSOs. The YSOs make up about half of 
the full sample of 816 variables.

\item We analysed the light curves of the variables in SFRs, after removal of likely Mira-types, and we classify them
as 106 eruptive variables, 39 faders, 45 dippers, 162 short term variables, 65 long-term periodic variables 
($P>100$~days) and 24 eclipsing binaries (EBs) and EB candidates. Individual YSOs may display more than 1 type of variability
and the low amplitude variation on short timescales seen in normal YSOs is common in every category.

\item Spectroscopic follow up of a substantial subset of the variables with eruptive light curves is presented in the 
companion paper (Paper II), confirming that the great majority of systems with eruptive light curves are indeed eruptive 
variables with signatures of strong accretion similar to those seen in EXors or FUors, or a mixture of the two. The 2 epochs 
of VVV $JHK_{\rm s}$ multi-colour data indicate that extinction is not the main cause of the variability in systems with
eruptive light curves, though there is a trend for the sources to become bluer when brighter, similar to EXors. The faders 
show a wider range of colour behaviour, with more examples of ``bluer when fainter''  than the other categories. 

\item Unsurprisingly, very few of the EBs showed significant magnitude or colour changes between the two multi-colour 
epochs. Amongst the STVs and dippers, we observe large colour and magnitude changes that appear to differ from the reddening vector, although this conclusion seems to depend on the selection of objects from these classes. We observed changes consistent with the reddening vector in a few
systems. It is possible that the flux changes in some of the STVs are caused by extinction, e.g. as has been observed in some 
binary YSOs with a circumbinary disc \citep{2014Windemuth, 2015Rice}.

\item The STVs typically have lower amplitudes than any other category except the EBs and EB candidates. Moreover, the
STVs have a bluer distribution of spectral indices than the full YSO sample, including a high proportion of class II YSOs
with relatively low extinction. It is likely that in many of the STVs with periods P$<15$~days and $K_{\rm s}$ amplitudes 
not far above 1 magnitude the light curve is rotationally modulated by unusually prominent bright or dark spots on the 
photosphere.

\item Variables in the eruptive and fader categories tend to have higher amplitudes than the full YSO sample over the 4 year
period, with mean $\Delta K_{\rm s}$ = 1.72 and 1.95~mag, respectively, compared to 1.56~mag for the full YSO sample.

\item It is reasonable to suppose that variable accretion is the cause of photometric variability in a proportion of the 
faders and some of the long-term periodic variables and STVs, although spectroscopic confirmation is limited in these categories as yet. In the periodic variables the accretion rate would presumably be modulated by a companion body \citep{2012Hodapp}. Adding these together with the $\sim$100 YSOs with eruptive light curves suggests that the sample contains 
between 100 and 200 YSOs in which the variability is caused by large changes in accretion. If we take the lower end of this 
range, this increases the number of probable eruptive variable YSOs available for study by a factor of 5. Some of these systems 
have $K_{\rm s}$ amplitudes not far above the 1 mag level, below which variability due to spots and smaller changes in accretion rate and 
extinction becomes much more common. However, we see no clear argument for a higher threshold given that the amplitudes have a 
continuous distribution that can be be influenced in individual YSOs by the mass and luminosity of the central protostar as much 
as the change in accretion rate \citep{1991Calvet}.

\item As a whole, the high amplitude variables are a very red sample, dominated by embedded YSOs (i.e. systems with Class
I or flat spectrum SEDs that are typically not observable at optical wavelengths). This contrasts with the optical selection
of the classical FUors and EXors, the majority of which have a class II or flat spectrum classification. While FUors are
often discussed as systems with a remnant envelope, only 3 or 4 deeply embedded eruptive variables were known prior to
this study and our recent UGPS 2-epoch study.

\item Variables with eruptive light curves tend to have the reddest SEDs. The spectral indices in the eruptive category
indicate that this type of variability is at least one order of magnitude more common among class I YSOs than class II YSOs. This demonstrates that eruptive variability is either much more common or recurs much more frequently amongst YSOs at earlier 
stages of pre-MS evolution, when average accretion rates are higher. We hope this result will inform ongoing efforts to develop 
a theoretical framework for the phenomenon. 

\item For the full sample of YSOs, the r.m.s. variability is higher at earlier evolutionary (SED) classes for time intervals
longer than 25~days, and reaches a maximum at 250-350 days for all SED classes. The full duration of the outbursts
in eruptive systems is typically 1 to 4 years. Some variables with eruptive light curves show more than 1 outburst. If some of the faders are eruptive variables in decline 
then a similar or slightly longer duration would apply. 

\item At time intervals shorter than 25 days the evolutionary dependence is reversed, with class II YSOs showing higher 
amplitudes than flat spectrum and class I YSOs. This suggests that the shorter timescale variations are dominated by rotational 
modulation by spots on the photosphere (which are more readily observed in class II systems than embedded YSOs) whereas 
accretion variations usually take place on longer timescales.

\item The 1 to 4 year duration of the eruptions is between that of FUors ($>10$ years) and EXors (weeks to months)
A small but growing number of eruptive YSOs with these intermediate durations have been found in recent years, some of
which have a mixture of the spectroscopic characteristics of FUors and EXors. This has led to the recent concept that FUor and
EXors may simply be part of a range of different eruptive behaviours on different timescales, all driven by large
variations in accretion rate. Until now it was unclear whether these recent discoveries were rare exceptions, but it now seems clear that they are not. In fact we find that YSOs with intermediate outburst durations outnumber short 
EXor-like outbursts and are now the majority of known eruptive systems, A much longer duration survey would be required to 
determine the incidence of FUor-like outbursts amongst embedded systems. In paper II we propose a new class of eruptive variable
to describe YSOs with eruptive outbursts of intermediate duration, which are usually optically obscured
class I or flat spectrum YSOs and display a variety of the EXor-like and/or FUor-like spectroscopic
signatures of strong accretion.

\item We investigated the intrinsically red {\it Spitzer} sources of \citet{2008Robitaille}, specifically those objects classified by the authors as likely YSOs. We find 2059 such objects in the area covered by VVV. From these 51 are found in the list of high amplitude variables in SFRs from this work, with 26 of them being classified as eruptive variables. After correcting for incompleteness imposed by the strict selection criteria
in the VVV sample, we estimate that high amplitude variability is observed from 6.8$\%$ (when considering 2010-2012 data) up to 9$\%$ (when considering 2010-2014 data) of class I YSOs. More importantly, the incidence of eruptive variability amongst class I YSOs rises to 3.4--4.6$\%$. This estimate agrees with those inferred from observations of class I YSOs by \citet{2009Evans} and \citet{2009Enoch}. However, the agreement might be a coincidence considering that their observations are likely tracing long-duration, FUor-like outbursts, rather than those detected in our study.

\item YSOs are the commonest type of high amplitude infrared variable detected by the VVV survey. We estimate a 
completeness-corrected source density of 7 deg$^{-2}$ in the mid-plane of quadrant 4, in the approximate mean magnitude
range $11<K_{\rm s}<16$. These YSOs are detected at typical distances of a few kpc, there being no very nearby
SFRs in the area surveyed; they are therefore likely to be intermediate-mass YSOs. If we were able to detect such objects out to the far edge of the Galactic disc the source density 
would rise to perhaps $\sim 40$ deg$^{-2}$. This confirms our previous suggestion in \citet{2014Contreras} that 
high amplitude YSO variables have a higher source density and average space density than Mira variables. EBs are very common
at low amplitudes and they may have a comparable space density to YSOs at $K_{\rm s}$ amplitudes of 1 to 1.6 mag (an 
approximate upper limit for EBs in {\it Kepler}). YSOs are more numerous at higher amplitudes and may well be more 
numerous for all amplitudes over 1~mag if the eruptive phenomenon extends to the lower part of the stellar 
Initial Mass Function.

\end{itemize}

\section*{Acknowledgments}

This work was supported by the UK's Science and Technology Facilities Council, grant numbers ST/J001333/1, ST/M001008/1 and ST/L001403/1.

We gratefully acknowledge the use of data from the ESO
Public Survey program 179.B-2002 taken with the VISTA 4.1m telescope and
data products from the Cambridge Astronomical Survey Unit. Support for
DM, and CC  is provided by the Ministry of Economy, Development, and 
Tourism’s Millennium Science Initiative through grant IC120009, awarded to
the Millennium Institute of Astrophysics, MAS. DM is also supported by the Center for 
Astrophysics and Associated Technologies PFB-06, and Fondecyt Project No. 1130196.
This research has made use of the SIMBAD database, operated at CDS, Strasbourg, France; 
also the SAO/NASA Astrophysics data (ADS). A.C.G. was supported by the Science Foundation of Ireland, grant 13/ERC/I2907. We also acknowledge the support of CONICYT REDES project No. 140042 ``Young variables and proper motion in the Galactic plane. Valparaiso-Hertfordshire collaboration''

C. Contreras Pe\~{n}a was supported by a University of Hertfordshire PhD studentship in the earlier stages of this research.

We thank Janet Drew for her helpful comments on the structure of the paper.

\bibliographystyle{mn2e}
\bibliography{ref.bib}

\appendix

\section{Eclipsing Binaries}\label{appen:EBs}

In order to compare the YSO and EB space densities, we look at the measured source density
of EBs detected in Galactic disc fields by OGLE-III \citep{2013Pietrukowicz} and estimate their distances with the help of a recent 
analysis of the physical properties of {\it Kepler} eclipsing binaries \citep{2014Armstrong}.


Shallow surveys such as the Automated All-Sky Survey \citep[ASAS][]{2006Paczynski} are sensitive only to the more luminous high amplitude 
EBs, which are mainly early type Algol systems, similar to the high amplitude EBs in the General Catalogue of Variable Stars (Samus et 
al.2010). However, simulations by \citet{2004Lopezmor} find that EBs seen by deeper surveys such as OGLE-III will be dominated by later 
type systems. \citet{2014Armstrong} 
analysed the large sample of EBs discovered by {\it Kepler}, providing temperature and radius estimates for the full sample. They note 
that their results are most accurate for cases where the temperatures of the primary and the secondary are very different, which is 
fortunate because this is characteristic of high amplitude EBs, in which a hotter main sequence star is eclipsed by a cooler giant star.
Inspection of their results shows that high amplitude EBs fall into 2 groups: EBs with early type primaries and EBs with F or early 
G-type primaries. Simple calculations assuming Planckian SEDs indicate that only the latter group can produce eclipses deeper than 1 
magnitude in the $K_{\rm s}$ passband (and this group also dominates in the OGLE $I$ passband) because early type stars emit little of 
their total flux in the infrared. The same calculations indicate that no systems with $K_{\rm s}$ amplitudes above 1.6 mag exist
in the {\it Kepler} sample, whereas YSOs with higher amplitudes are common in our sample, see Sect. \ref{vvv:sec_physmec}. Applying this approach in the
optical indicates that EBs should exist with amplitudes up to 3 mag in the $I$ passband. This limit is confirmed by the OGLE-III sample of 
\citet{2013Pietrukowicz}, thereby giving confidence that our limit of 1.6~mag at $K_{\rm s}$ is robust. Indeed our VVV sample of 72 EBs contains no systems with higher amplitude
than this. We can assume that the EBs 
from \citet{2013Pietrukowicz} are mainly composed of F or early G-type stars, based on the {\it Kepler} results, together with the fact that 
later type primaries are not expected given the need for these systems to have produced a post-main sequence star within the lifetime of the 
Galactic disc.
\citet{2013Pietrukowicz} reports the discovery of 11589 EBs in area of 7.21 deg$^{2}$ across different fields on the Galactic midplane. The
catalogue of EBs is reported to be 75$\%$ complete to $I$=18.
Correcting for the 75\% completeness indicates a source density of 2143 EBs deg$^{-2}$, from which only $\sim1.6\%$ display 
$\Delta I > 1$~magnitude. Our earlier calculation (assuming Planckian SEDs) from the results of \citet{2014Armstrong} suggests that this drops to 0.6\% at $K_{\rm s}$. 
This implies $\sim$12 high amplitude EBs per deg$^{2}$. 
We can estimate the extinction to the EBs from their observed $V-I$ colours and then roughly estimate their typical distances either using the absolute I magnitudes of F to early G-stars or by converting
extinction to distance using the red clump giant branch (as in Sect. \ref{vvv:sec_alphaclass}). Both approaches indicate that the 
EBs from \citet{2013Pietrukowicz} are typically at heliocentric distances of a few kpc, which is similar to the estimated distances to our 
YSOs based on radial velocities, literature distances to the SFRs and SED fitting (see paper II). Given that the surface 
density of EBs ($\sim$12 deg$^{-2}$) is similar to
the surface density of the VVV YSO population sampled in this study ($\sim$7 deg$^{-2}$), the similar distances suggest that the high amplitude 
EB and YSO populations may have similar average space densities, within an order of magnitude. The YSO population is certainly more
numerous at $K_{\rm s}$ amplitudes over 1.6~mag. If the high amplitude YSO population 
extends well below a solar mass, which is very possible, then they would probably be more numerous even at our $\Delta K_{\rm s}$=1~mag 
threshold. However, the VVV and OGLE-III data tell us only that both populations are substantial. The number of YSOs rises steeply up to the 
sensitivity limit and a similar trend is obtained in EBs if we assume that a significant fraction of the unclassified variables shown in 
Fig. \ref{vvv:nonsfrs_prop} are EBs.

\section{Mid-infrared colour and magnitude variability}\label{apenn:mid-ir}

Photometry from the WISE and NEOWISE missions in the W1 (3.4 $\mu$m) and W2 
(4.6 $\mu$m) passbands are available at 4 epochs for the whole sky, with 
simultaneous photometry in the two filters. The WISE satellite scanned the 
whole sky twice in 2010, at epochs separated by 6 months, and the extended 
mission, NEOWISE, has repeated the process in 2014. For any given sky location, 
each epoch is composed of multiple scans taken over a period of several
days. For all the high amplitude variables, we downloaded photometry for all 
these scans from the AllWISE Multiepoch Photometry table (for the 2010 data) 
and NEOWISE-R Single Exposure L1b Source Table (for the 2014 data), both of
which are archived in IRSA.

Inspection of the W1 and W2 photometry indicated that the uncertainties 
typically become larger for saturated stars and for faint stars in crowded 
Galactic fields. For this analysis we therefore considered only sources with 
$7 < W1 < 11$, and $6 < W2 < 11$ (defining these cuts with the AllWISE Source 
Catalog). We combined the data from the multiple scans into the 4 widely 
separated epochs by binning the scans into groups (epochs) with a maximum
baseline of 20 days and computing the medians of $W1$, $W2$
and $(W1-W2)$ at each epoch.

In most cases the WISE data do not sample the full peak to trough variation
in the VVV light curves, so changes in W1 and W2 fluxes are often small. 
We define $\Delta W1$, $\Delta W2$ and $\Delta (W1-W2)$ for each source as 
the largest difference observed at any two epochs (not necessarily the same
pair of epochs for each quantity). Then considering each type of VVV light 
curve category, we present the median of these quantities
in Table \ref{table:wise}. We see that both magnitude and colour variability are typically 
larger for faders than the other categories, as is the case in $K_{\rm s}$. 
Eruptive sources, dippers and lpv-YSOs have similar magnitude changes, while 
STVs have smaller changes (as in $K_{\rm s}$). EBs typically show negligible
changes since the eclipses are very rarely sampled.

\begin{table}
\begin{center}
\begin{tabular}{@{}l@{\hspace{0.15cm}}c@{\hspace{0.2cm}}c@{\hspace{0.2cm}}c@{\hspace{0.2cm}}c@{\hspace{0.2cm}}}
Category & Median & Median & Median & No. in \\
	 & $\Delta W1$ & $\Delta W2$ & $\Delta (W1-W2)$ & sample\\ 
\hline
Faders  &   1.33   & 0.93 & 0.37 & 12  \\
Eruptive &  0.72   & 0.56 & 0.26 & 21  \\
LPV-YSOs &  0.62   & 0.67 & 0.13 & 23   \\
STVs &      0.50   & 0.38 & 0.10 & 11  \\
Dippers &   0.59 & 0.62 & 0.27 & 9   \\
\end{tabular}
\caption{Colour and megnitude changes measured by WISE.}\label{table:wise}
\end{center}
\end{table}

The WISE data are less useful than the $JHK_{\rm s}$ data for investigating the
physical cause of the photometric variability in $K_{\rm s}$ because the 
extinction vs wavelength relation appears to depend strongly on environment
\citep{2014Koenig}. Also, extinction is lower in the 
WISE passbands than at $K_{\rm s}$ so extinction may have less effect on 
the $W1-W2$ colour than even a modest change in accretion rate. 

In most STVs and LPV-YSOs we can confidently state the extinction is not the 
main cause of the measured W1-W2 colour changes because the trend over the 4 
epochs is "bluer when fainter", or a negligible colour change, or there is no 
clear trend. In the eruptive and fader categories about half the variables 
(10/21 and 7/12) have a "redder when fainter" trend that might be due to 
variable accretion, variable extinction, or a combination of the two (in some 
sources, 1 of the 4 epochs does not follow the trend of the other 3 epochs). 
The remainder of the  eruptive variables and faders show a different trend or 
no clear result. In some sources with a "redder when fainter" trend in the 
WISE data, the 2 epochs of $JHK_{\rm s}$ data appear to rule out extinction 
as the cause of variability. In dippers, the majority (5/9) sources have a 
"redder when fainter" trend that might be consistent with extinction. 

These results are not very informative because of the limited sampling of
the WISE data. In most cases $K_{\rm s}-W1$, $K_{\rm s}-W2$ changes cannot be
investigated because there are very few sources for which two of the WISE 
epochs are contemporaneous with VVV $K_{\rm s}$ epochs. 

Fortunately, there are some sources with clear long term trends in the 2010-2014 
VVV light curves for which we can usefully compare WISE data from 2010 and 2014.
Four examples of such sources are VVVv270, VVVv631, VVV118 and VVVv562, all of 
which are members of the spectroscopic sub-sample that are discussed individually 
in Paper II. To bring together those sources here, we note that
all four have unusually large amplitudes in $K_{\rm s}$, $W1$
and $W2$ ($2.6<\Delta K_{\rm s}<4.2$~mag). The first two show a rising trend (classed 
as eruptive light curves) whereas the other two show a declining trend: VVVv562 is a 
fader but VVVv118 is an eruptive variable with several brief eruptions superimposed on a 
long-term decline. In every case the trend is similar in the 3 wavelengths but the 
amplitude is larger in $K_{\rm s}$ than in $W1$ and $W2$, typically by a factor 
between 1 and 2. Three sources have $\Delta W1 \approx \Delta W2$ (VVVv118, VVVv562, 
VVVv631), whereas in VVVv270 $\Delta W1 \approx 2 \Delta W2$.

The spectroscopic evidence in paper II indicates that these four variables are all 
bona fide eruptive variables driven by episodic accretion, so the different amplitudes 
in the different filters presumably reflect differing changes in the luminosity of the 
different regions of the accretion disc responsible for most of the emission at each 
wavelength, as well as the effect of temperature changes on the flux emitted 
by each region at different wavelengths.


\section{Table 2}

In here we present the full version of Table \ref{table:vvvpar}. 

\onecolumn

\begin{landscape}
\clearpage

\pagestyle{empty}

\setlength\LTleft{0pt}
\setlength\LTright{0pt}
\setlength\topmargin{10pt}
\setlength\textwidth{702pt}
\setlength\textheight{38pc}


\begin{longtable}{@{}l@{\hspace{0.15cm}}c@{\hspace{0.15cm}}c@{\hspace{0.15cm}}c@{\hspace{0.15cm}}c@{\hspace{0.15cm}}c@{\hspace{0.15cm}}c@{\hspace{0.15cm}}c@{\hspace{0.15cm}}c@{\hspace{0.15cm}}c@{\hspace{0.15cm}}c@{\hspace{0.15cm}}c@{\hspace{0.15cm}}c@{\hspace{0.15cm}}c@{\hspace{0.15cm}}c@{\hspace{0.15cm}}c@{\hspace{0.15cm}}c@{\hspace{0.15cm}}c@{\hspace{0.15cm}}c@{\hspace{0.15cm}}l@{\hspace{0.15cm}}c@{}}
\caption{Parameters of the high-amplitude variables from VVV.}\\
\hline
Object ID & VVV Designation        & $\alpha$  & $\delta$  & $l$ & $b$ & $Z$ & $Z_{err}$ & $Y$ & $Y_{err}$ &  $J$ & $J_{err}$ & $H$ & $H_{err}$ & $K_{\rm s}$ & $K_{\rm s,err}$ &  $\Delta K_{\rm s}$ & $\alpha_{class}$ & SFR & Class & Period\\
 &  & (J2000) & (J2000) & (degrees) & (degrees) & (mag) & (mag) & (mag) & (mag) & (mag) & (mag) & (mag) & (mag) & (mag) & (mag) & (mag) & & & & (days)\\
\hline
\endfirsthead
\multicolumn{21}{c}{\tablename\ \thetable\ -- \textit{Continued from previous page}} \\
\hline
Object ID & VVV Designation        & $\alpha$  & $\delta$  & $l$ & $b$ & $Z$ & $Z_{err}$ & $Y$ & $Y_{err}$ &  $J$ & $J_{err}$ & $H$ & $H_{err}$ & $K_{\rm s}$ & $K_{\rm s,err}$ &  $\Delta K_{\rm s}$ & $\alpha_{class}$ & SFR & Class & Period\\
 &  & (J2000) & (J2000) & (degrees) & (degrees) & (mag) & (mag) & (mag) & (mag) & (mag) & (mag) & (mag) & (mag) & (mag) & (mag) & (mag) & & & & (days) \\
\hline
\endhead
\hline 
\multicolumn{21}{r}{\textit{Continued on next page}} \\
\endfoot
\hline
\endlastfoot
VVVv1 & VVV J114135.16-622055.51 & 11:41:35.16 & -62:20:55.51  &  294.92603 & -0.56770  &  -- & -- & -- & -- & 17.99 & 0.06 & 15.95 & 0.02 & 14.44 & 0.01 & 1.34 & -0.29 & y & STV &   72\\ 
VVVv2 & VVV J114412.94-623449.09 & 11:44:12.94 & -62:34:49.09  &  295.28005 & -0.71146  &  -- & -- & -- & -- & -- & -- & 18.78 & 0.23 & 15.71 & 0.03 & 2.52 &  1.22 & y & STV & --\\ 
VVVv3 & VVV J115113.03-623729.29 & 11:51:13.03 & -62:37:29.29  &  296.07199 & -0.55784  &  13.17 & 0.01 & 12.93 & 0.01 & 12.90 & 0.01 & 12.70 & 0.01 & 12.24 & 0.01 & 2.21 & -- & n & Known & --\\ 
VVVv4 & VVV J115808.69-630708.60 & 11:58:08.69 & -63:07:08.60  &  296.95057 & -0.86785  &  -- & -- & -- & -- & 18.23 & 0.08 & 16.62 & 0.04 & 15.32 & 0.02 & 1.10 & -0.24 & y & STV & --\\ 
VVVv5 & VVV J115959.68-622613.20 & 11:59:59.68 & -62:26:13.20  &  297.02026 & -0.15716  &  17.69 & 0.02 & 16.62 & 0.01 & 15.87 & 0.01 & 15.25 & 0.01 & 13.53 & 0.01 & 1.30 & -- & n & LPV & --\\ 
VVVv6 & VVV J115937.81-631109.77 & 11:59:37.81 & -63:11:09.77  &  297.12836 & -0.89953  &  19.02 & 0.05 & 18.08 & 0.04 & 16.80 & 0.02 & 15.95 & 0.02 & 15.50 & 0.02 & 1.02 & -- & n & EB &    1.64\\ 
VVVv7 & VVV J120202.67-623615.60 & 12:02:02.67 & -62:36:15.60  &  297.28472 & -0.27538  &  -- & -- & -- & -- & -- & -- & -- & -- & 17.22 & 0.12 & 1.60 &  2.39 & y & Eruptive & --\\ 
VVVv8 & VVV J120059.11-631636.18 & 12:00:59.11 & -63:16:36.18  &  297.29582 & -0.95838  &  -- & -- & -- & -- & -- & -- & -- & -- & 16.86 & 0.09 & 1.38 &  0.64 & y & Eruptive & --\\
\hline
\end{longtable}

\end{landscape}
\label{lastpage}
\end{document}